\documentclass[twocolumn]{aastex631}
\usepackage{showyourwork}
\usepackage{amsmath}
\usepackage{listings}
\usepackage{xcolor}

\definecolor{codegreen}{rgb}{0,0.6,0}
\definecolor{codegray}{rgb}{0.5,0.5,0.5}
\definecolor{codepurple}{rgb}{0.58,0,0.82}
\definecolor{backcolour}{rgb}{0.95,0.95,0.92}

\defcitealias{Lohner-Bottcher2019}{L19}
\defcitealias{Lohner-Bottcher2018}{L18}

\newcommand{\ms}{{\rm m\ s}^{-1}}
\newcommand{\cms}{\ {\rm cm\ s}^{-1}}

\newcommand{\grass}{\texttt{GRASS}}
\newcommand{\PSUAA}{Department of Astronomy \& Astrophysics, 525 Davey Laboratory, The Pennsylvania State University, University Park, PA, 16802, USA}
\newcommand{\PSUCEHW}{Center for Exoplanets and Habitable Worlds, 525 Davey Laboratory, The Pennsylvania State University, University Park, PA, 16802, USA}
\newcommand{\PSETI}{Penn State Extraterrestrial Intelligence Center, 525 Davey Laboratory, The Pennsylvania State University, University Park, PA, 16802, USA}
\newcommand{\PSUStats}{Center for Astrostatistics, 525 Davey Laboratory, The Pennsylvania State University, University Park, PA, 16802, USA}
\newcommand{\PSUICDS}{Institute for Computational and Data Sciences, The Pennsylvania State University, University Park, PA 16802, USA}
\newcommand{\UNIGE}{Observatoire Astronomique de l’Université de Genève, Chemin Pegasi 51, 1290 Versoix, Switzerland}

\definecolor{ebf}{rgb}{0.4, 0.0, 0.6}

\definecolor{mlp}{rgb}{0.294, 0.612, 0.827}

\definecolor{cedit}{rgb}{1.0, 0.25, 0.25}

\usepackage{natbib}
\shorttitle{GRASS II: Simulating Granulation Mitigation}
\shortauthors{M.L. Palumbo III et al.}

\graphicspath{{./}{figures/}}
\begin{document}
\tabletypesize{\small}

\setlength{\textfloatsep}{0.1cm}

\title{GRASS II: Simulations of Potential Granulation Noise Mitigation Methods}

\author[0000-0002-4677-8796]{Michael L. Palumbo III}
\affiliation{\PSUAA}
\affiliation{\PSUCEHW}

\author[0000-0001-6545-639X]{Eric B. Ford}
\affiliation{\PSUAA}
\affiliation{\PSUCEHW}
\affiliation{\PSUICDS}
\affiliation{\PSUStats}

\author[0009-0002-0987-652X]{Elizabeth B. Gonzalez}
\affiliation{\PSUAA}
\affiliation{\PSUCEHW}

\author[0000-0001-6160-5888]{Jason T. Wright}
\affiliation{\PSUAA}
\affiliation{\PSUCEHW}
\affiliation{\PSETI}

\author[0000-0002-3212-5778]{Khaled Al Moulla}
\affiliation{\UNIGE}

\author[0000-0002-7386-8578]{Rolf Schlichenmaier}
\affiliation{Leibniz-Institut f\"ur Sonnenphysik (KIS), Sch\"oneckstr. 6, 79104 Freiburg, Germany}

\correspondingauthor{Michael L. Palumbo III}
\email{palumbo@psu.edu}

\begin{abstract}
We present an updated version of \grass\ (the GRanulation And Spectrum Simulator, \citealt{Palumbo2022}) which now uses an expanded library of 22 solar lines to empirically model time-resolved spectral variations arising from solar granulation. 
We show that our synthesis model accurately reproduces disk-integrated solar line profiles and bisectors, and we quantify the intrinsic granulation-driven radial-velocity (RV) variability for each of the 22 lines studied. We show that summary statistics of bisector shape (e.g., bisector inverse slope) are strongly correlated with the measured anomalous, variability-driven RV at high pixel signal-to-noise ratio (SNR) and spectral resolution. Further, the strength of the correlations vary both line by line and with the summary statistic used. These correlations disappear for individual lines at the typical spectral resolutions and SNRs achieved by current EPRV spectrographs;  so we use simulations from \grass\ to demonstrate that they can, in principle, be recovered by selectively binning lines that are similarly affected by granulation. In the best-case scenario (high SNR and large number of binned lines), we find that a $\lesssim$30$\%$ reduction in the granulation-induced root mean square (RMS) RV can be achieved, but that the achievable reduction in variability is most strongly limited by the spectral resolution of the observing instrument. Based on our simulations, we predict that existing ultra-high-resolution spectrographs, namely ESPRESSO and PEPSI, should be able to resolve convective variability in other, bright stars.
\end{abstract}


\section{Introduction} \label{intro}

Intrinsic stellar variability is one of the chief obstacles limiting the detection of small, rocky planets in the habitable zones of Sun-like stars (i.e., ``Earth twins'') with the radial-velocity (RV) method \citep{Crass2021}. Various strategies have been devised for coping with different sources and manifestations of this intrinsic variability. For stellar pulsations ($p$-modes), \citet{Chaplin2019} devised theoretically-motivated observing strategies designed to mitigate residual $p$-mode amplitudes to the sub-$10\cms$ level; these strategies have since been widely implemented in RV surveys (e.g., \citealt{Blackman2020}, \citealt{Gupta2021}) and explicitly validated by \citet{Gupta2022}. For magnetic activity on rotation timescales, various works have employed Gaussian Processes (GPs; e.g., \citealt{Haywood2014}; \citealt{Rajpaul2015,Rajpaul2016}; \citealt{Gilbertson2020}; and references therein), statistically-motivated methods (e.g., \citealt{CollierCameron2019, Zhao2020}), and neural networks (e.g., \citealt{deBeurs2022, Liang2024}) to disentangle the velocity contributions of spots and faculae from those of planets with the aid of various activity indicators. Additionally, models such as \texttt{SOAP-GPU} \citep{Zhao2023} have been developed to numerically forward model activity-driven variability. However, aside from integrating over several hours, astronomers lack an effective strategy for mitigating the noise introduced by granules that 
has been definitively demonstrated and implemented in extant extremely precise radial velocity (EPRV) surveys \citep{Cegla2019a, Crass2021}. Indeed, two recent works have shown that supergranulation, the manifestation of convection on larger spatial and longer temporal scales than granulation (see reviews such as \citealt{Rieutord2010} and \citealt{Cegla2019b}), is the dominant cause of RV variability during solar minimum \citep{Lakeland2024} and that (super)granulation will preclude the blind detection of Earth-twin exoplanets even if variability from magnetic activity can be perfectly mitigated \citep{Meunier2023}. \par 

As RV spectrographs have begun to achieve instrumental precisions and stabilities at and below the $\sim$1 $\ms$ level, granulation has become a greater cause for concern in the EPRV community. Recognizing that stellar oscillations and granulation would constitute large sources of RV noise problematic for discovery and characterization of Earth-mass planets with the HARPS spectrograph, \citet{Dumusque2011} evaluated various observations strategies in order to determine which ones optimally averaged out noise from these processes. Using simulated RV time series generated from observationally-derived stellar velocity power spectra, \citet{Dumusque2011} determined that averaging multiple observations per night spaced apart by hours more effectively mitigates granulation noise than a single long or multiple consecutive exposures. They conclude, quite optimistically, that ``granulation phenomena and oscillation modes will not prevent us from finding Earth-like planets in habitable regions.'' \par 

Following the \citet{Dumusque2011} study, \citet{Meunier2015} took a differing but complementary approach for assessing the impact of (super)granulation on the measurement of precise RVs. Tiling a stellar hemisphere with simulated granules of varying sizes, intensities, and velocities given relations for these quantities derived from empirical laws and studies of hydrodynamical (HD) simulations (see \S2.2 of \citealt{Meunier2015} and references therein), they allowed the granules to probabilistically evolve in time. Over a simulated 12-year time span, they note a root mean square (RMS) RV of $\sim$0.8 $\ms$ and a photometric RMS of 67 ppm. Simulating observation strategies for mitigating this variability, \citet{Meunier2015} show that the RMS from granulation cannot be sufficiently reduced for averaging timescales commensurate with the granule lifetime ($\sim$10-$15$ minutes in the case of the Sun). After 30 minutes of averaging, they report an RMS RV of about $\sim$50 $\cms$, which they state is in conflict with the claim made by \citet{Dumusque2011} that granulation (but not supergranulation) can be largely averaged out over these timescales. Nevertheless, they do report that performing multiple measurements over the course of a night is more successful in reducing the granulation RMS than back-to-back observations, but not to the degree reported by \citet{Dumusque2011}. \par

The results presented in both \citet{Dumusque2011} and \citet{Meunier2015} consistently suggest that spreading observations out over a night (or over several nights in the case of supergranulation; see \S4 of \citealt{Meunier2015}) more effectively mitigates the impact of granulation than consecutive exposures. However, they disagree somewhat about the absolute effectiveness of their best-case scenario observation strategies, with \citet{Meunier2015} painting a notably more pessimistic picture. Considering more recent results, including \citet{Meunier2023} and \citet{Lakeland2024}, it does appear that \citet{Dumusque2011} underestimated the amplitude and impact of granulation on precise RV measurements. \par 

It is important to note, though, that the methods used for simulating and interpreting stellar velocities in these works differ fundamentally from how velocities are measured in reality. Whereas \citet{Dumusque2011} and \citet{Meunier2015} directly simulate disk-integrated stellar velocities, in practice RVs are inferred from the observed Doppler shifting of absorption lines in stellar spectra. These measured RVs will differ from the somewhat fictitious RVs studied by \citet{Dumusque2011} and \citet{Meunier2015} because lines form over a range of heights in stellar atmospheres, tracing different portions of the full 3D atmospheric velocity field with varying sensitivities as a function of height. \par

To address this complicating reality, a series of papers \citep{Cegla2013, Cegla2018, Cegla2019a} used 3D magnetohydrodynamic (MHD) simulations coupled with detailed radiative transfer modeling to faithfully reproduce stellar magnetoconvection and absorption profiles for the Fe \textsc{I} 6302 \AA\ line (see also \citealt{Cegla2019b} for a detailed review). Through this series of papers, the authors show that perturbations in the shapes of their model disk-integrated line profiles encode information about spurious Doppler shifts from granulation. More specifically, they find strong correlations between the apparent RV of the line studied and various summary statistics designed to describe the bisector curve, including, e.g., bisector inverse slope (BIS, \citealt{Queloz2001}). \par  

\begin{figure*}[!htb]
    \epsscale{1.155}
    \plotone{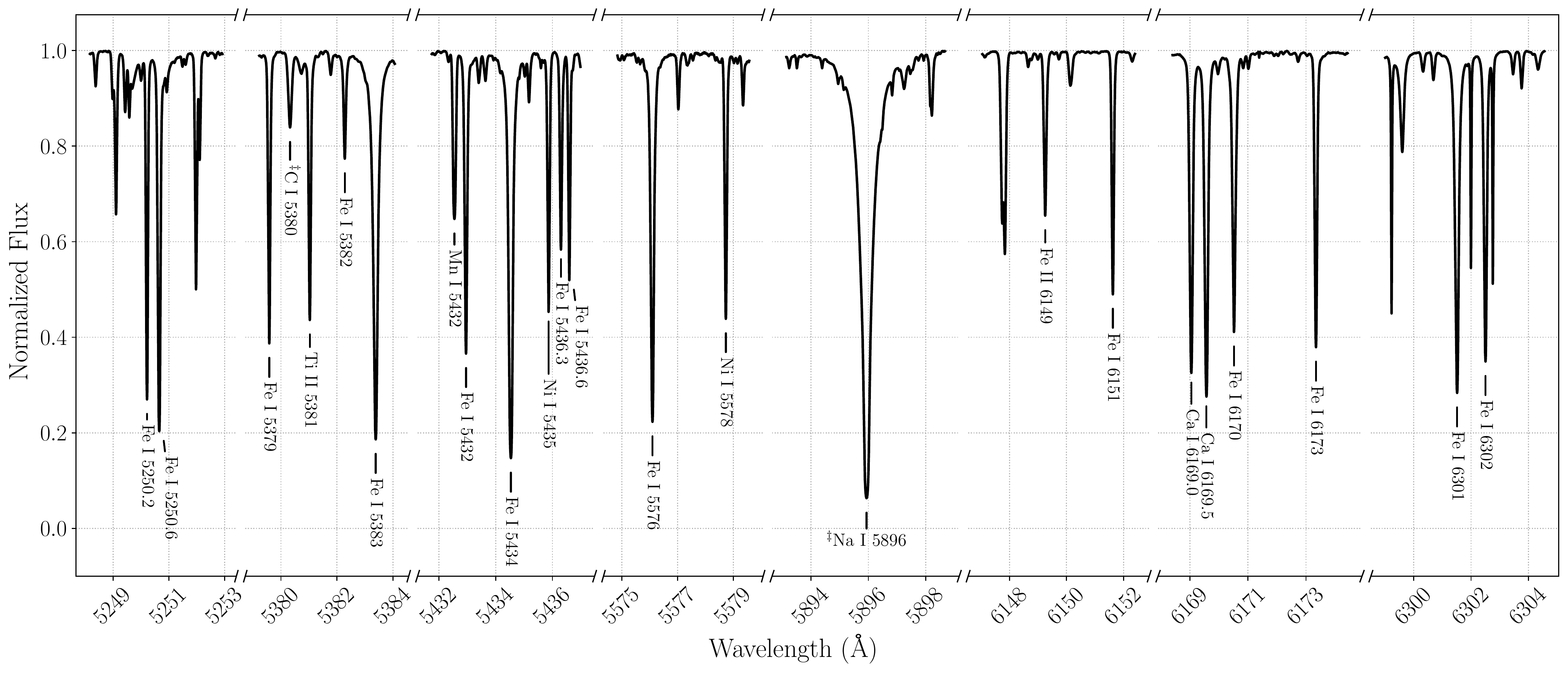}
    \caption{Collage of continuum-normalized, disk-center solar spectra originally observed for \citet{Lohner-Bottcher2018}, \citet{Stief2019}, and \citet{Lohner-Bottcher2019} with annotations for the lines used as input in this work. Lines denoted with a double dagger (C \textsc{I} 5380 \AA\ and Na \textsc{I} 5896 \AA) were included in the convective-blueshift analysis presented in \citet{Lohner-Bottcher2019}, but are excluded from the analyses presented in this work, as explained in \S\ref{methods}. Other lines are of telluric origin, significantly blended, and/or very shallow. Note the breaks in the wavelength axis; eight spectral regions spanning only a few \AA\ each were observed. Spectroscopic properties and parameters for these lines are tabulated in Table~\ref{tab:line_props}.}
    \label{fig:spectra}
\end{figure*}

In principle, such correlations are promising: precise measurements of individual bisector asymmetries predict the anomalous (i.e., granulation-induced) velocity shift, and could consequently be used to mitigate velocities from granulation. However, the authors of \citet{Cegla2019a} consider only one absorption line in their work and consequently caution that the optimal bisector summary statistic (or the optimal regions of the bisector for measuring such summary statistics) will likely vary among lines. They further warn that the typical spectral resolutions and line spread function (LSF) sampling (i.e., pixelization) of extant spectrographs, together with the effects of photon noise (see \citealt{Povich2001}), will preclude measuring sufficiently precise bisectors for such a mitigation technique to be viable. \par 

In order to address the limitations of and complement previous studies of granulation in the context of EPRVs, \citet{Palumbo2022} presented the GRanulation And Spectrum Simulator (\grass), an open-source computation tool which uses time- and disk-resolved solar spectra first observed for \citet{Lohner-Bottcher2018, Lohner-Bottcher2019} and \citet{Stief2019} to \textit{empirically} model changes in line shapes created by granulation. Validating this tool, \citet{Palumbo2022} showed that \grass\ reproduces the time-averaged line profile and bisector of the Fe \textsc{I} $5434$ \AA\ line observed in the disk-integrated Sun by \citet{Reiners2016b}, as well as the degree of variability expected from granulation. \par 

In this work, we present a large update to the \grass\ software (v2.0), and use it to explore potential avenues for mitigating granulation noise in EPRV measurements. In \S\ref{methods}, we describe the solar observations which are used to synthesize spectra (\S\ref{data}) and provide an overview of the procedure used by \grass\ to construct disk-integrated line profiles from the input solar observations (\S\ref{synthesis_procedure}). In \S\ref{sec:three}, we validate our model for line synthesis by comparing our synthetic disk-integrated line profiles to the time-averaged profiles observed in the IAG Solar Atlas \citep{Reiners2016b}. In \S\ref{sec:mitigation}, we characterize the line-by-line variability in each of the solar lines studied in this work, and simulate avenues for recovering the line-shape vs. anomalous RV correlation with limitations imposed by finite spectral resolution and SNR. In \S\ref{discussion}, we interpret and discuss our results in the context of previous studies concerned with the impact of granulation on the measurement of precise RVs, comment on future avenues for demonstrating the mitigation techniques considered in this work, and highlight various limitations of \grass\ and the scope of this study. Finally in \S\ref{conclusions}, we briefly summarize our findings and emphasize that granulation, strictly speaking, encodes coherent \textit{signals}, rather than simple noise, in the shape of stellar lines. \par 

\section{\grass\ Software Description, Methods, and Updates} \label{methods}

\grass\ uses disk- and time-resolved observations of solar lines to synthesize time series model stellar (i.e., disk-integrated) spectra with variability from granulation. \grass\ and supporting documentation are publicly available on GitHub\footnote{\url{https://github.com/palumbom/GRASS}}. Tagged version releases corresponding to \cite{Palumbo2022} and this work are archived on both GitHub and Zenodo \citep{palumbo_iii_2023_8271530}. In this section, we briefly describe the solar observations used as input to \grass\ (\S\ref{data}), provide an overview of the procedure used by \grass\ to synthesize disk-integrated spectra (\S\ref{synthesis_procedure}), and detail our process for measuring velocities from lines (\S\ref{subsec:measure_vel}). Additional, software-specific implementation details for \grass\ v2.0 (including the input data pre-processing, stellar grid tiling procedure, and the GPU implementation) are discussed in detail in Appendix~\ref{app:implementation}. \par 

\subsection{Summary of Simulation Input Data} \label{data}
\vspace{-7.5mm}
\centerwidetable
\begin{deluxetable*}{ccccccccc}
\tabletypesize{\footnotesize}
\tablecaption{Parameters for spectroscopic lines in this work. Quantities denoted with a dagger ($^\dagger$) are from \citet{Lohner-Bottcher2019}. All other quantities are from the NIST Atomic Spectra Database \citep{NISTASD}. Rows marked with a double dagger ($^\ddagger$) are excluded from the analyses presented in this work, as explained in \S\ref{methods}. The number of significant figures reported matches those given in the NIST Atomic Spectra Database or \citealt{Lohner-Bottcher2019} (for daggered quantities). \label{tab:line_props}}

\tablehead{\colhead{Species} & \colhead{Air Wavelength} & \colhead{Energy} & \colhead{$\log(gf)$} & \multicolumn{2}{c}{Spectroscopic Term} & & \multicolumn{2}{c}{Energy Level (eV)} \\
\cline{5-6}
\cline{8-9}
& (\AA) & (eV) & & Lower & Upper & & Lower & Upper}

\startdata
Fe I & 5250.2084$^\dagger$ & 2.36085268 & -4.938 &  $3d^6 4s^2 \ a \ ^5D_0$ & $3d^6 (^5 D) 4s 4p (^3 P^o)\ z \ ^7D^o_1$ & & 0.12126572 & 2.48211840 \\
Fe I & 5250.6453$^\dagger$ & 2.36065641  & -2.181 & $3d^7 (^4P) 4s\ a \ ^5P_2 $ & $3d^6 (^5D) 4s 4p (^1P^o) \ y \  ^5P^o_3$ & & 2.19786640 & 4.55852281 \\
Fe I & 5379.5734 & 2.30408106  & -1.514 & $3d^6 4s^2 \ b \ ^1G_4$ & $3d^7 (^2G)4p \ z \ ^1H^o_5$ & & 3.69459719 & 5.99867825 \\
$^\ddagger$C I & 5380.3308 & 2.30375701 & -1.62 &  $2s^2 2p 3s \ ^1P^o_1$  & $2s^2 2p 4 p \ ^1P_1$ & & 7.68476777 & 9.98852478   \\ 
Ti II & 5381.0216 & 2.30346096  & -1.921 & $3d^3 \ b \  ^2D_{3/2}$ & $3d^2 (^3F) 4p \ z  \ ^2F^*_{5/2}$ & & 1.56577729 & 3.86923825 \\
Fe I & 5382.2562 & 2.30293259 & - & $3d^6 4s^2 \ a \ ^5D_0$ & $3d^6 4s^2 \ a \ ^3P_1$  & & 0.12126572 & 2.42419831 \\ 
Fe I & 5383.3680 & 2.30245699 & 0.645 & $3d^7 (^4F) 4p \ z\ ^5G^o_5$ & $3d^7 (^4F) 4d \ e\ ^5H_6$ & & 4.31247059 & 6.61492758 \\ 
Mn I & 5432.546 & 2.281614 & -3.795 & $3d^5 4s^2 \ a \ ^6S_{5/2}$ & $3d^5 (^6S) 4s 4p (^3P^o) \ z\ ^8P^o_{5/2}$ & & 0.000000 & 2.281614 \\ 
Fe I & 5432.9470$^\dagger$ & 2.28144563 & -1.02 & $3d^7 (^4F) 4p \ z \ ^5G^o_2$ & $3d^7 (^4F) 4d \ g\ ^5F_2$  & & 4.44562726 & 6.72707289 \\
Fe I & 5434.5232$^\dagger$ & 2.28078418 & -2.122 & $3d^7(^4F) 4s \ a \ ^5F_1$ & $3d^6 (^5D)4s 4p (^3P^o) \ z\ ^5D^o_0$ & & 1.01105568 & 3.29183986 \\
Ni I & 5435.858 & 2.2802243 & -2.60 & $3d^8 (^3P) 4s^2 \ ^3P_0$ & $3d^8 (^3F) 4s 4p (^3P^o) \ ^3D^o_1$ & & 1.9858928 & 4.2661171  \\ 
Fe I & 5436.2946 & 2.28004101 & -1.51 & $3d^7 (^4F) 4p \ z \ ^3G^o_5$ & $3d^7 (^4F) 4d \ f \ ^5G_4$ & & 4.38646275 & 6.66650376 \\ 
Fe I & 5436.5876 & 2.27991815 & -2.964 & $3d^6 4s^2 \ a\ ^3P_2$ & $3d^6 (^5D) 4s 4p (^1P^o) \ y \ ^5P^o_3$ & & 2.27860466 & 4.55852281  \\ 
Fe I & 5576.0881$^\dagger$ & 2.22288021 & -0.94 & $3d^6 (^5D) 4s 4p (^3P^o) \ z \ ^5F^o_1$ & $3d^6 (^5D) 4s (^6D) 5s \ e \ ^5D_0$ & & 3.43018998 & 5.65307019 \\
Ni I & 5578.718 & 2.2218325 & -2.64 & $3d^8 (^1D) 4s^2 \ ^1D_2$ & $3d^9 (^2D) 4p\ ^1D^o_2$ & & 1.6764334 & 3.8982659 \\ 
$^\ddagger$Na I & 5895.92424 & 2.102297177 & -0.194 & $2p^6 3s \ ^2S_{1/2}$ &  $2p^6 3p\ ^2P^o_{1/2}$ & & 0.000000000 & 2.102297177 \\ 
Fe II & 6149.2460 & 2.01569257 & -2.8 & $3d^6 (^3D) 4s \ b \ ^4D_{1/2}$ & $3d^6 (^5D) 4p \ z \ ^4P^o_{1/2}$ & & 3.88919250 & 5.90488507  \\
Fe I & 6151.6170 & 2.01491569 & -3.299 & $3d^7 (^4P) 4s \ a \ ^5P_3$ &  $3d^7 (^4F) 4p \ y\ ^5D^o_2$ &  & 2.17594512 & 4.19086081 \\ 
Ca I & 6169.042 & 2.0092246 & -0.54 & $3p^6 3d 4s \ ^3D_2$ & $3p^6 4s 5p \ ^3P^o_1$ & & 2.5229867 & 4.5322113 \\ 
Ca I & 6169.563 & 2.0090547 & -0.27 & $3p^6 3d 4s \ ^3D_3$ & $3p^6 4s 5p \ ^3P^o_2$ & & 2.5256821 & 4.5347368 \\ 
Fe I & 6170.5056 & 2.00874785 & - & $3d^7 (^4F) 4p \ y \ ^3D^o_2$ & $3d^7 (^4F) 4d \ e \ ^3P_2$ & & 4.79546566 & 6.80421351 \\ 
Fe I & 6173.3344$^\dagger$ & 2.00782751 & -2.880 & $3d^7 (^4P) 4s \ a \ ^5P_1$  & $3d^7 (^4F) 4p \ y \ ^5D^o_0$ & & 2.22271209 & 4.23053960 \\
Fe I & 6301.4996 & 1.96699083 & -0.718 & $3d^6 (^5D) 4s 4p (^3P^o) \ z\ ^5P^o_2$ & $3d^6 (^5D) 4s (^6D) 5s \ e \ ^5D_2$ & & 3.65369332 & 5.62068415 \\
Fe I & 6302.4932 & 1.96668075 & - & $3d^6 (^5D) 4s 4p (^3P^o) \ z\ ^5P^o_1$ & $3d^6 (^5D) 4s (^6D) 5s \ e \ ^5D_0$ & & 3.68638944 & 5.65307019  \\
\enddata
\end{deluxetable*}

In \citet{Palumbo2022}, we used disk- and time-resolved observations of the Fe I 5434.5 \AA\ line from \citet{Lohner-Bottcher2019} to produce synthetic time-resolved, disk-integrated line profiles. In this work, we use additional spectra originally presented in \citet{Lohner-Bottcher2018}, \citet{Stief2019}, and \citet{Lohner-Bottcher2019} to synthesize other disk-integrated lines. The observed spectral regions include 22 solar lines which are sufficiently unblended and deep to use as models for spectral variability from granulation. Example disk-center spectra from \citet{Lohner-Bottcher2018}, \citet{Stief2019}, and \citet{Lohner-Bottcher2019} are shown in Figure~\ref{fig:spectra} with annotations for the lines used in this work. A tabulation of the atomic parameters for these lines is provided in Table~\ref{tab:line_props}. We below provide a brief summary of the processing procedure for these data; a more comprehensive description is provided in Appendix~\ref{sub:preprocess}. \par

LARS, the instrument that provided input spectra for this work, is described in detail in \citet{Lohner-Bottcher2017}. The observing scheme and reduction procedure used to obtain these spectra is summarized in \citet{Palumbo2022} and explained in greater detail in \citet{Lohner-Bottcher2018}, \citet{Stief2019}, and \cite{Lohner-Bottcher2019}. In brief, spectra were observed at 41 locations along the North-South and East-West axes of the Sun; only quiet-Sun regions were observed. The spectra have $R\sim 700{,}000$, and are calibrated on an absolute wavelength scale with accuracy $\sim$0.02 m\AA\ (or about 1 $\ms$). On short timescales (e.g., over the duration of the $\gtrsim$20-minute observations baselines), \citet{Lohner-Bottcher2017} report that the instrument instability is the largest source of uncertainty in the positions of lines; generally this error is at the level of a few$\cms$, but can be larger in some cases. In this work, we quantify the impact of this uncertainty by repeatedly synthesizing many realizations of our synthetic spectra, from which we measure sample means and standard deviations for the RV variability of lines. This process is described further in \S\ref{subsec:line_by_line}. \par

As in \citet{Palumbo2022}, we measure line bisectors and widths as a function of depth into the line for all observed lines for each 15-second bin, which together we refer to as the ``input data,'' following the nomenclature established in \citet{Palumbo2022}. These input data losslessly encode the temporal variability in the shape of each line and are used as the basis of our stellar surface simulation and integration described in the following section, \S\ref{synthesis_procedure}. Our process for measuring these input data from the solar spectra, which has changed slightly from \citet{Palumbo2022}, is described in detail in \S\ref{sub:preprocess}. This pre-processing of the solar observations was performed once, and the resulting data products are downloaded from Zenodo \citep{palumbo_iii_michael_l_2023_8271417} by \grass\ upon installation. \par   

\subsection{Overview of \grass\ Synthesis Procedure} \label{synthesis_procedure}

Owing to the expanded library of solar templates lines available to \grass, we have slightly updated and optimized the procedure it uses to create synthetic, disk-integrated spectra from these input data. Here, we describe the synthesis procedure used by \grass, highlighting changes and updates to the synthesis procedure detailed in \S3.2.1 of \citet{Palumbo2022}. \par 

Prior to the spectral synthesis step, \grass\ first tiles a model stellar grid and computes the requisite weights and rotational velocities (see \S\ref{subsec:model_grid}). These quantities are computed once and stored in memory, since they are assumed to be unchanging in time. Following the initial tiling and geometric computations, \grass\ uses the input data to reconstruct disk-resolved line profiles in each tile on the model stellar surface for each time step of the simulation. As in \citet{Palumbo2022}, \grass\ selects the input data with the closest $\mu$ value and matching directional axis for each stellar surface tile. In the event that a certain axis and $\mu$ position lack data for a given template line, the next-nearest input data are used. \par 

The temporal variability in each disk-resolved line profile is directly encoded from the time-series input observations. However, as described in \S3.2.3 of \citet{Palumbo2022}, we do apply a random phase offset to the input data in each stellar surface tile. This random phase offset ensures that the disk-resolved line profiles in adjacent tiles do not unrealistically move in concert. Because some template lines were observed contemporaneously (e.g., Fe \textsc{I} 5432 \AA\ and Fe \textsc{I} 5434 \AA\ are in the same observed spectral region, and so were always observed simultaneously; see Figure~\ref{fig:spectra}), the input data for these lines is, by default, kept in-phase within each stellar surface tile. The applied phase offsets also lead to the suppression of $p$-modes in the final disk-integrated spectrum, since the oscillations in the input data are added \textit{incoherently} (see \S3.2.4 of \citealt{Palumbo2022}). \par 

Similarly to \grass\ v1.0 \citep{Palumbo2022}, the disk-integrated flux as a function of time and wavelength is then calculated as the (weighted) sum over the individual tile intensities, where (as noted in Appenidx~\ref{subsec:model_grid}) tile weights are given by the product of the limb-darkened continuum intensity and the projected tile area (analogous to Equation 18.3 of \citealt{Gray2008}). As in \cite{Palumbo2022}, these summations are performed in place to avoid excess memory allocation (i.e., the disk-resolved spectra computed for each tile are not retained), which would become quite unwieldy for larger spectra. \par 

\subsection{Measurement of Velocities from Spectra} \label{subsec:measure_vel}

As in \citet{Palumbo2022}, we compute CCFs in order to measure velocities from individual lines and spectra. In implementation, \texttt{EchelleCCFs} \citep{EchelleCCFs} was used to cross correlate the considered line profile(s) with a Gaussian template mask centered at the known rest-frame wavelength of said line and width (in units of velocity) given by the speed of light divided by the spectral resolution. The mask was projected onto the line profile in steps of 100 $\ms$. The extremum of the resulting cross-correlation function (CCF) was fit with a Gaussian function via non-linear least squares, and the RV was taken as the mean of the best-fit Gaussian.  \par 

Our velocity-measurement procedure was tested by injecting known Doppler shifts into synthetic spectra; we find that the recovered velocities are accurate, with errors at the $\lesssim$1$\cms$ level (compare to the expected amplitude of granulation noise at a few to several tens of$\cms$). We also tested other combinations of mask shapes (tophat vs.\@ Gaussian) and functional forms for fitting the CCF peak (quadratic vs.\@ Gaussian) and found that the chosen combination of Gaussian mask with Gaussian fit performed the best. \par 

\section{\grass\ Model validation} \label{sec:three}

In \citet{Palumbo2022}, the line synthesis procedure used by \grass\ was validated by comparison of time-averaged line profiles to the IAG Solar Atlas \citep{Reiners2016b}. In the following subsections, we re-perform this validation for each template line used in this work (\S\ref{validation}), and also discuss how effectively these template lines are able to model the shapes and behaviors of other lines (\S\ref{subsec:template_models}).

\begin{figure*}[!htb]
    \gridline{\fig{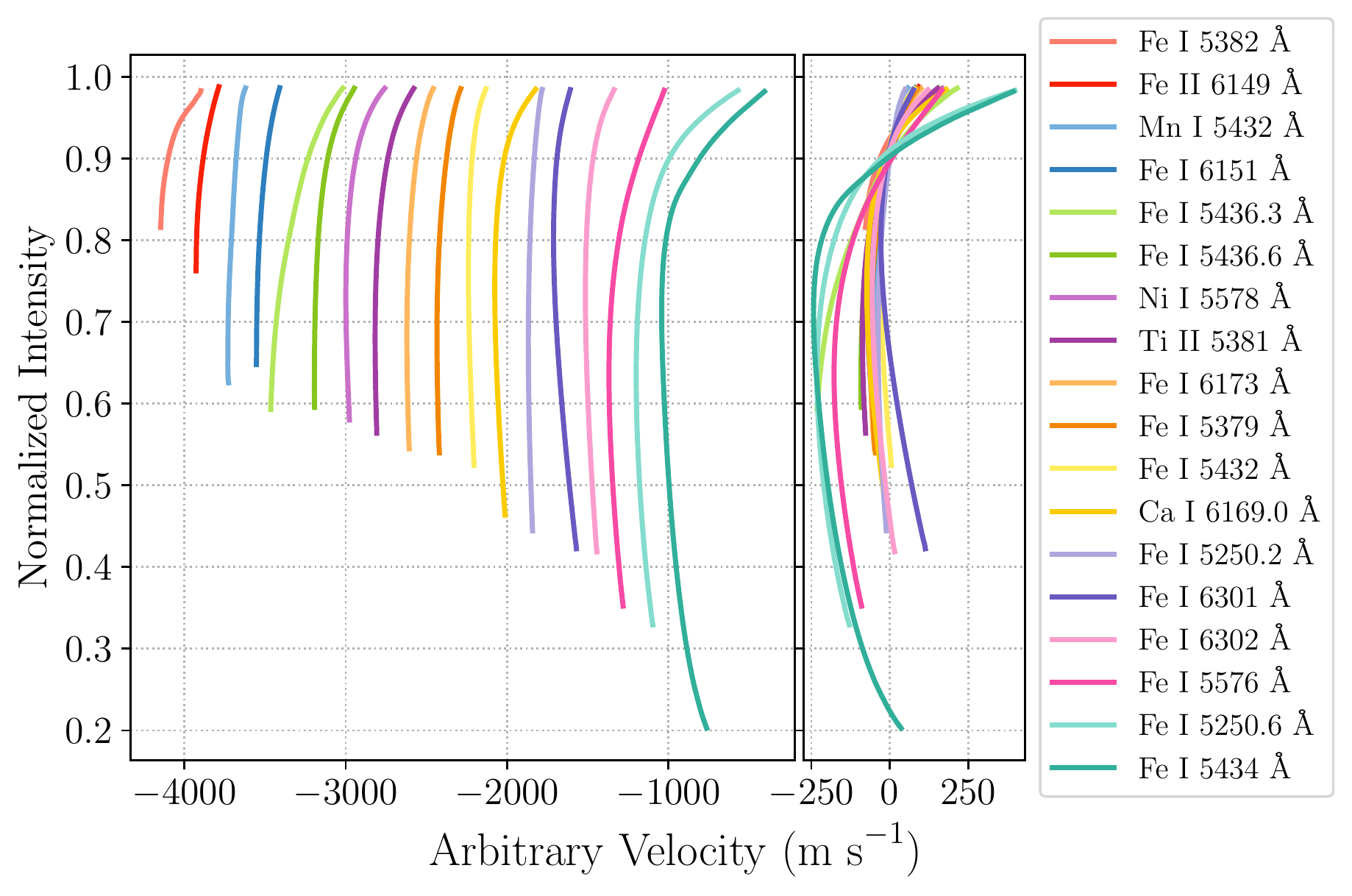}{0.5\textwidth}{}
              \fig{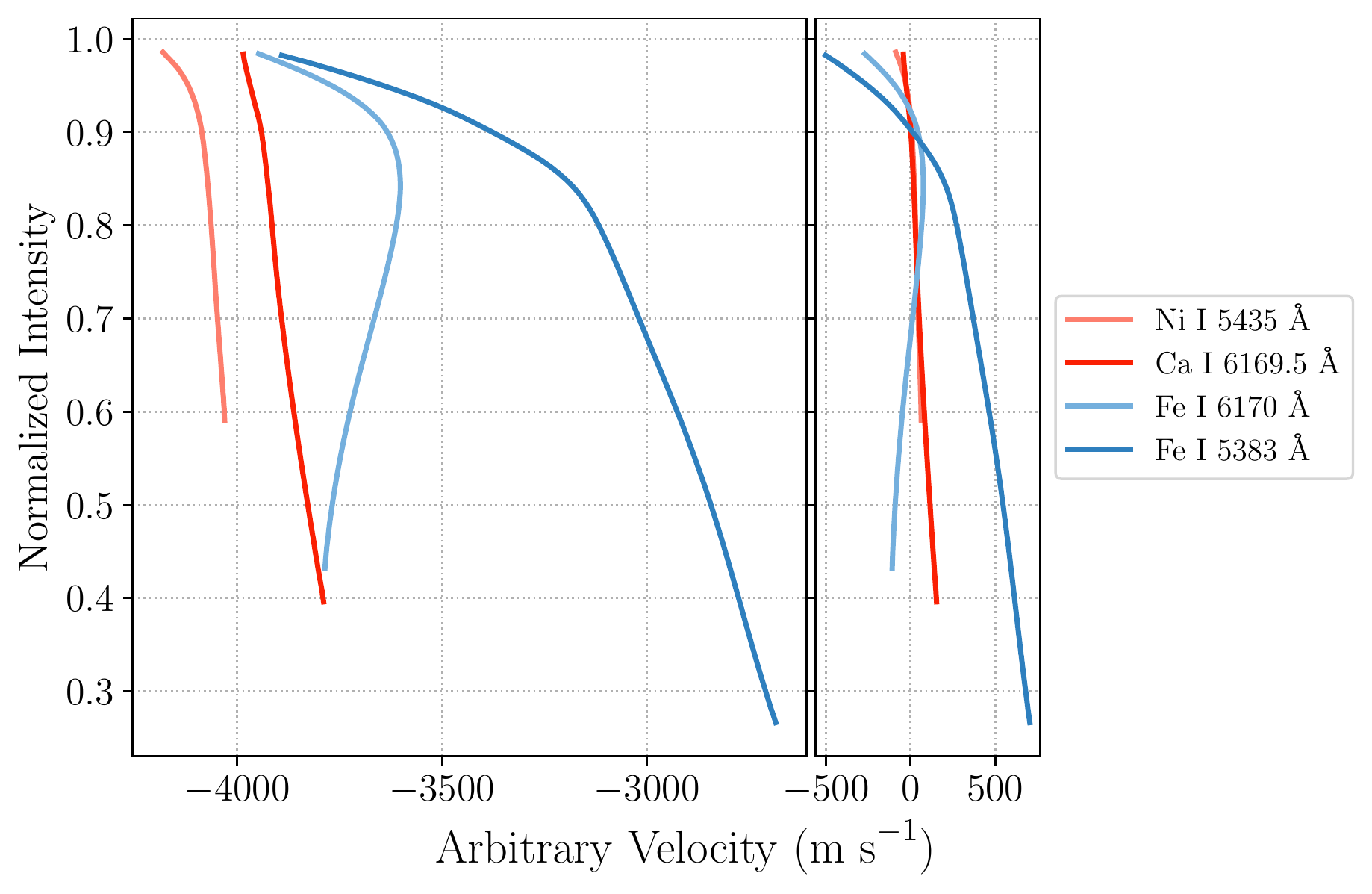}{0.5\textwidth}{}}
    \vspace{-7mm}
    \caption{Ensemble of synthetic disk-integrated bisectors for the 22 lines analyzed in this work; compare to Figure 17.14 of \citet{Gray2008}. These bisectors were measured from significantly oversampled spectra synthesized at $R \sim 700{,}000$ to ensure smoothness. Classical ``C''-shaped bisectors are shown in the left-hand figure, and bisectors of deeply blended lines are shown in the right-hand figure. The velocity offsets in both panels of each figure are arbitrary and for illustrative purpose only. As in Figure 17.14 of \citet{Gray2008}, the bisectors are composited at right in each figure to show their (dis)similarity.}
    \label{fig:bisector_dossier}
\end{figure*}

\subsection{Disk-Integrated Line Profiles and Bisectors} \label{validation}

In \citet{Palumbo2022}, we verified that \grass\ could accurately reproduce disk-integrated line profiles and bisectors using disk-resolved spectra as input. Here, we validate the synthesis for the other lines presented in this work. Similarly to \citet{Palumbo2022}, we compare time-averaged synthetic line profiles from \grass\ to line profiles observed in the IAG Solar Atlas \citep{Reiners2016b} in order to semi-quantitatively assess our synthesis accuracy. \par 

To perform this comparison, we degrade the spectral resolution of the IAG Solar Atlas to the nominal LARS resolution of $R \sim 700{,}000$ via convolution with a Gaussian LSF given by:

\begin{equation}
    {\rm LSF}(\lambda) = \frac{1}{2\pi \sigma(\lambda) } \exp{\left(-\frac{\lambda^2}{2\sigma(\lambda)^2} \right)}
\end{equation}

\noindent where $\sigma(\lambda)$ is given by

\begin{equation}
    \sigma(\lambda) = \lambda / (2 \sqrt{2 \ln 2} \cdot R), 
\end{equation}

\noindent and $\lambda$ is the wavelength sampled by a given pixel. To compute residuals, we then perform a flux-conserving interpolation onto a common grid of wavelengths with $\sim$4 pixels per LSF FWHM (following the algorithm of \citealt{Carnall2017}). Rather than computing bisectors directly from the line profiles, we instead measure bisectors from individual-line CCFs. These CCF bisectors are smoother (and consequently easier to compare) than bisectors measured directly from line profiles. The resulting line profiles and bisectors are discussed individually in Appendix~\ref{app:iag}. An example set of comparisons is shown in Figure~\ref{fig:iag_bis_first}; the full set of comparison figures is available from the online journal. The full ensemble of reproduced bisectors are shown in Figure~\ref{fig:bisector_dossier} and discussed further in \S\ref{subsec:template_models}. \par 

In general, we find that we are able to synthesize line profiles that are faithful to those seen in the IAG Atlas with error within about 1\% flux. Deviations are usually caused by blends in the wings of the lines of interest, which are modeled out in the pre-processing stage for the LARS data (see \S\ref{sub:preprocess}). Likewise, bisectors are most accurate where the first derivative of the line profile is largest, but tend to deviate in the top 20\% or so of line, owing to blends and/or our imperfect modeling of the line wings in the preprocessing stage, as well as at the very bottom of the line, where interpolation noise and measurement error become large. Below $\sim$80\% of the continuum, the velocity errors in the line bisectors are generally within a couple of $\ms$ and within 1 $\ms$ in the best cases. For lines that have blends in their wings, the velocity errors are larger toward the continuum (above 10 $\ms$ in some cases). As noted in \citet{Palumbo2022}, deviations in the cores of lines (particularly the deeper lines) could arise from chromospheric activity captured during the observation of the IAG Atlas during a period of heightened solar activity in 2014 \citep{Hathaway2015, Reiners2016b}. Specific comments for each line are given in Appendix~\ref{app:iag}, as well as Section~3 and Appendix~A of \cite{Lohner-Bottcher2019}. \par  

\subsection{Use of Template Lines as Generalized Line Models} \label{subsec:template_models}

Only a limited number of lines were observed by \citet{Lohner-Bottcher2018, Lohner-Bottcher2019}. Since thousands of absorption lines are generally used to measure velocities from spectra, v1.0 of \grass\ \citep{Palumbo2022} modeled lines of differing depths by truncating the deep Fe I 5434 \AA\ line bisector (see \S3.2.2 of \citealt{Palumbo2022}). This approach was motivated by the heuristic presented in \citet{Gray2008}, wherein it is noted that the bisectors of shallower lines tend to reflect the shape of the upper portions of bisectors of deeper lines (see their Figure 17.14). For comparison, we have overplotted the synthetic-disk integrated bisectors produced in this work in Figure~\ref{fig:bisector_dossier}. \par 

It is readily apparent that the four blend bisectors at right in Figure~\ref{fig:bisector_dossier} cannot be superimposed as readily as the classical ``C''-shaped bisectors at left, with the potential exception of Ni \textsc{I} 5435 \AA\ and Ca \textsc{I} 6169.5 \AA. Further, even in the left-hand plot of Figure~\ref{fig:bisector_dossier}, there is notable variance between the bisectors, especially in the case of the three deepest lines: Fe \textsc{I} 5576 \AA, Fe \textsc{I} 5250.6 \AA, and Fe \textsc{I} 5434 \AA. This imperfect correspondence is not entirely surprising, given that the limited number of lines considered were not cherry-picked for the similarities in their bisectors (as is the case in Figure~17.14 of \citealt{Gray2008}). Rather, the lines and spectral regions observed for \citet{Lohner-Bottcher2019} and then used in this work were chosen based on their past usage in the heliophysics literature. \par 

Although the picture painted by Figure~17.14 of \cite{Gray2008} is certainly convenient, it is clear from Figure~\ref{fig:bisector_dossier} that the correspondence of bisectors values at given intensities/depths is not universal. This reality is not entirely surprising. Lines do not form at a single representative ``formation height'', but rather over a run of heights that differs in accordance with the properties of the stellar atmosphere and the atomic/ionic properties of a line-forming species. Consequently, a given continuum-normalized intensity cannot be neatly mapped to a singular physical height in the solar atmosphere across differing lines. For the results presented in this study, we only use the 22 template lines to reconstruct their own disk-integrated profiles. Future works could explore which template lines used herein are best-suited to model variability in other lines not observed by LARS, but such a study is beyond the scope of the analysis presented in this work (see also the discussion in \S\ref{subsec:disc_less_variable} and \S\ref{subsec:discuss_diagnostics}). \par 

\section{Simulations of Granulation Mitigation} \label{sec:mitigation}

As is visually apparent from Figure~\ref{fig:bisector_dossier}, granulation affects lines \textit{differentially}. Consequently, measured RV variability will differ line by line, and the amount of granulation noise measured in RV observations will differ depending on the set of lines used. In the following subsections, we examine and quantify this line-by-line variability for the set of solar lines modeled in this study (\S\ref{subsec:line_by_line}), explore correlations between line-shape summary statistics and anomalous RVs (\S\ref{subsec:shape_corr}), and simulate how these correlations might be used for mitigation given the constraints of existing instrumentation (\S\ref{subsec:aggregate}). \par 

\subsection{Line-by-line Variability} \label{subsec:line_by_line}

It has been widely shown that the measured extent of variability manifests differentially between lines and sets of lines. In the case of individual lines, \cite{Elsworth1994} and \cite{Palle1999} report different RMS RVs for the K \textsc{I} 7699 \AA\ line and the Na \textsc{I} D$_1$ and D$_2$ lines, respectively. Considering sets of lines, \citet{AlMoulla2022, AlMoulla2024} showed that the measured RMS RV varies among sets of solar lines binned by formation temperature. As suggested by these studies, simply measuring RVs from a minimally variable set of lines may constitute one potentially viable, albeit limited, method for granulation mitigation. To quantify the different levels of variability in lines, we synthesized many disk-integrated time series for each of the 22 lines considered in this work, from which we measured the RMS of the individual line RVs. \par 

\begin{figure*}[!htb]
    \gridline{\fig{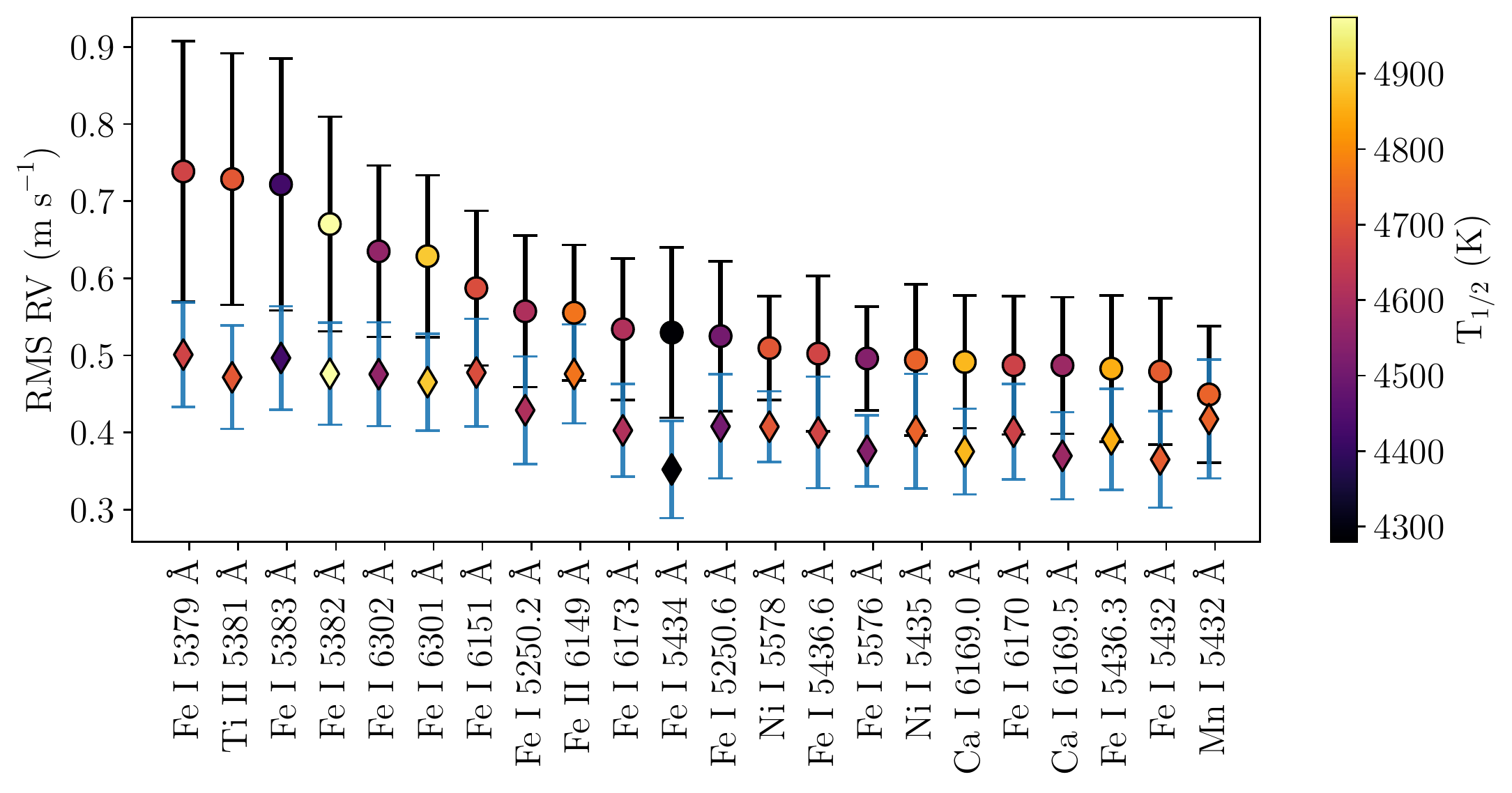}{0.585\textwidth}{}
              \fig{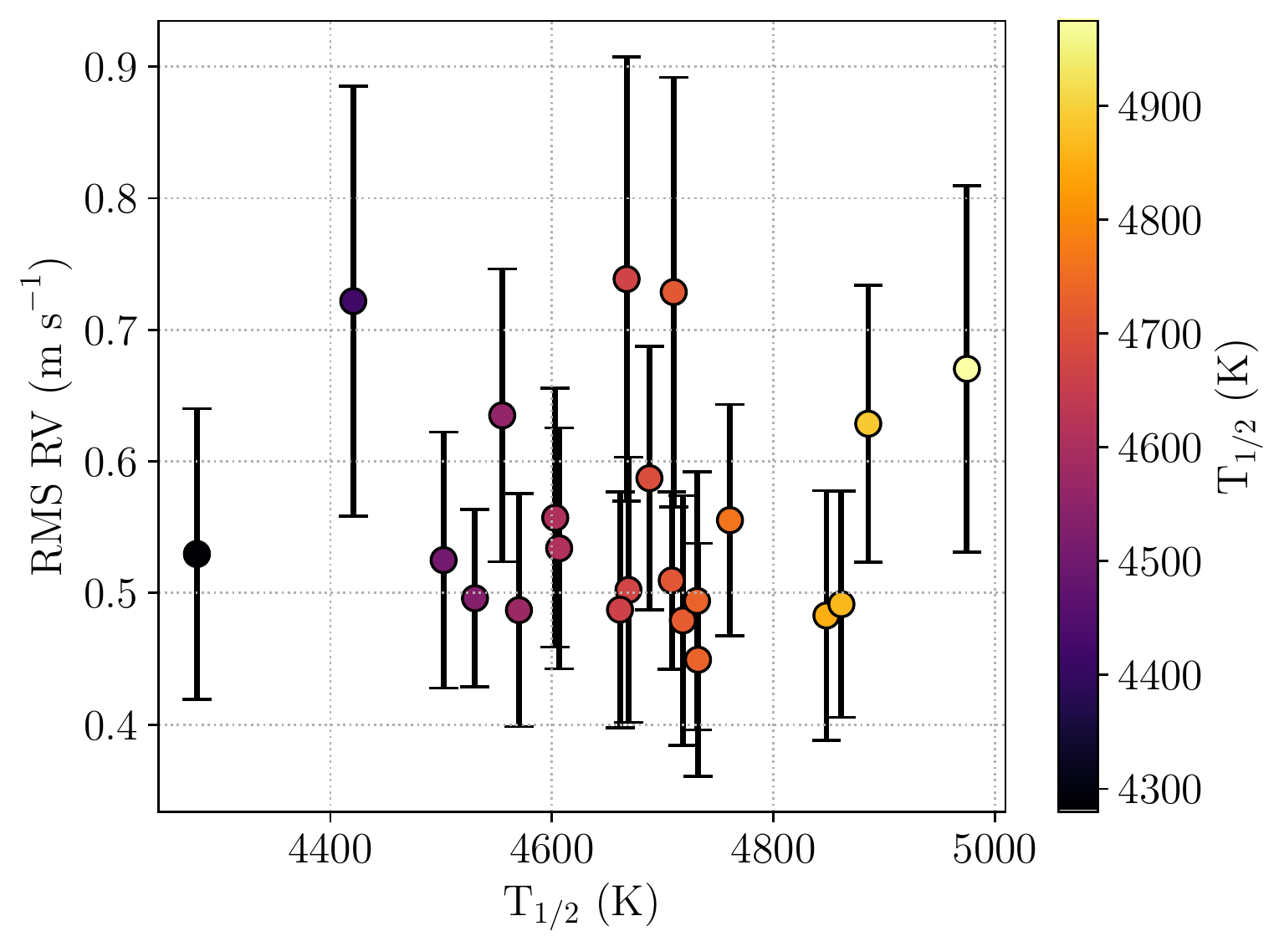}{0.415\textwidth}{}}
    \vspace{-5mm}
    \caption{Line-by-line differences in granulation-induced variability. \textit{Left:} Colored circles indicate the mean RMS RV measured from time-series of individual disk-integrated line profiles, following from the procedure outlined in \S\ref{subsec:line_by_line}. The colored diamonds show the mean residual RMS RV after using the tuned BIS to correct granulation velocities, as detailed in \S\ref{subsec:shape_corr} and \S\ref{subsubsec:tuned_bis}. In both cases, the error bars are the $1\sigma$ width of the RMS RV distributions measured from many trials, as described in \S\ref{subsec:line_by_line}. \textit{Right:} The amount of variability (expressed as the RMS RV) exhibits no clear correlation with $T_{1/2}$ (see \citealt{AlMoulla2022}), or any other single quantity, including: wavelength, line depth, and potential of the lower and upper states of each transition (not shown). Although there is an appreciable difference between the most and least variable lines, it is not immediately apparent how the variability of a line could be predicted from the formation or atomic properties of a line.}
    \label{fig:rms_ladder}
\end{figure*}

The procedure for determining these individual line RMS values is as follows: \grass\ was used to generate $40$-minute disk-integrated time series with a time resolution of $15$ seconds for each considered line profile (see Appendix~\ref{sub:preprocess} for details on the temporal cadence and baseline of the input data); velocities were measured from these line profiles as described in \S\ref{subsec:measure_vel} (see also \S3.3 of \citealt{Palumbo2022}); lastly, an RMS was measured for the resulting RV time series. This process was repeated many times with different realizations of the synthetic line profiles to yield a robust mean and standard deviation for each RMS RV. \par 

The resulting RMS values are shown in descending order in the left-hand panel of Figure~\ref{fig:rms_ladder} and are reported in Table~\ref{tab:rms}. There is a notable difference in the RMS between the most and least variable lines. At the high end, the Fe \textsc{I} 5379 \AA\ line exhibits a $\sim$75$\cms$ RMS. At the low end, Mn \textsc{I} 5432 \AA\ has a $\sim$45$\cms$ RMS. Given this nearly $\sim$30$\cms$ difference between the most and least variable lines, a fruitful granulation mitigation strategy might consistent of identifying the least variable lines and measuring RVs from only these lines, rather than from a larger line list also including highly variable lines. To enable such an approach, one would need to know the intrinsic variability of each line for a given star. If such a quantity can not be measured directly from spectra (owing to e.g., limitations from SNR, sampling, etc.), then predicting variability from a proxy variable may be the next-best approach. However, we note no significant trend connecting the RV variability of a line to its depth, wavelength, or the potential of the lower/upper state of its transition. \par 

Recently, \citet{AlMoulla2022} and \citet{AlMoulla2024} used solar and stellar data, respectively, to show that the measured RMS RV has a dependence on the formation temperature of the lines used. Following the methods of \citet{AlMoulla2022}, we also computed their $T_{1/2}$ parameter (the temperature at the height in the atmosphere corresponding to 50\% of a line's cumulative contribution function; see Figure 1 of \citealt{AlMoulla2022}) at the central wavelength of each line. The RMS RVs for each line are plotted against their $T_{1/2}$ in the right-hand panel of Figure~\ref{fig:rms_ladder}. As with the other variables, we note no simple trend connecting a line's variability to its $T_{1/2}$. Excluding the two lines with the highest variability (around $T_{1/2} \sim 4700$ K), one might plausibly deduce hints of the same ``high-low-high'' trend of RMS with temperature noted by \citet{AlMoulla2022} and \citet{AlMoulla2024}. However, given the small number of lines considered, the significance of such a trend in our data is questionable at best. We discuss these results and prospects for studying line-by-line variability further in \S\ref{subsec:disc_less_variable}. \par 

\subsection{Mitigation via Bisector Diagnostic Correlations} \label{subsec:shape_corr}

\begin{figure*}[!htb]
    \epsscale{1.14}
    \plotone{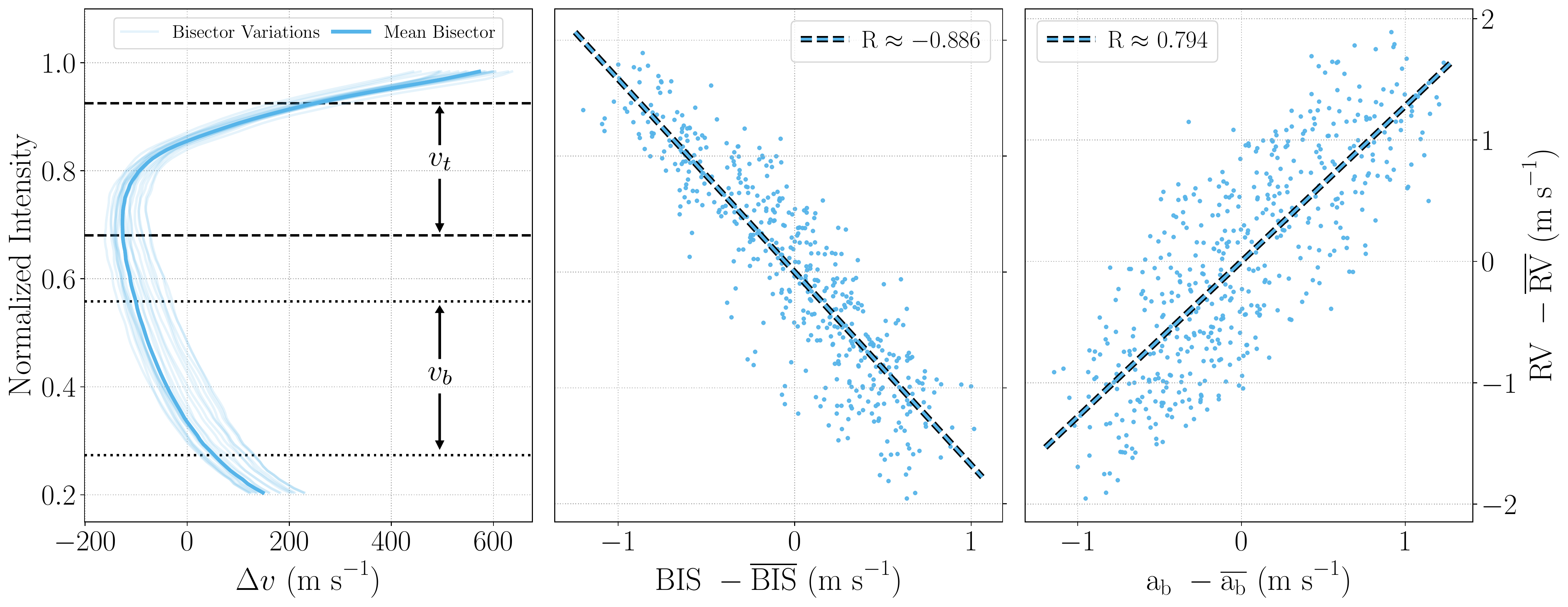}
    \caption{Example correlations between bisector shape summary statistic and anomalous, granulation-induced velocity shift for a synthetic Fe \textsc{I} 5434 \AA\ time series. \textit{Left:} Time-averaged line bisector (opaque blue curve) and bisector variations (transparent blue curves). The amplitude of the bisector variations are exaggerated 50$\times$ for visual effect. The regions bound by the horizontal dashed lines and dotted lines represent the $v_t$ and $v_b$ regions used in the calculation of BIS, respectively. \textit{Middle:} Correlation between BIS and apparent RV excursion with best-fit relation. Each point corresponds to one snapshot of the synthetic time series. \textit{Right:} Same as middle panel, but for bisector amplitude $a_b$.}
    \label{fig:rv_bis}
\end{figure*}

Using disk-integrated Fe \textsc{I} 6302 \AA\ profiles derived from MHD simulations, \citet{Cegla2019a} showed notable correlations between various line-shape summary statistics and the anomalous (i.e.,\@ granulation-induced) RV measured. An example of such a correlation between the bisector inverse slope (BIS) and velocity for synthetic Fe \textsc{I} 5434 \AA\ profiles produced by \grass\ are shown in Figure~\ref{fig:rv_bis}. Note that the amplitude of the bisector variations (shown in the left-hand panel of Figure~\ref{fig:rv_bis}) were amplified by $50\times$ in order to make these variations (which are at the sub-$\ms$ scale) visible on the scale of the mean bisector (which spans multiple hundreds of $\ms$ from the red-most to the blue-most point). \par

Notably, the simulations performed by \citet{Cegla2019a} considered only one line (the Fe \textsc{I} 6302 \AA\ line) under an artificially strong vertical magnetic field, which they note inhibits convection (and therefore the observed variability of their line) relative to what is observed in the quiet Sun. To supplement the analysis presented \citet{Cegla2019a}, we here perform a similar bisector analysis for the 22 lines studied in this work. To perform this analysis, we first used \grass\ to synthesize time series for each of the 22 template lines considered, and then measured various bisector-shape summary statistics and RVs for each time snapshot of each simulation. RVs were measured from CCFs as in \S\ref{subsec:line_by_line}, and bisector shape diagnostics were measured from the CCF bisectors. As in \S\ref{validation}, we measured bisectors from CCFs, rather than directly from line profiles, in order to produce smoother curves. In later sections (namely \S\ref{subsec:aggregate}), we will also measure bisector diagnostics for CCFs computed from many lines; by measuring CCF bisectors for individual lines, we can directly compare the performance of the bisector diagnostics on lines individually and in aggregate. Finally, mitigated RVs were measured by subtracting off the RVs predicted by a linear fit between the raw measured RVs and each bisector diagnostic. This procedure was repeated many times for each line using many different realizations of the synthetic profile time series to measure robust averages for the correlation coefficients and improved RMS RVs. \par 

\begin{deluxetable}{ccc}
\tablecaption{RMS RV and 1$\sigma$ variability in RMS RV for each line considered in this work; the data in this table correspond to the values shown in Figure~\ref{fig:rms_ladder}. Velocity quantities are rounded to $\cms$ precision. \label{tab:rms}}

\tablehead{\colhead{Line} & \colhead{Mean RMS RV} & \colhead{$\sigma$ RMS RV} \\ 
(\AA) & ($\ms$) & ($\ms$)}
\startdata
Fe I 5250.2 &     0.56 &         0.10 \\
Fe I 5250.6 &     0.52 &         0.10 \\
  Fe I 5379 &     0.74 &         0.17 \\
 Ti II 5381 &     0.73 &         0.16 \\
  Fe I 5382 &     0.67 &         0.14 \\
  Fe I 5383 &     0.72 &         0.16 \\
  Mn I 5432 &     0.45 &         0.09 \\
  Fe I 5432 &     0.48 &         0.10 \\
  Fe I 5434 &     0.53 &         0.11 \\
  Ni I 5435 &     0.49 &         0.10 \\
Fe I 5436.3 &     0.48 &         0.10 \\
Fe I 5436.6 &     0.50 &         0.10 \\
  Fe I 5576 &     0.50 &         0.07 \\
  Ni I 5578 &     0.51 &         0.07 \\
 Fe II 6149 &     0.56 &         0.09 \\
  Fe I 6151 &     0.59 &         0.10 \\
Ca I 6169.0 &     0.49 &         0.09 \\
Ca I 6169.5 &     0.49 &         0.09 \\
  Fe I 6170 &     0.49 &         0.09 \\
  Fe I 6173 &     0.53 &         0.09 \\
  Fe I 6301 &     0.63 &         0.10 \\
  Fe I 6302 &     0.64 &         0.11 \\
\enddata
\end{deluxetable}
\vspace{-4mm} 
These initial simulations and correlations were constructed for line profiles synthesized at $R = 700{,}000$ with ${\rm SNR} = \infty$ (i.e., no additional noise, photon or otherwise, was simulated in the synthesized spectra). We acknowledge that these simulation parameters are certainly not realistic; this initial analysis was performed in order to determine the relative usefulness of each bisector shape diagnostic in a hypothetical best-case scenario where granulation is the sole source of RV noise. The effects of various limitations imposed by more realistic observing conditions (including spectral resolution and photon noise) are evaluated in \S\ref{subsec:aggregate}. \par

The considered bisector shape diagnostics include: bisector inverse slope (BIS), bisector curvature ($C$), and bisector span (or amplitude, $a_b$). Definitions for these diagnostics are given below; see also Figure~6 of \citet{Cegla2019a} for an excellent illustrative demonstration of these shape diagnostics. The bisector inverse slope (BIS, \citealt{Queloz2001}) is given by: 

\begin{equation} \label{eq:BIS}
    {\rm BIS} = v_t - v_b,
\end{equation}

\noindent where $v_t$ is the average velocity of the bisector between $10\%$ and $40\%$ of the line depth, and $v_b$ is the average velocity between $55\%$ and $90\%$. The bisector curvature ($C$, \citealt{Povich2001}) is given by:

\begin{equation} \label{eq:curve}
    C = (v_3 - v_2) - (v_2 - v_1),
\end{equation}

\noindent where $v_3$, $v_2$, and $v_1$ regions are bounded at $20\%$-$30\%$, $40\%$-$55\%$, and $75\%$-$95\%$ of the line depth, respectively. Lastly, the bisector amplitude ($a_b$, \citealt{Livingston1999}) is given by:

\begin{equation}
    a_b = v_{\rm min} - v_{0}
\end{equation}

\noindent where $v_{\rm min}$ is the blue-most velocity in the bisector curve and $v_0$ is the velocity in the line core (the bottom of the bisector). In practice, we calculate $v_0$ as the mean velocity in the bottom few points of the bisector, excluding the bottom-most measurement which is often highly erroneous to due to interpolation noise. \par 

We present and analyze the results of these simulations below in \S\ref{subsubsec:default_bis}, where correlation coefficients and corrected RMS RVs were calculated using the default definitions of the bisector diagnostics given above. In \S\ref{subsubsec:tuned_bis} we follow \citep{Cegla2019a} and iteratively ``tune'' the definitions of these diagnostics  to maximize the correlation for each line. \par 

\subsubsection{Default Bisector Diagnostics} \label{subsubsec:default_bis}

The results of the granulation mitigation simulations described above are tabulated in Table~\ref{tab:correlation}. For each line and bisector diagnostic, the mean Pearson correlation coefficient ($R$), mean RMS RV, 1$\sigma$ variation in RMS RV, and percent improvement (rounded to the nearest whole number) over the RMS RVs reported in Table~\ref{tab:rms} are shown. The reported improvements should be taken as approximate and representative; a $1\%$ change in the RMS RV would amount to $\lesssim$1$\cms$ at the $\sim$30-70$\cms$ velocity scale of granulation. \par 

The best-performing diagnostics are able to remove, optimistically, $\sim$25-33\% of the intrinsic RV variability measured for each line. However, the level of correction achieved by each diagnostic was inconsistent across the considered lines. E.g., although BIS performed quite well for Fe \textsc{I} 5434 \AA, this diagnostic's performance varied widely for other lines. Likewise, the bisector amplitude performed well for a small handful of lines, but failed to achieve any meaningful improvement for several lines. Interestingly, the bisector curvature performed poorly across the board, with the notable exception of Fe \textsc{I} 5383 \AA. \par 

This variance is not surprising: \citet{Cegla2019a} inferred that the BIS performed well when the $v_t$ and $v_b$ regions from which this quantity is calculated bound the bend of the ``C'' and bottom of the bisector (corresponding to the line core), respectively, for their synthetic Fe \textsc{I} 6302 \AA\ line. This picture is generally corroborated by our results: BIS generally performs best for lines with classical ``C''-shaped bisectors whose bend and core happen to be captured by the canonical $v_t$ and $v_b$ regions of $10\%$-$40\%$ and $55\%$-$90\%$ depth. Indeed, we find that BIS performs well for our Fe \textsc{I} 6302 \AA\ line simulations, removing $\sim$24\% of the line's intrinsic variability. \par 

In the context of \citet{Cegla2019a}, the observed variance of the bisector amplitude is perhaps harder to explain. Yet, on closer inspection it is clear that the bisector amplitude tends to fail to predict the anomalous RV for lines whose bisectors deviate from the classical ``C'' shape. Two salient examples are Fe \textsc{I} 6170 \AA\ (which has a ``/''-shaped bisector owing to line blends; see the online journal figure set and Appendix~\ref{app:iag})
and Fe \textsc{I} 5436.3 \AA\ (which has a bisector corresponding to only the top-most portion of the classical ``C'' shape; see the online journal figure set and Appendix~\ref{app:iag}).
In the latter case, the bisector amplitude hardly varies from $0$, since the blue-most point and the core of the bisector are, in most snapshots in any given synthetic time series, one and the same. \par

The generally poor performance of the bisector curvature $C$, might suggest that increasing the number of averaging regions (three, compared to two for BIS) does not necessarily improve sensitivity to changes in bisector asymmetry. In some part, though, the poor performance of $C$ may also be happenstance: the canonical averaging regions may not be ideally placed to probe changes in bisector curvature for every line. In the following section, we explore this possibility by adjusting the averaging regions used in both BIS and $C$ in order to maximize their diagnostic power for each line. \par 

\subsubsection{Optimized Bisector Diagnostics} \label{subsubsec:tuned_bis}
As noted by \citet{Cegla2019a}, these bisector diagnostics (except for $a_b$) are not ``tailored'' to the properties of a given line: although the $v_t$ region happens to capture the kink in the ``C''-shaped curve of the Fe \textsc{I} 5434 \AA\ line (see Figure~\ref{fig:rv_bis}), it may not for another line. To address this fact \citet{Cegla2019a} iteratively varied the averaging regions used in the definitions of BIS and the bisector curvature $C$ in order to optimize their correlations with the measured RVs of their Fe \textsc{I} 6302 \AA\ line profiles. We have performed this tuning for each of the 22 lines studied in this work. \par 

To tune the bisector diagnostics, we first drew values for the bounding intensities (which we refer to as $b_1$, $b_2$, $b_3$, and $b_4$ for BIS; and $c_1$, $c_2$, $c_3$, $c_4$, $c_5$, and $c_6$ for $C$) from uniform random distributions. As in \citet{Cegla2019a}, we required that each averaging region spanned at least 5\% in depth, and we additionally restricted the bounding intensities to fall between 15\% and 95\% of the given line depth (in order to avoid the regions near the continuum and the core of line, where bisector measurement error is largest). Draws which violated these restrictions were rejected and re-drawn. For each accepted draw, we repeatedly calculated the Pearson $R$ between the tuned BIS/tuned $C$ and measured RV using many different realizations of each line time series. For a given draw, we retain the median Pearson $R$, and then choose the optimized bounding intensities as those with the highest median Pearson $R$. The optimized values and the corresponding improved RMS RVs are given in Table~\ref{tab:tuned_params} and plotted as the colored diamonds with blue error bars in Figure~\ref{fig:rms_ladder}. \par

Generally, tuning BIS is successful: the correlation strengths are improved, and the anomalous RVs for each line are better mitigated than in the case of the canonical BIS definition. Optimistically, the tuned BIS removes $\sim$25-30\% of the intrinsic RV variability, depending on the line. Like for BIS, the bisector curvature $C$ was successfully tuned for most lines. In fact, the tuned $C$ vastly outperformed the canonical $C$, which generally failed to produce any significant improvement over the intrinsic RMS with the exception of the Fe \textsc{I} 5383 \AA\ line. Optimistically, the tuned $C$ diagnostics remove $\sim$20-30\% of the intrinsic RV variability. At best, however, the tuned $C$ diagnostics match the performance of the tuned BIS. As previously mentioned in \S\ref{subsubsec:default_bis}, this would suggest that increasing the number of averaging regions does not increase the performance of a diagnostic. \par 

It is notable, and quite visually apparent from Figure~\ref{fig:rms_ladder}, that the lines with greater intrinsic variability see the largest absolute reduction in their RMS RV. Less variable lines see a reduction that is less significant. Interestingly, the full span in intrinsic (i.e., unmitigated) RMS RVs is about $\sim$30$\cms$ (from $\sim$75$\cms$ to $\sim$45$\cms$), whereas the span in mitigated RMS RVs is only about $\sim$15$\cms$ (from $\sim$50$\cms$ to $\sim$35$\cms$). This decreased span is interesting, and might suggest that simple linear correlations between bisector diagnostic and anomalous RV can only correct granulation velocities to some limit. This, and other possibilities, are discussed further in \S\ref{subsec:discuss_diagnostics}. \par 

\subsubsection{The Mn I 5432 \AA\ Line} \label{subsubsec:manganese}

Compared to all other lines considered in our analysis of bisector diagnostics, the Mn \textsc{I} 5432 \AA\ is an interesting outlier: the tuning of both BIS and $C$ failed for this line. The standard BIS diagnostic produced a $\sim$5\% improvement over the intrinsic RMS RV, compared to $\sim$7\% for the tuned BIS. Likewise, the standard $C$ produced a $\sim$6\% improvement, compared to $\sim$7\% for the tuned $C$. We do not consider these changes between the tuned and un-tuned diagnostics to be significant. On first inspection, it is not immediately clear why the tuning failed. Mn \textsc{I} 5432 \AA\ is among the shallowest studied lines, which might suggest that the averaging regions are too small or too close to each other to meaningfully diagnose changes in bisector shape. Yet, both the BIS and tuned BIS performed well for other shallow lines (e.g., Fe \textsc{I} 5382 \AA), suggesting that the modest depth of Mn \textsc{I} 5432 \AA\ is not the implicating factor. \par 

It is also plausible that the failure of the bisector diagnostic tuning is related to the intrinsic variability of this line: Mn \textsc{I} 5432 is the least variable line examined in this work (Table~\ref{tab:rms} and Figure~\ref{fig:rms_ladder}), and so it might follow that the shape of this line is \textit{minimally} perturbed by granulation. Lending credence to this narrative, it has been previously recognized in the literature that the properties of the Mn \textsc{I} 5395 and 5432 \AA\ lines are somewhat peculiar: these lines produce anomalous abundances \citep{Booth1984, Scott2015} and are curiously sensitive to global solar activity, despite forming in the photosphere \citep{Doyle2001, Livingston2007}. \citet{Vitas2009} demonstrate that this peculiar behavior is a consequence of the hyperfine structure of the Mn atom \citep{Murakawa1955}, which drives significant non-thermal broadening of the Mn \textsc{I} 5395 and 5432 \AA\ lines. It is this broadening, claim \citet{Vitas2009}, that makes these lines less susceptible to ``smearing'' by convective motions. The minimal variability we observe for the Mn \textsc{I} 5432 \AA\ line suggests consistency with these previous works; the failure of the diagnostic tuning can then be understood as a simple reflection of the fact that there is minimal line-shape variability to optimize on. \par

\centerwidetable
\movetabledown=10mm
\begin{rotatetable*}
\begin{deluxetable*}{c|cccc|cccc|cccc}
\tabletypesize{\small}
\tablecaption{Summary of granulation mitigation potential for individual lines using various diagnostics of bisector shapes. For each line and bisector shape diagnostic, $R$ is the median Pearson correlation coefficient between the measured line RV and considered bisector diagnostic. The corrected RMS RV and 1$\sigma$ variability in the RMS RV are reported, and the percent improvement (rounded to the nearest whole number) over the ``intrinsic'' RMS RVs reported in Table~\ref{tab:rms} are shown for each line and bisector shape diagnostic. \label{tab:correlation}}

\tablehead{\colhead{Line} & \multicolumn{4}{c}{Bisector Inverse Slope} & \multicolumn{4}{c}{Bisector Amplitude} & \multicolumn{4}{c}{Bisector Curvature}  \\
\cline{1-13}
& $R$ & RMS RV & $\sigma$ RMS RV & Imprv.\@ & $R$ & RMS RV & $\sigma$ RMS RV & Imprv.\@ & $R$ & RMS RV & $\sigma$ RMS RV & Imprv.\@  \\
(\AA) &  & ($\ms$) & ($\ms$) & (\%)&  & ($\ms$) & ($\ms$) & (\%)&  & ($\ms$) & ($\ms$) & (\%)}

\startdata
Fe I 5250.2 &              -0.603 &               0.43 &               0.07 &                  23 &          0.552 &          0.45 &          0.07 &             19 &          -0.175 &           0.54 &           0.10 &               3 \\
Fe I 5250.6 &              -0.597 &               0.42 &               0.07 &                  20 &          0.284 &          0.49 &          0.09 &              5 &          -0.008 &           0.52 &           0.10 &               0 \\
  Fe I 5379 &              -0.669 &               0.54 &               0.07 &                  27 &          0.643 &          0.55 &          0.08 &             26 &           0.022 &           0.73 &           0.17 &               1 \\
 Ti II 5381 &              -0.722 &               0.50 &               0.07 &                  32 &          0.656 &          0.53 &          0.08 &             27 &           0.141 &           0.71 &           0.14 &               3 \\
  Fe I 5382 &              -0.667 &               0.49 &               0.07 &                  27 &              - &          0.66 &          0.13 &              2 &           0.345 &           0.63 &           0.11 &               6 \\
  Fe I 5383 &              -0.701 &               0.50 &               0.07 &                  30 &          0.610 &          0.56 &          0.08 &             22 &           0.607 &           0.56 &           0.08 &              22 \\
  Mn I 5432 &              -0.258 &               0.43 &               0.08 &                   5 &         -0.252 &          0.43 &          0.08 &              5 &          -0.281 &           0.42 &           0.08 &               6 \\
  Fe I 5432 &              -0.558 &               0.39 &               0.06 &                  19 &          0.613 &          0.36 &          0.06 &             24 &           0.116 &           0.47 &           0.09 &               2 \\
  Fe I 5434 &              -0.710 &               0.36 &               0.06 &                  33 &          0.636 &          0.40 &          0.07 &             25 &           0.354 &           0.48 &           0.10 &               9 \\
  Ni I 5435 &              -0.525 &               0.41 &               0.07 &                  16 &          0.250 &          0.48 &          0.09 &              2 &          -0.037 &           0.49 &           0.10 &               0 \\
Fe I 5436.3 &              -0.569 &               0.39 &               0.06 &                  18 &         -0.338 &          0.45 &          0.08 &              6 &           0.048 &           0.48 &           0.09 &               1 \\
Fe I 5436.6 &              -0.470 &               0.44 &               0.08 &                  13 &          0.729 &          0.34 &          0.04 &             33 &          -0.018 &           0.50 &           0.10 &               0 \\
  Fe I 5576 &              -0.633 &               0.38 &               0.05 &                  25 &          0.514 &          0.42 &          0.05 &             17 &           0.307 &           0.46 &           0.06 &               7 \\
  Ni I 5578 &              -0.547 &               0.42 &               0.05 &                  17 &          0.469 &          0.44 &          0.05 &             13 &          -0.075 &           0.50 &           0.07 &               1 \\
 Fe II 6149 &              -0.466 &               0.49 &               0.06 &                  13 &          0.386 &          0.50 &          0.07 &             10 &           0.099 &           0.55 &           0.09 &               2 \\
  Fe I 6151 &              -0.533 &               0.49 &               0.07 &                  17 &          0.752 &          0.38 &          0.04 &             36 &          -0.044 &           0.58 &           0.10 &               2 \\
Ca I 6169.0 &              -0.604 &               0.38 &               0.06 &                  22 &          0.560 &          0.39 &          0.06 &             20 &           0.052 &           0.48 &           0.08 &               1 \\
Ca I 6169.5 &              -0.616 &               0.37 &               0.06 &                  24 &          0.312 &          0.46 &          0.08 &              6 &           0.077 &           0.48 &           0.09 &               2 \\
  Fe I 6170 &              -0.491 &               0.42 &               0.06 &                  15 &          0.203 &          0.48 &          0.08 &              3 &           0.219 &           0.47 &           0.08 &               4 \\
  Fe I 6173 &              -0.622 &               0.40 &               0.06 &                  24 &          0.451 &          0.46 &          0.07 &             12 &          -0.089 &           0.52 &           0.09 &               1 \\
  Fe I 6301 &              -0.661 &               0.46 &               0.06 &                  27 &          0.659 &          0.46 &          0.06 &             27 &           0.294 &           0.59 &           0.09 &               6 \\
  Fe I 6302 &              -0.634 &               0.48 &               0.07 &                  24 &          0.588 &          0.51 &          0.08 &             20 &           0.155 &           0.62 &           0.11 &               2 \\
\enddata
\end{deluxetable*}
\end{rotatetable*} \centerwidetable
\movetabledown=10mm
\begin{rotatetable*}
\begin{deluxetable*}{c|cccc|cccc|cccccc|cccc}
\tablecaption{Optimized regions for bisector diagnostics (BIS and $C$) and corresponding improvement in RMS RV. As in Table~\ref{tab:correlation}, the improvement is relative to the ``intrinsic'' RMS of each line and rounded to the nearest whole percent. The parameters $b_1$ and $b_2$ refer to the depths bounding the $v_t$ region; $b_3$ and $b_4$ bound the $v_b$ region (see Equation~\ref{eq:BIS}). Similarly for $C$: $c_1$ and $c_2$ bound $v_1$, and so on (see Equation~\ref{eq:curve}). \label{tab:tuned_params}}
\tabletypesize{\small}
\tablehead{\colhead{Line} & \multicolumn{8}{c}{Tuned BIS} & \multicolumn{10}{c}{Tuned $C$} \\ 
\cline{1-19}
 & \colhead{$b_1$} & \colhead{$b_2$} & \colhead{$b_3$} & \colhead{$b_4$} & R & RMS RV & $\sigma$ RMS RV & Imprv. & \colhead{$c_1$} & \colhead{$c_2$} & \colhead{$c_3$} & \colhead{$c_4$} & \colhead{$c_5$} & \colhead{$c_6$} &  R & RMS RV & $\sigma$ RMS RV & Imprv. \\ 
(\AA) & (\%) & (\%) & (\%) & (\%) &  &($\ms$) & ($\ms$) & (\%) & (\%) & (\%) & (\%) & (\%)& (\%)& (\%) &  &($\ms$) & ($\ms$) & (\%)}
\startdata
Fe I 5250.2 &  17 &  35 &  79 &  93 &           -0.667 &           0.43 &           0.07 &              23 &  18 &  42 &  73 &  82 &  83 &  92 &             -0.598 &             0.45 &             0.07 &                20 \\
Fe I 5250.6 &  15 &  66 &  70 &  93 &           -0.673 &           0.41 &           0.07 &              22 &  30 &  42 &  61 &  66 &  69 &  74 &             -0.584 &             0.44 &             0.08 &                15 \\
  Fe I 5379 &  43 &  60 &  73 &  79 &           -0.767 &           0.50 &           0.07 &              32 &  44 &  71 &  72 &  81 &  82 &  90 &             -0.714 &             0.52 &             0.07 &                29 \\
 Ti II 5381 &  19 &  60 &  75 &  95 &           -0.785 &           0.47 &           0.07 &              35 &  24 &  30 &  30 &  41 &  56 &  93 &              0.765 &             0.51 &             0.07 &                30 \\
  Fe I 5382 &  25 &  50 &  66 &  91 &           -0.730 &           0.48 &           0.07 &              29 &  19 &  35 &  79 &  88 &  89 &  95 &             -0.681 &             0.50 &             0.07 &                26 \\
  Fe I 5383 &  26 &  52 &  71 &  86 &           -0.760 &           0.50 &           0.07 &              31 &  26 &  67 &  77 &  84 &  87 &  94 &             -0.738 &             0.49 &             0.07 &                32 \\
  Mn I 5432 &  70 &  78 &  86 &  93 &            0.363 &           0.42 &           0.08 &               7 &  35 &  49 &  50 &  72 &  85 &  91 &             -0.387 &             0.42 &             0.08 &                 7 \\
  Fe I 5432 &  21 &  67 &  84 &  92 &           -0.666 &           0.36 &           0.06 &              24 &  26 &  65 &  68 &  77 &  78 &  84 &             -0.627 &             0.39 &             0.06 &                20 \\
  Fe I 5434 &  20 &  41 &  45 &  93 &           -0.773 &           0.35 &           0.06 &              34 &  17 &  43 &  61 &  68 &  69 &  74 &             -0.725 &             0.37 &             0.06 &                30 \\
  Ni I 5435 &  22 &  79 &  82 &  94 &           -0.651 &           0.40 &           0.07 &              18 &  33 &  73 &  78 &  84 &  84 &  90 &             -0.647 &             0.39 &             0.07 &                20 \\
Fe I 5436.3 &  23 &  41 &  57 &  91 &           -0.646 &           0.39 &           0.06 &              19 &  27 &  34 &  72 &  81 &  83 &  91 &             -0.626 &             0.40 &             0.06 &                17 \\
Fe I 5436.6 &  32 &  50 &  72 &  81 &           -0.642 &           0.40 &           0.07 &              20 &  38 &  43 &  66 &  80 &  81 &  87 &             -0.595 &             0.42 &             0.07 &                17 \\
  Fe I 5576 &  20 &  61 &  75 &  88 &           -0.687 &           0.38 &           0.05 &              25 &  30 &  45 &  75 &  86 &  87 &  93 &             -0.651 &             0.38 &             0.05 &                25 \\
  Ni I 5578 &  33 &  56 &  62 &  77 &           -0.624 &           0.41 &           0.05 &              20 &  40 &  78 &  80 &  86 &  87 &  92 &             -0.595 &             0.40 &             0.04 &                21 \\
 Fe II 6149 &  18 &  56 &  77 &  94 &           -0.537 &           0.48 &           0.06 &              15 &  39 &  46 &  73 &  78 &  82 &  88 &             -0.503 &             0.48 &             0.07 &                14 \\
  Fe I 6151 &  36 &  49 &  64 &  90 &           -0.623 &           0.48 &           0.07 &              19 &  25 &  46 &  68 &  80 &  81 &  88 &             -0.595 &             0.49 &             0.07 &                16 \\
Ca I 6169.0 &  22 &  63 &  74 &  95 &           -0.661 &           0.38 &           0.06 &              23 &  38 &  56 &  76 &  82 &  83 &  91 &             -0.588 &             0.40 &             0.06 &                18 \\
Ca I 6169.5 &  22 &  42 &  74 &  87 &           -0.680 &           0.37 &           0.06 &              24 &  22 &  56 &  81 &  87 &  87 &  94 &             -0.624 &             0.38 &             0.06 &                23 \\
  Fe I 6170 &  17 &  77 &  82 &  89 &           -0.596 &           0.40 &           0.06 &              18 &  21 &  51 &  68 &  76 &  77 &  82 &             -0.530 &             0.42 &             0.06 &                14 \\
  Fe I 6173 &  20 &  51 &  61 &  83 &           -0.695 &           0.40 &           0.06 &              24 &  31 &  50 &  77 &  82 &  84 &  90 &             -0.648 &             0.41 &             0.06 &                23 \\
  Fe I 6301 &  21 &  64 &  74 &  91 &           -0.705 &           0.46 &           0.06 &              26 &  33 &  42 &  80 &  87 &  87 &  94 &             -0.675 &             0.47 &             0.07 &                25 \\
  Fe I 6302 &  18 &  56 &  79 &  92 &           -0.698 &           0.48 &           0.07 &              26 &  22 &  62 &  70 &  78 &  79 &  89 &             -0.651 &             0.50 &             0.08 &                22 \\
\enddata
\end{deluxetable*}
\end{rotatetable*} 
\subsection{Granulation Signal Aggregation} \label{subsec:aggregate}

\begin{figure*}[!htb]
    \gridline{\fig{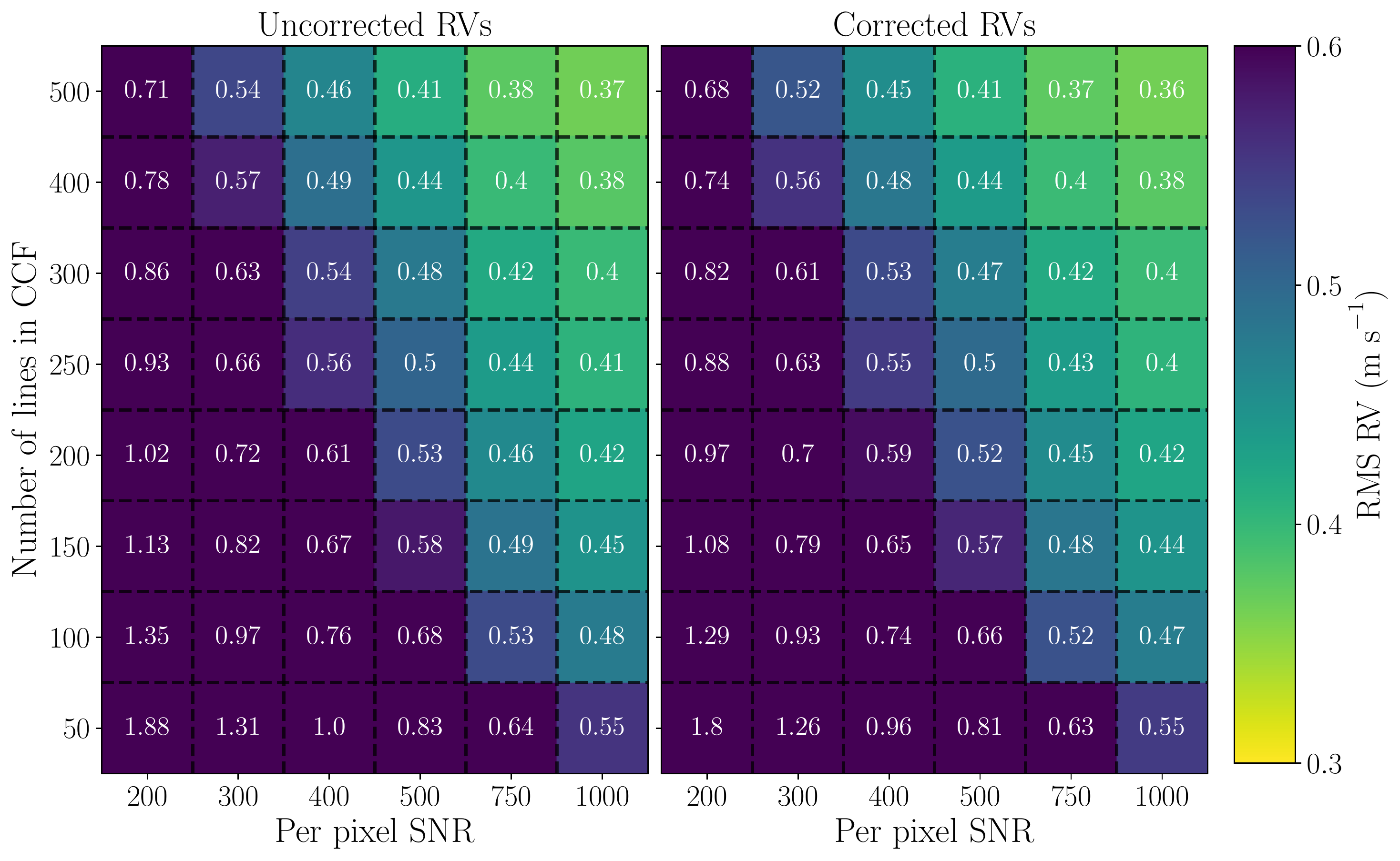}{0.51\textwidth}{$R \sim 98{,}000$ (KPF)}
              \fig{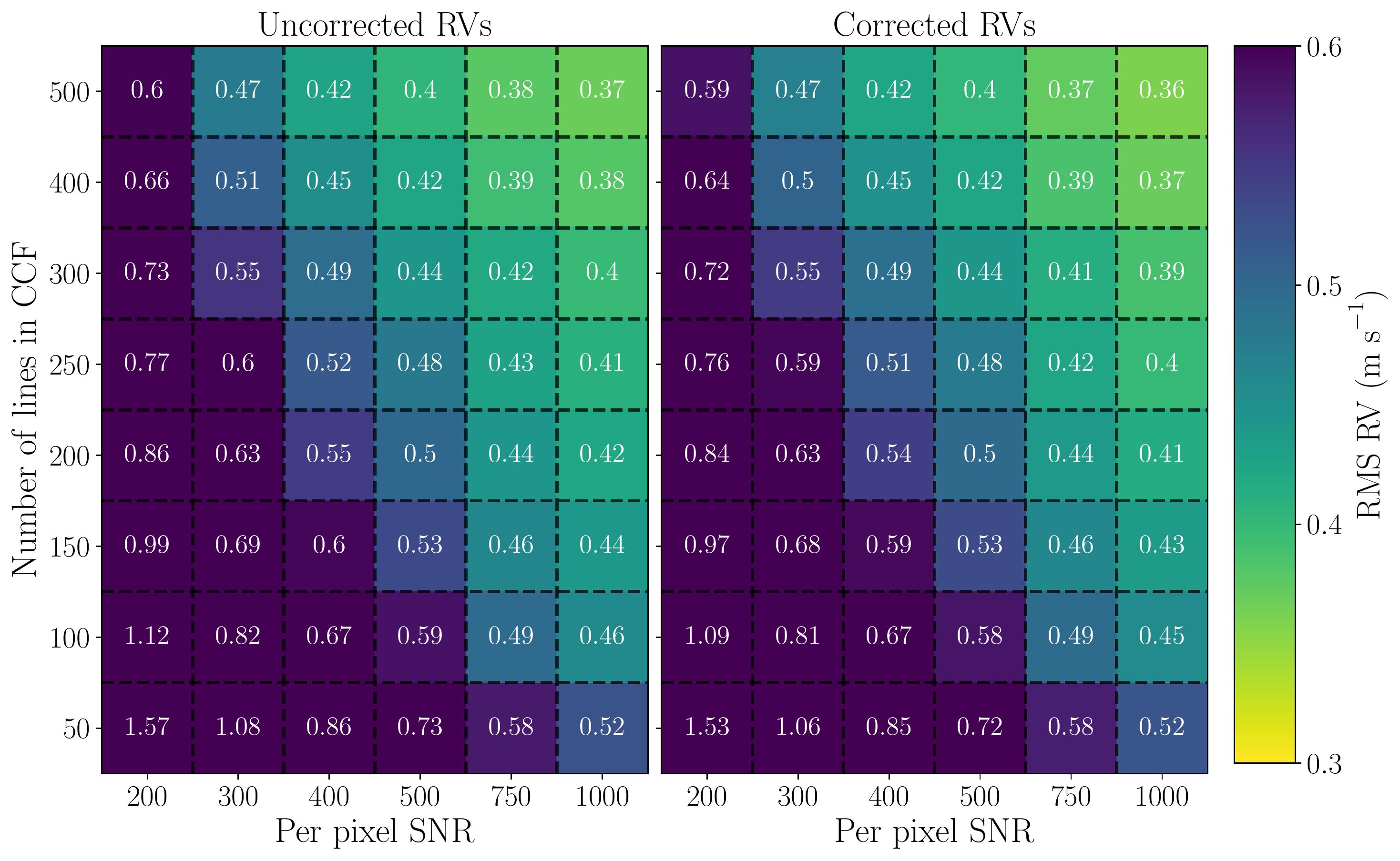}{0.51\textwidth}{$R \sim 120{,}000$ (NEID)}}
    \gridline{\fig{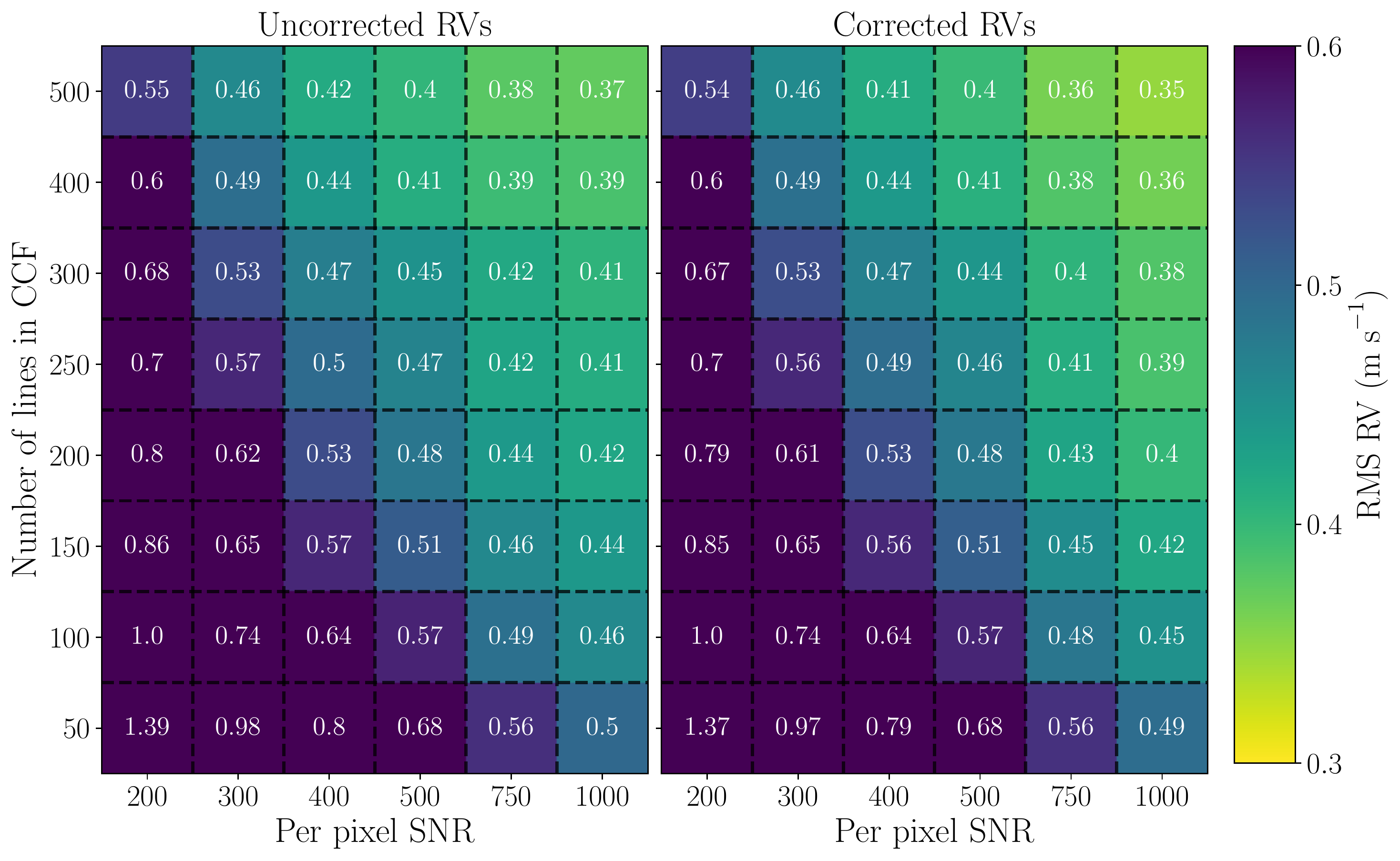}{0.51\textwidth}{$R \sim 137{,}000$ (EXPRES)}
              \fig{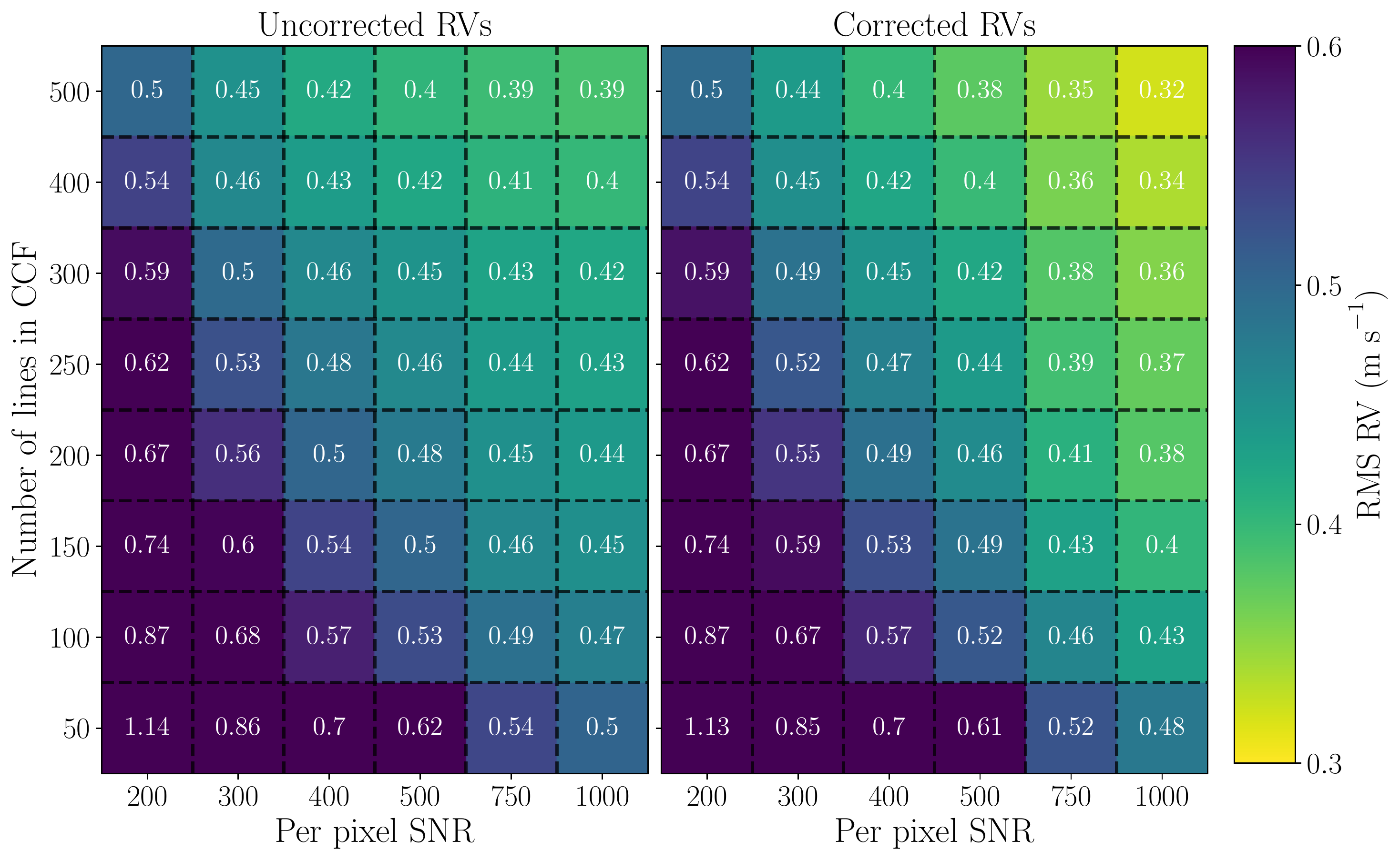}{0.51\textwidth}{$R \sim 190{,}000$ (ESPRESSO UHR)}}
    \gridline{\fig{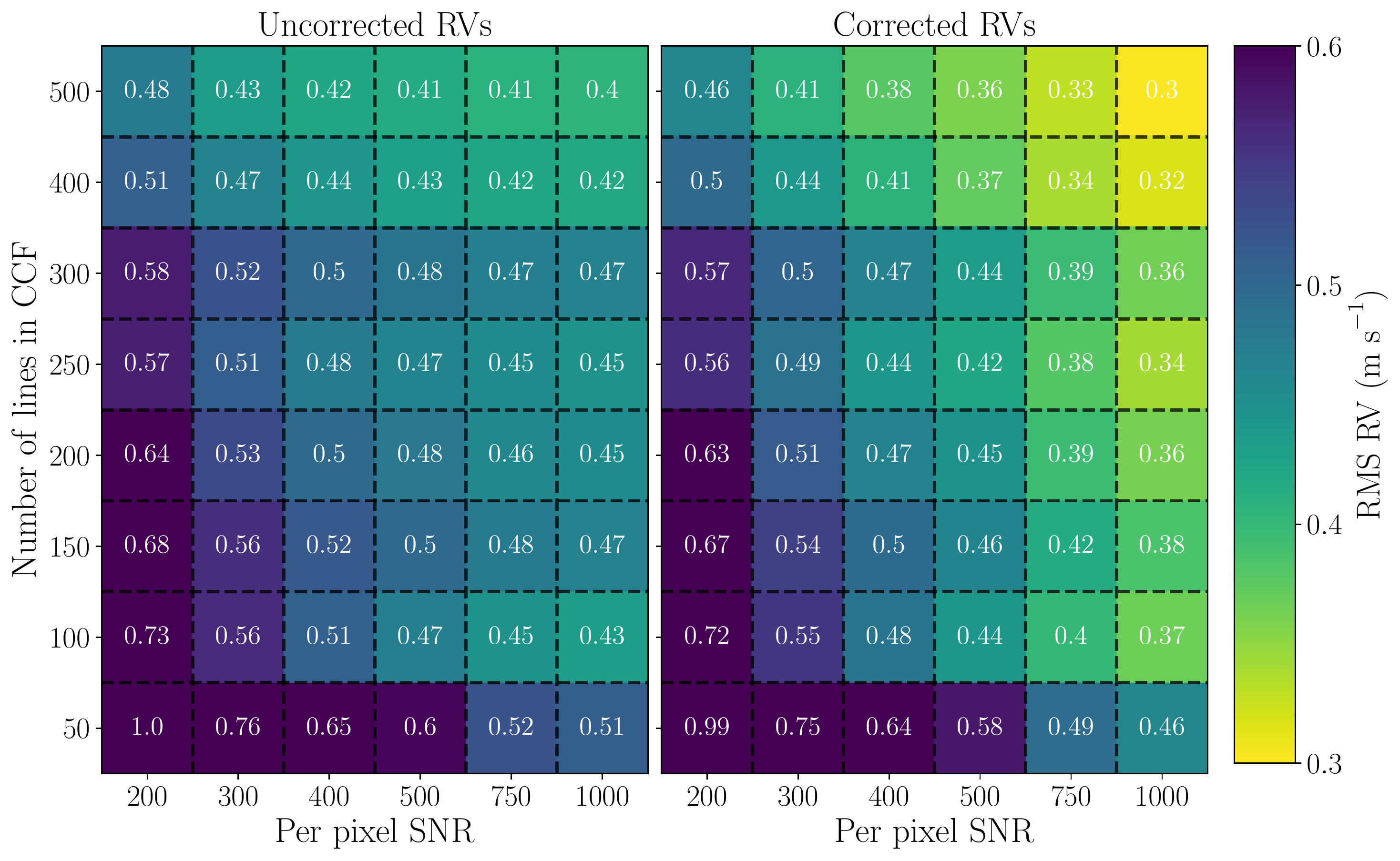}{0.51\textwidth}{$R \sim 270{,}000$ (PEPSI)}
              \fig{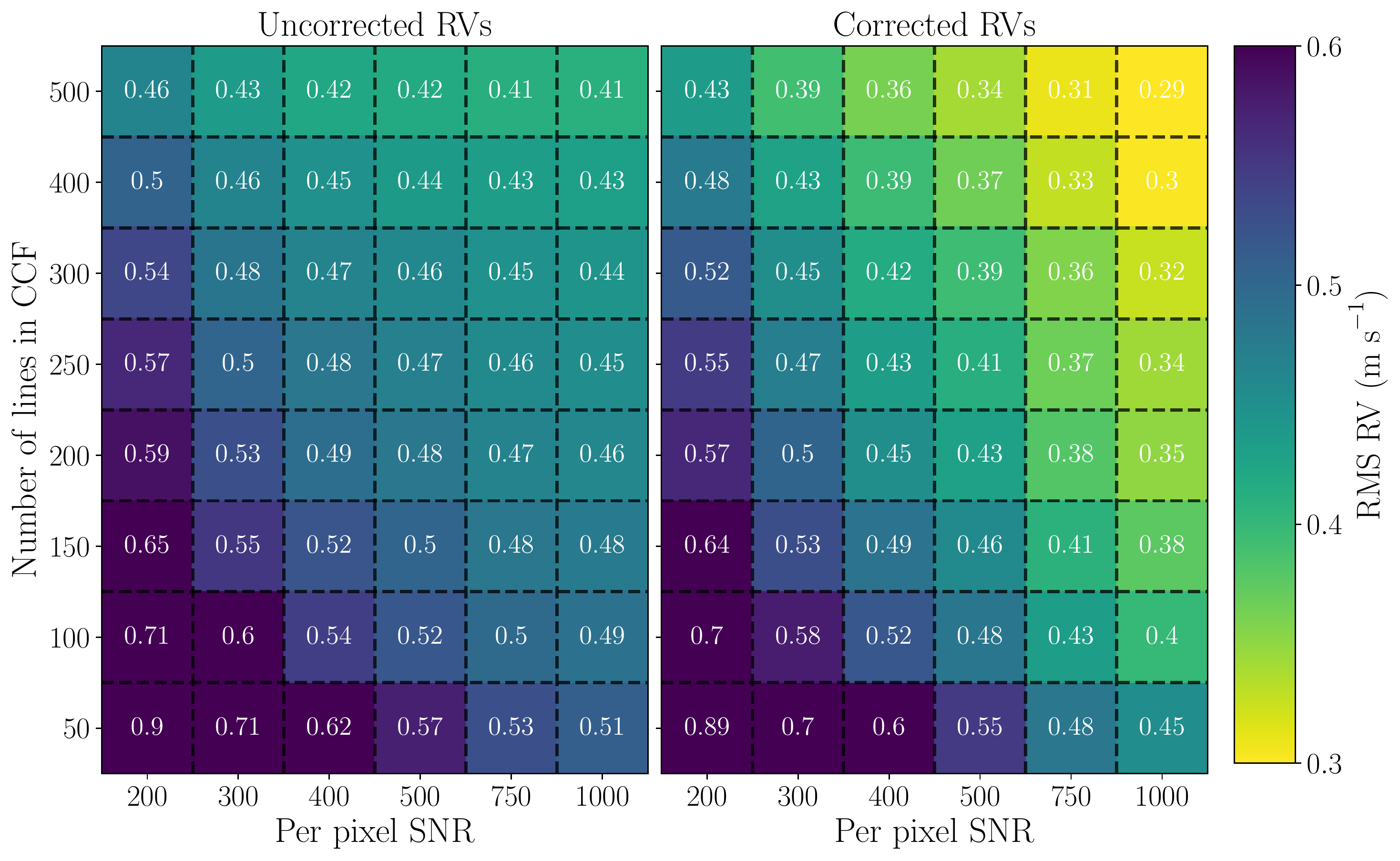}{0.51\textwidth}{$R \sim 350{,}000$ (Fictional)}}
    \caption{Results of the shape-based granulation mitigation simulation described in \S\ref{subsec:aggregate}. Each set of plots corresponds to simulations of spectra observed at spectral resolutions approximately representative of those of existing instruments, except for the bottom-right panel which is meant to represent a fictionalized ultra-high-resolution instrument as the best-case, fiducial scenario. The RMS RVs reported in left-hand panels of each figure correspond to measurements performed without any granulation mitigation. The RMS RVs in the right-hand panels were calculated from RVs corrected using the procedure described in \S\ref{subsubsec:simulation_aggregate}. The reported RMS RVs are the mean of many trials; the typical errors on the mean are at the $\lesssim$1.5\% level (not shown). Note that the step sizes for SNR and number of lines are not of fixed size. Unsurprisingly, both the raw and mitigated RMS RVs improve as the SNR and number of lines are increased. Notably, though, the achievable mitigation is strongly dependent on spectral resolution - only the three highest resolution simulations achieve statistically significant reductions in the RMS RV (see Figure~\ref{fig:improvement} and associated text).}
    \label{fig:aggregate}
\end{figure*}

In practice, it is not feasible to measure precise bisector shapes (or RVs) for individual spectral lines for stars other than the Sun, owing to limitations imposed principally by photon noise, spectral resolution, and detector sampling. These limiting factors were not accounted for in the analysis presented in \S\ref{subsec:shape_corr}; re-performing these simulations at $R\sim120{,}000$ and per-pixel ${\rm SNR} \sim 400$ (values representative of those optimistically achieved with current EPRV surveys, such as those conducted with EXPRES or NEID), none of the considered bisector diagnostics retain their correlation with the measured line velocity (not shown), and we find RMS RVs well exceeding the $\ms$ level. This imprecision is not surprising, given that photon noise dominates compared to the variability produced by granulation in this regime. \par 

Of course, velocities are not measured from individual lines in RV surveys. Typically forward-modeling (e.g., \citealt{Petersburg2020}), a CCF-based approach (e.g., \citealt{Pepe2002}), or a line-by-line method (e.g., \citealt{Dumusque2018}) is used to aggregate velocity information across lines in order to measure a precise RV. Schematically, it may be helpful to think of this methodology in terms of the central limit theorem: each line contains some amount of information about the bulk velocity of the star with some error (e.g., from photon noise). By measuring velocities from many lines together, one can estimate the ``true'' bulk velocity as the (weighted) sample mean of the individual line velocities. \par 

However, this picture begins to fall apart when one considers the effects of granulation on spectral lines. In reality, bisector shapes (e.g., Figure~\ref{fig:bisector_dossier}) and bisector variability (e.g., Figure~\ref{fig:rms_ladder}) will differ across lines. As a result, the bisector of a CCF constructed from many disparate lines will not be representative of an underlying ``true'' bisector common to each line. Therefore, any underlying correlation between individual line profile (or individual-line CCF) bisector shape and velocity (e.g., Figure~\ref{fig:rv_bis}) is not retained in the case of the spectrum-CCF bisector shape and velocity. Indeed, one recent work \citep{Sulis2023} examined ESPRESSO CCFs (which are measured from thousands of disparate lines) for two bright stars (a G0V and F7V), and was unable to find significant correlations between CCF bisector curvatures and RVs, as in \citet{Cegla2019a} and this work. \par 

\subsubsection{Simulations of Binned Granulation Signals} \label{subsubsec:simulation_aggregate}

In principle, it may be possible to overcome this present limitation by using much more selective line lists in CCF measurements. I.e., constructing CCFs for families of lines that share very similar bisectors. To test this possibility, we used \grass\ to synthesize time series consisting of many copies of the Fe \textsc{I} 5432 \AA\ line (a moderately deep Fe \textsc{I} line with a classical ``C''-shaped bisector that achieved a fractional reduction in its RMS RV representative of the median improvement among the lines studied in this work; see Table~\ref{tab:tuned_params}). Only the central wavelengths of these copies were varied; their convective blueshifts were identical, as well as the depths of the disk-resolved line profiles (the disk-integrated line depths varied $\lesssim$1\% due to the wavelength scaling of the rotational broadening; see, e.g., \citealt{Carvalho2023}). The copies were initially placed 4 \AA\ apart, such that they were totally unblended. To mitigate any impact from aliasing with the discretized grid of wavelengths, the position of each line was perturbed by a (known) random offset drawn from a Gaussian distribution with width 0.1 \AA. Because these lines were synthesized as copies of one another with known absolute convective blueshifts, they share a common time-averaged bisector shape and they also are perturbed in an identical manner in each snapshot of the simulation. \par 

\begin{figure*}[!htb]
    \gridline{\fig{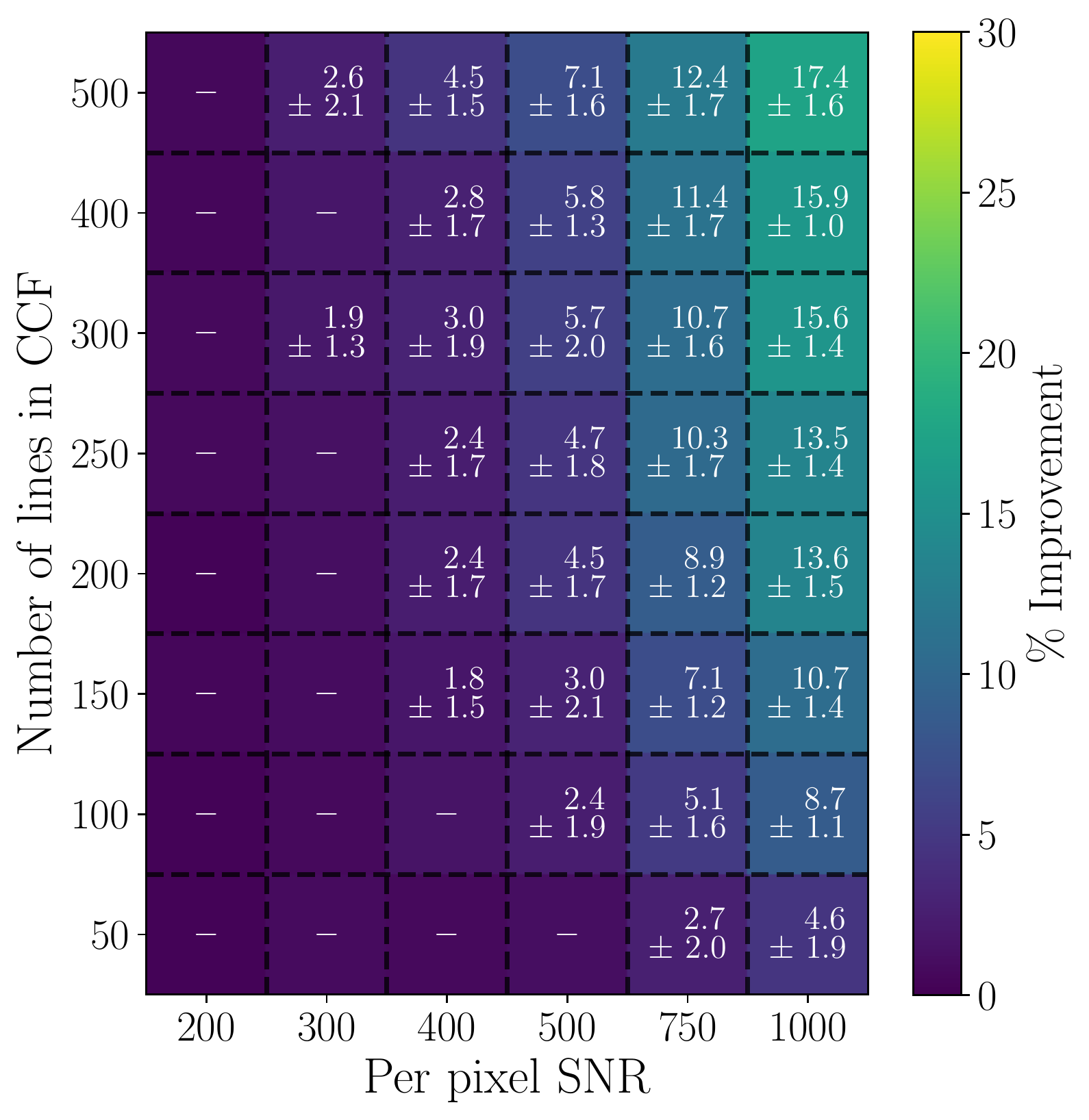}{0.33\textwidth}{$R \sim 190{,}000$ (ESPRESSO UHR)}
             \fig{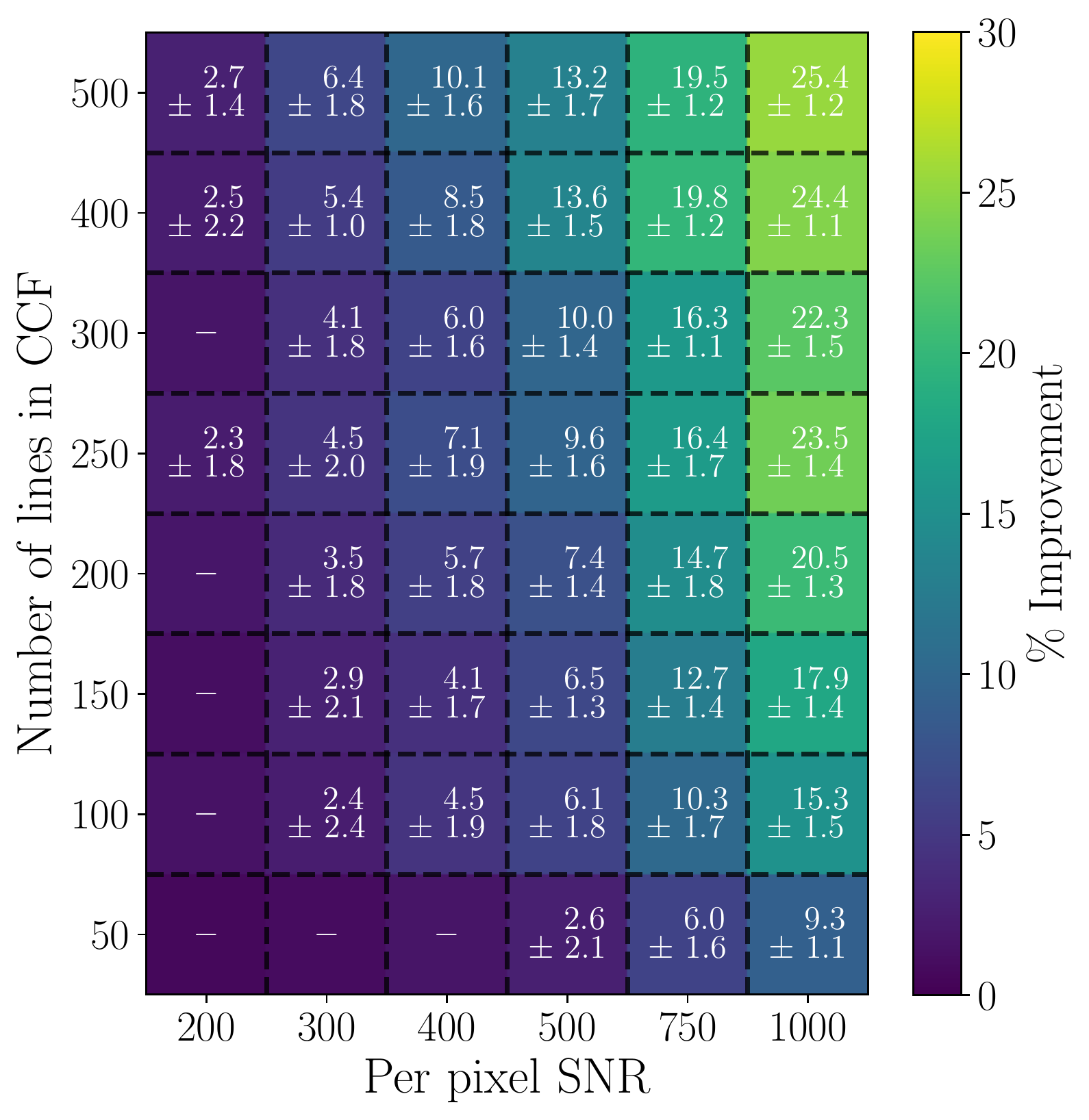}{0.33\textwidth}{$R \sim 270{,}000$ (PEPSI)}
              \fig{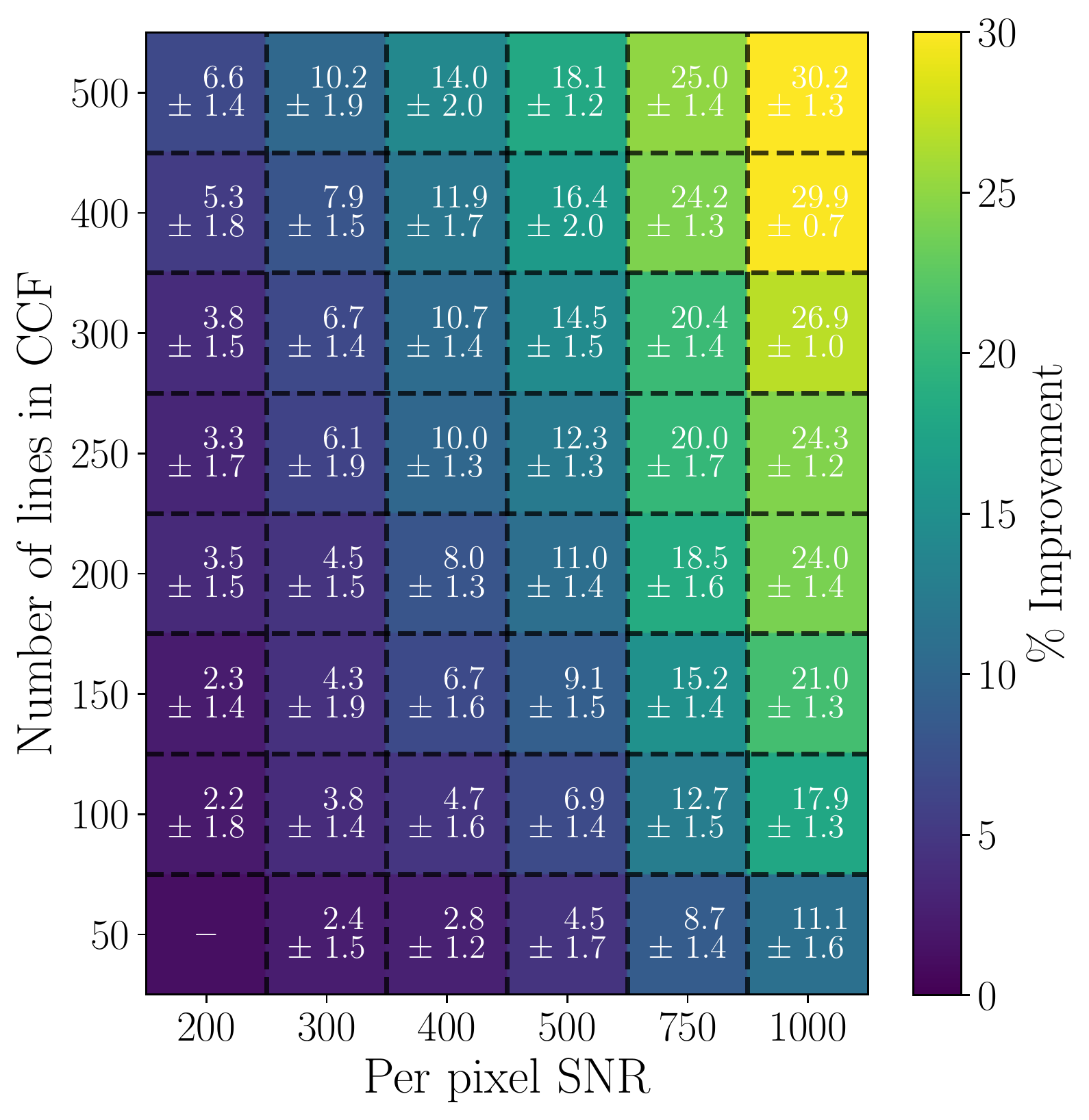}{0.33\textwidth}{$R \sim 350{,}000$ (Fictional)}}
    \caption{Percent improvement in RMS RV for a subset of the simulations described in \S\ref{subsubsec:simulation_aggregate} and shown in Figure~\ref{fig:aggregate}. Combinations of SNR and number of binned lines denoted with a ``-'' showed no significant improvement in RMS RV; none of the three lowest-resolution simulations ($R \sim 98{,}000$, $R \sim 120{,}000$, and $R \sim 137{,}000$) achieved significant improvements in RMS RV for any combination of SNR and number of binned lines and so are not shown. It is plausible that ESPRESSO in UHR mode and PEPSI could demonstrate a retrieval (and correction of) convective variability for a bright star if enough similarly-varying lines could be identified and binned, as discussed in \S\ref{subsubsec:practical}.}
    \label{fig:improvement}
\end{figure*}

Then, varying the spectral resolution, SNR, and number of identical lines considered, we calculated CCFs from which we measured velocities and bisector shape diagnostics. The resolutions considered are meant to reflect the (approximate) median resolutions of existing spectrographs: KPF ($R \sim 98{,}000$ -- \citealt{Gibson2020}), NEID ($R \sim 120{,}000$ -- \citealt{Schwab2016}), EXPRES ($R \sim 137{,}000$ -- \citealt{Jurgenson2016}), the ultra-high resolution (UHR) mode of ESPRESSO ($R \sim 190{,}000$ -- \citealt{Pepe2021}), and PEPSI ($R \sim 270{,}000$ -- \citealt{Strassmeier2015}). For the sake of comparison, we also ran simulations at $R \sim 350{,}000$ to probe a fictionalized, best-case scenario as a fiducial. In all cases, the sampling was set at $\sim$4 pixels per LSF FWHM (i.e., twice the number of pixels necessary to satisfy the Nyquist criterion), and the width of the Gaussian CCF mask was set as the speed of light divided by the resolution. We found that changing the number of pixels per LSF FWHM to two or six pixels did not significantly impact our results. Velocities were measured from the resulting CCFs as described in \S\ref{subsec:line_by_line}. This process was performed repeatedly in order to measure sample means and standard deviations. The results of this exercise are visualized in Figure~\ref{fig:aggregate} for several spectral resolutions. \par

First, we consider the results of the simulation where no granulation mitigation was performed (the left-hand panels of the plots in Figure~\ref{fig:aggregate}). Across all spectral resolutions, the observed RMS RV is dominated by photon noise when the SNR and number of lines used in the CCF are low. In the most optimistic scenario, where the number of lines is large and the SNR is high, photon noise is essentially entirely averaged out and the ``intrinsic'' variability of the line from granulation entirely dominates the RMS RV. It is notable that the higher spectral resolution simulations do not necessarily perform better when no granulation mitigation is attempted, except in the most extreme scenarios where the SNR and number of lines are quite low. Beginning at moderate SNR and number of lines (e.g., SNR $\sim$ 400 and 200 lines) and up to the highest SNR and number of lines, the measured RMS RVs improve by only several$\cms$ at most. \par 

For each combination of SNR, spectral resolution, and number of lines, we also attempted to construct a correlation between the measured velocities and tuned CCF BIS (since the tuned BIS generally outperformed all other considered diagnostics). The linear fit to these quantities was then used to correct the measured velocities, from which the final RMS RV was measured and reported. The results of this procedure are shown in the right-hand panels of Figure~\ref{fig:aggregate}, and the percent improvements in RMS RVs (with propagated errors) are shown in Figure~\ref{fig:improvement} for the subset of simulations that achieved meaningful improvement. Notably, significant mitigation via bisector diagnostic correlation can only be achieved at higher spectral resolution. At the three lowest considered resolutions ($R \leq 137{,}000$), no scientifically significant improvement in the RMS RV is achieved, even at the maximum considered SNR and number of lines. At $R \sim 190{,}000$, the improvement in RMS RVs is at the $\lesssim$20\% level in the best-case scenario, perhaps approaching detectability. At the highest considered resolutions, the most optimistic improvement in the RMS RV is consistent with that achieved in the full-resolution, noiseless simulation laid out in \S\ref{subsubsec:tuned_bis}. We discuss some considerations for attempting the observational retrieval of such signals below. \par 

\subsubsection{Practical Considerations for Existing Instruments} \label{subsubsec:practical}

The analysis presented in \S\ref{subsubsec:simulation_aggregate} and shown in Figures~\ref{fig:aggregate} and \ref{fig:improvement} suggests that existing instruments might might be able to capture (and correct for) granulation-driven variations in line shape. Whether this can be performed \textit{in reality} will depend on a few factors. First, signal must be binned across lines which are perturbed by convection in a \textit{similar-enough} manner. Some initial work \citep{Dravins2017a, Dravins2017b} has already successfully shown that ``similar'' Fe lines can be binned in order to enable retrievals of time-averaged spatially resolved stellar line profiles during planet transits. Additionally, HD simulations performed by \citet{Dravins2023} show that different absorption lines fluctuate in phase (with some dependence on line strength and spectral region), suggesting that it is indeed plausible that convective variability can be coherently binned across lines. \par 

Second, the requisite SNR of the unbinned spectra may change depending on the number of lines that are available for binning. For the science case demonstrated in \citet{Dravins2017b}, binning only 26 lines was sufficient; resolving convective variability in other stars will require much greater precision. \citet{Stenflo1977} list over 400 particularly clean Fe lines in the 400-686 nm range in the solar spectrum, so it is optimistically plausible that multiple hundred lines could be available for binning, even after accounting for practical details such as loss due to detector gaps and lines falling in the edges of orders. It is apparent from Figure~\ref{fig:improvement} that an implausible number of lines ($>$500) would need to be binned to compensate for even moderate SNRs. We caution that observations should (at least initially) target the highest achievable SNR (therefore favoring brighter stars), since the number of lines available for binning is not precisely known at present. \par 

Finally, the resolution of the observing instrument appears to be the strongest constraint on the detectability of convective variability. As discussed in \S\ref{subsubsec:simulation_aggregate}, the three lowest-resolution simulations achieved no reduction in RMS RV regardless of the SNR or number of binned lines. Optimistically, ESPRESSO in UHR mode has sufficient resolving power to detect this variability; prospects for PEPSI are slightly more promising. Additional gains can be made beyond the resolution of PEPSI, though increasing the resolution much further may offer diminishing returns (in addition to the growing challenge of reaching adequate SNR at such high resolutions). We discuss some considerations for the design of future instruments in \S\ref{subsubsec:overcome_obs}. \par 

We caveat that these simulations are \textit{not} survey or truly realistic observation simulations and should not be interpreted as such. Notably, the simulated SNR is divorced from exposure time, and the instrument LSFs are modeled as simple Gaussians. Mirror size, throughput, limiting magnitudes of the observed stellar sample, etc.\@ vary greatly across instruments, and it is beyond the scope of this study to consider these myriad factors. Instead, we emphasize that these simulations were carried out in order to assess how photon noise and spectral resolution interact and affect attempts to measure and mitigate granulation \textit{signals} encoded in the shapes of lines. As shown by this exercise, retrieving these signals given realistic observing and engineering constraints will be quite challenging. Despite this challenge, we believe that recovering these granulation signals may be possible at present, and perhaps a requirement for achieving $\sim$10$\cms$ wholesale RV precision, as discussed further in the following section. \par 

\section{Discussion} \label{discussion}

Recent studies, particularly \citet{Meunier2023} and \citet{Lakeland2024}, have shown that granulation and supergranulation will pose significant barriers to the detection of Earth-mass exoplanets, even in very magnetically inactive stars. Assessments of optimal observing strategies for mitigating granulation and supergranulation (particularly \citealt{Dumusque2011} and \citealt{Meunier2015}) agree that binning multiple observations over timescales larger than the typical (super)granule lifetime perform better than consecutive long exposures, but disagree on the absolute effectiveness of brute-force binning alone as a mitigation strategy. As an alternative to a ``beating-down-the-noise'' approach, \citet{Cegla2019a} used MHD-driven Sun-as-a-star simulations to show that various summary statistics capturing changes in individual line bisectors strongly correlate with the anomalous, granulation-induced velocity shift. To complement these previous studies, we have presented and used v2.0 of the \grass\ tool to empirically synthesize stellar line profiles with perturbations from granulation. In the following subsections, we discuss the various insights into granulation mitigation that simulations from \grass\ have offered. We additionally comment on what further work will be needed to enable and implement these mitigation methods in EPRV surveys. \par 

\subsection{Need for Broader Studies of Line-by-Line Variability} \label{subsec:disc_less_variable}

In \S\ref{subsec:line_by_line}, we used \grass\ to empirically measure the intrinsic, granulation-driven variability for the 22 solar lines shown in Figure~\ref{fig:spectra}. As shown in Figure~\ref{fig:rms_ladder}, the most variable line exhibits a $\gtrsim$70$\cms$ RMS RV, and the least variable line $\gtrsim$40$\cms$ RMS RV. That different lines exhibit different degrees of variability is unsurprising; past works have shown that measured RVs change with the sets of lines used to measure velocities (e.g., \citealt{Meunier2017, AlMoulla2022, AlMoulla2024}). Lines that trace global activity have been the focus of many studies and good progress has been made toward understanding the physical mechanisms underpinning their ability to trace this activity (e.g., \citealt{Vitas2009, Wise2022, Cretignier2024}). \par 

In comparison to studies of global activity, the magnetoconvective variability of individual lines has been somewhat understudied. Studies utilizing HD and MHD codes (e.g., \citealt{Dravins2017a, Cegla2019a, Dravins2023}) have made important contributions to our understanding of this ``microvariability'' (borrowing the language of \citealt{Dravins2023}), but are limited by computational costs and our current lack of disk-resolved line profiles for other stars to use as a basis for validation. A growing number of works have catalogued and investigated the absolute convective blueshift of lines both in the Sun and in other stars. Studies of the Sun have noted trends in convective blueshift with line depth (e.g., \citealt{Reiners2016b}), temperature (\citealt{AlMoulla2022}), and wavelength (\citealt{Ellwarth2023a}). Studies of other stars have shown that the gradient of convective velocities increases with stellar effective temperature (e.g., \citealt{Gray2009} and \citealt{Liebing2021}). Though it is possible to estimate an approximate granulation noise amplitude with simple stellar scaling relations \citep{Dalal2023}, previous observational studies have thus far not probed the temporal variability in the blueshifts of individual lines, a key focus of this work. \par 

Looking beyond the characterization of the 22 lines presented in this work, future studies should attempt to characterize the convective jitter in individual lines \textit{en masse}. Such works could use existing solar data sets (e.g., solar observations from HARPS-N, NEID, and/or KPF) to empirically measure these jitters, though the lower spectral resolutions of such instruments may make direct measurement difficult for individual lines. Separating the influence of global and/or localized magnetic activity may also prove quite challenging. Despite these challenges, achieving a synoptic understanding of stellar velocity fields, especially the convective velocity field, will be a key goalpost toward the detection of Earth analogs. \par 
 
\subsection{Promise and Limitations of Correction with Bisector Diagnostics} \label{subsec:discuss_diagnostics}

\citet{Cegla2019a} found that various bisector diagnostics (particularly the bisector inverse slope BIS, bisector amplitude $a_b$, and the bisector curvature $C$) could be used to remove 50-60\% of the RV noise from granulation in their synthetic Fe \textsc{I} 6302 \AA\ profiles. They noted, however, that the strength of the correlations used to achieve this reduction would likely change line-to-line and star-to-star. To expand on this study, we have performed this same exercise for the 22 solar lines studied in this work. Following \citet{Cegla2019a}, we used both the canonical definitions of these bisector diagnostics (given in \S\ref{subsec:shape_corr}) and definitions tuned to maximize the correlation for each given line and diagnostic. \par 

As \citet{Cegla2019a} predicted, the canonical definitions of BIS and $C$ need to be tuned to each line to maximize their predictive power. In general, we found that the best-performing bisector diagnostics could be used to improve the RMS RV for a given line by $\sim$25-35\% (see Table~\ref{tab:tuned_params}). \citet{Cegla2019a} were able to correct their observed variability at the $50$-$60$\% level in their most optimistic scenario, but it should be noted that their raw RMS RV was $\sim$10$\cms$, artificially low likely as a result of the enhanced magnetic field in their simulations (see discussion in \S2.3 of \citealt{Cegla2019a}). Although our best-case fractional improvement in line RMS RV is generally lower than that found by \citet{Cegla2019a}, our absolute improvement is relatively large. E.g., we achieve in excess of $20$ $\cms$ for some lines using the optimized bisector diagnostics. \par 

Interestingly, none of the diagnostics were able to correct the RMS RV of any line to below $\sim$30$\cms$, suggesting that not all of the observed RV variability is driven by changes in line asymmetry. Pure Doppler shifts of bisectors will produce no change in the considered bisector diagnostics (modulo small measurement errors owing to pixelization), and so the remaining RV variability likely constitutes an upper limit on the net RV shift introduced by granulation. Of course, the bisector-diagnostic correlations are imperfect, and so some asymmetry-driven variability likely remains. Instrumental jitter probably also contributes to this remaining variability (see \S\ref{data} and \citealt{Lohner-Bottcher2017}). It is possible to envision more advanced methods for correcting the asymmetry-driven variability (which would enable tighter constraints on the purely shift-driven granulation noise), but limitations introduced by finite sampling and photon noise will probably necessitate some degree of averaging within the bisector (as in the definitions of BIS and $C$). \par

If the residual, uncorrected RMS RV observed is indeed created primarily by pure shifts in bisector position, then another strategy may need to be devised to cope with this variability. One plausible avenue is in binning or smoothing of measurements over time. Although binning alone will likely not suffice as a granulation mitigation strategy \citep{Meunier2015}, in reality some combination of diagnostic-based granulation mitigation and averaging could be employed. As \citet{Cegla2019b} note in their conclusion, the path forward likely lies in a combination of empirically and theoretically motivated strategies. \par 

\subsubsection{Overcoming Observational Constraints} \label{subsubsec:overcome_obs}

The correlations between bisector diagnostic and anomalous RV will not be observable for \textit{individual} stellar lines given constraints introduced by the spectral resolution and typical SNR achieved in current instrumentation. These same limitations introduce large errors in the RVs measured from individual lines; in practice this is overcome by binning signal across lines via a forward-modeling, CCF, or line-by-line approach. However, as currently implemented, these techniques are not well-suited for measuring changing convective velocities: whereas a wholesale motion of the star will create a uniform Doppler shift in every line, changes in convection will manifest differentially in each line. \par 

Although it is beyond the scope of this work to devise new methods for binning granulation signals across lines, we carried out an exercise in \S\ref{subsec:aggregate} to show that, in principle, these signals can be aggregated and retrieved. Synthesizing spectra consisting of varying numbers of \textit{identical} lines with only differing central wavelengths at varying SNRs (owing to photon noise alone) and spectral resolutions, we found that correlations between CCF BIS and RV could be retrieved under reasonably realistic (if slightly difficult) conditions. Of the existing suite of high-resolution RV spectrographs, it is most plausible that ESPRESSO (in UHR mode) or PEPSI could resolve the sub-$\ms$ stellar line-shape variations that characterize granulation-driven jitter, assuming an adequate number of similarly varying lines can be identified and binned. As discussed in \S\ref{subsubsec:practical}, a recent work based on HD simulations \citep{Dravins2023} demonstrated that subsets of lines do indeed vary in phase, motivating our claim that variability in granulation signals would need to be coherently binned across lines. Line lists derived from or informed by such HD simulations could provide a valuable starting point for observational studies of convective variability in other stars.   \par 

Looking forward to future generations of instruments, particularly in the coming age of extremely large telescopes (ELTs), we emphasize that spectral resolution is fundamental to resolving the \textit{signals} that granulation encodes in the shapes of spectral lines. Though the velocities that fall out of these changes in line shape and position are largely treated and discussed as noise within the EPRV community and literature at present, it is important to recognize that mitigating their impact on the measurement of$\cms$-precise bulk motions of stars likely lies in treating them as signals that vary somewhat from line to line. As shown in \S\ref{subsec:aggregate} and Figure~\ref{fig:aggregate}, the spectral resolution of an instrument determines its ability to resolve perturbations in the shapes of lines. In order to resolve this variability, future instruments should target higher resolutions: ESPRESSO in UHR mode (at median $R \sim 190{,}000$) currently sets the bar for future instruments to surpass. As \citet{Dravins2017b} argue, such ultra-high resolution instruments would enable extremely precise and novel studies of both stars and planets. \par 

\subsection{Limitations and Caveats of \grass}

As discussed in \S6.2 of \citet{Palumbo2022}, \grass\ has limitations that should be considered both in the context of this study and before using \grass\ in future works. Principally, \grass\ uses \textit{solar} data to construct line profiles and spectra. The time-average shape of bisectors is known to be highly sensitive to the structure of the stellar atmosphere, and is therefore a strong function of the stellar spectral type and luminosity class \citep{Gray2008}. An independent limitation of \grass\ is imposed by the length of the time series used as input to the simulation. These observations are described in greater detail in \S2 of \citet{Lohner-Bottcher2019}; but, in brief, the minimum time baseline for most observation was $\sim$40 minutes. Consequently, \grass\ is sensitive to frequencies corresponding to the maximum realistically expected lifetime of granules (see \citealt{Hirzberger1999}), but is completely insensitive to longer timescale phenomenon, namely supergranulation. Because of this limitation, we refrain from running simulations of smoothing and binning as in \citet{Meunier2015}, instead focusing on line-by-line variabilities and bisector-diagnostic-based mitigation techniques. \par 

\section{Concluding Remarks} \label{conclusions}

Compared to other drivers of intrinsic stellar variability, methods for mitigating granulation-driven RV variability are comparatively poorly developed. Because stellar granulation precipitates a $\sim$30-70$\cms$ RV noise source (depending on the line or lines measured), current and upcoming RV surveys will need to develop methods for mitigating the effects of granulation in order to detect true Earth twins, i.e., Earth-mass planets orbiting in the habitable zones of Sun-like stars. In this work, we: 

\begin{enumerate}
    \setlength\itemsep{0.1em}
    \item Present v2.0 of the GRanulation And Spectrum Simulator -- \texttt{GRASS} -- and document the changes and upgrades since v1.0 \citep{Palumbo2022};
    \item Verify that \grass\ \textit{empirically} reproduces the time-averaged, disk-integrated line profiles and bisectors observed by \citet{Reiners2016b};
    \item Quantify the line-by-line RV variability in 22 solar lines, showing that there is $\sim$30$\cms$ difference between the most variable and least variable lines;
    \item Show that diagnostics of bisector shape generally correlate with the granulation-induced RV (consistent with the MHD-driven simulations of \citealt{Cegla2019a}), and can be used to remove $25$-$35$\% of the measured granulation noise;
    \item Demonstrate that although these correlations can't be retrieved for individual lines at the resolutions and typical SNRs achieved by existing spectrographs, informed and selective binning of similar lines can overcome these limitations; 
    \item Show that on the basis of their ultra-high spectral resolution, ESPRESSO (in UHR mode) and PEPSI are the best-suited existing spectrographs to demonstrate a retrieval of such line-shape correlations;
    \item Argue that future spectrographs should target ultra-high resolutions of $R \gtrsim 190{,}000$ in order to resolve convective variability in lines;
    \item Emphasize that granulation encodes a \textit{signal} in the shapes of lines, and that future instrument builders should be mindful of the high spectral resolution needed to faithfully resolve them.
\end{enumerate}

\section{Acknowledgments}
\noindent We thank the anonymous referee for their thorough review which has immensely improved the clarity and content of this work. We thank W.\@ Schmidt for facilitating the access to the LARS data used in this work. M.L.P.\@ thanks R.\@ Rubenzahl and C.\@ Dedrick for their help tuning the visualizations presented in this work. M.L.P.\@ thanks A.\@ Reiners for helpful discussions regarding the stellar surface tiling prescription used. This research was supported by Heising-Simons Foundation Grant \#2019-1177. This work was supported by a grant from the Simons Foundation/SFARI (675601, E.B.F.). E.B.F. acknowledges the support of the Ambrose Monell Foundation and the Institute for Advanced Study. M.L.P. acknowledges the support of the Penn State Academic Computing Fellowship. The Center for Exoplanets and Habitable Worlds and the Penn State Extraterrestrial Intelligence Center are supported by the Pennsylvania State University and the Eberly College of Science. Computations for this research were performed on the Pennsylvania State University’s Institute for Computational and Data Sciences’ Roar supercomputer with additional computation supported in part by NSF award PHY \#2018280. This research has made use of NASA's Astrophysics Data System Bibliographic Services. \par

The Vacuum Tower Telescope at the Observatorio del Teide on Tenerife is operated by the Leibniz Institute for Solar Physics (KIS) Freiburg which is a public-law foundation of the state of Baden-W\"urttemberg and member of the Leibniz-Gemeinschaft. The installation and characterization of LARS at the VTT were funded by Leibniz-Gemeinschaft from 2011 to 2014 through the ``Pakt f\"ur Forschung und Innovation." The initial characterization of the laser frequency comb at the VTT was based on an agreement between Max-Planck-Institute for Quantum Optics (Garching) and KIS. The scientific exploitation of LARS was supported by Deutsche Forschungsgemeinschaft under grant Schm-1168/10 between 2016 and 2018. \par 

\software{Julia \citep{Julia},
          \grass\ (v2.0; \citealt{palumbo_iii_2023_8271530}), 
          EchelleCCFs (v0.1.11; \citealt{EchelleCCFs}), 
          Matplotlib (\citealt{Hunter2007}),
          JuliaGPU \citep{JuliaGPU}}

\bibliography{bib,misc}

\begin{thebibliography}{}
\expandafter\ifx\csname natexlab\endcsname\relax\def\natexlab#1{#1}\fi
\providecommand{\url}[1]{\href{#1}{#1}}
\providecommand{\dodoi}[1]{doi:~\href{http://doi.org/#1}{\nolinkurl{#1}}}
\providecommand{\doeprint}[1]{\href{http://ascl.net/#1}{\nolinkurl{http://ascl.net/#1}}}
\providecommand{\doarXiv}[1]{\href{https://arxiv.org/abs/#1}{\nolinkurl{https://arxiv.org/abs/#1}}}

\bibitem[{{Al Moulla} {et~al.}(2024){Al Moulla}, {Dumusque}, \&
  {Cretignier}}]{AlMoulla2024}
{Al Moulla}, K., {Dumusque}, X., \& {Cretignier}, M. 2024, \aap, 683, A106,
  \dodoi{10.1051/0004-6361/202348150}

\bibitem[{{Al Moulla} {et~al.}(2022){Al Moulla}, {Dumusque}, {Cretignier},
  {Zhao}, \& {Valenti}}]{AlMoulla2022}
{Al Moulla}, K., {Dumusque}, X., {Cretignier}, M., {Zhao}, Y., \& {Valenti},
  J.~A. 2022, \aap, 664, A34, \dodoi{10.1051/0004-6361/202243276}

\bibitem[{Besard {et~al.}(2019)Besard, Churavy, Edelman, \&
  De~Sutter}]{besard2019prototyping}
Besard, T., Churavy, V., Edelman, A., \& De~Sutter, B. 2019, Advances in
  Engineering Software, 132, 29

\bibitem[{Besard {et~al.}(2018{\natexlab{a}})Besard, Foket, \&
  De~Sutter}]{JuliaGPU}
Besard, T., Foket, C., \& De~Sutter, B. 2018{\natexlab{a}}, IEEE Transactions
  on Parallel and Distributed Systems, \dodoi{10.1109/TPDS.2018.2872064}

\bibitem[{Besard {et~al.}(2018{\natexlab{b}})Besard, Foket, \&
  De~Sutter}]{besard2018juliagpu}
---. 2018{\natexlab{b}}, IEEE Transactions on Parallel and Distributed Systems,
  \dodoi{10.1109/TPDS.2018.2872064}

\bibitem[{Bezanson {et~al.}(2017)Bezanson, Edelman, Karpinski, \& Shah}]{Julia}
Bezanson, J., Edelman, A., Karpinski, S., \& Shah, V.~B. 2017, SIAM Review, 59,
  65, \dodoi{10.1137/141000671}

\bibitem[{{Blackman} {et~al.}(2020){Blackman}, {Fischer}, {Jurgenson},
  {Sawyer}, {McCracken}, {Szymkowiak}, {Petersburg}, {Ong}, {Brewer}, {Zhao},
  {Leet}, {Buchhave}, {Tronsgaard}, {Llama}, {Sawyer}, {Davis}, {Cabot},
  {Shao}, {Trahan}, {Nemati}, {Genoni}, {Pariani}, {Riva}, {Fournier}, \&
  {Pawluczyk}}]{Blackman2020}
{Blackman}, R.~T., {Fischer}, D.~A., {Jurgenson}, C.~A., {et~al.} 2020, \aj,
  159, 238, \dodoi{10.3847/1538-3881/ab811d}

\bibitem[{{Booth} {et~al.}(1984){Booth}, {Blackwell}, \& {Shallis}}]{Booth1984}
{Booth}, A.~J., {Blackwell}, D.~E., \& {Shallis}, M.~J. 1984, \mnras, 209, 77,
  \dodoi{10.1093/mnras/209.1.77}

\bibitem[{{Carnall}(2017)}]{Carnall2017}
{Carnall}, A.~C. 2017, arXiv e-prints, arXiv:1705.05165,
  \dodoi{10.48550/arXiv.1705.05165}

\bibitem[{{Carvalho} \& {Johns-Krull}(2023)}]{Carvalho2023}
{Carvalho}, A., \& {Johns-Krull}, C.~M. 2023, Research Notes of the American
  Astronomical Society, 7, 91, \dodoi{10.3847/2515-5172/acd37e}

\bibitem[{{Cegla}(2019)}]{Cegla2019b}
{Cegla}, H.~M. 2019, Geosciences, 9, 114, \dodoi{10.3390/geosciences9030114}

\bibitem[{{Cegla} {et~al.}(2013){Cegla}, {Shelyag}, {Watson}, \&
  {Mathioudakis}}]{Cegla2013}
{Cegla}, H.~M., {Shelyag}, S., {Watson}, C.~A., \& {Mathioudakis}, M. 2013,
  \apj, 763, 95, \dodoi{10.1088/0004-637X/763/2/95}

\bibitem[{{Cegla} {et~al.}(2019){Cegla}, {Watson}, {Shelyag}, {Mathioudakis},
  \& {Moutari}}]{Cegla2019a}
{Cegla}, H.~M., {Watson}, C.~A., {Shelyag}, S., {Mathioudakis}, M., \&
  {Moutari}, S. 2019, \apj, 879, 55, \dodoi{10.3847/1538-4357/ab16d3}

\bibitem[{{Cegla} {et~al.}(2018){Cegla}, {Watson}, {Shelyag}, {Chaplin},
  {Davies}, {Mathioudakis}, {Palumbo}, {Saar}, \& {Haywood}}]{Cegla2018}
{Cegla}, H.~M., {Watson}, C.~A., {Shelyag}, S., {et~al.} 2018, \apj, 866, 55,
  \dodoi{10.3847/1538-4357/aaddfc}

\bibitem[{{Chaplin} {et~al.}(2019){Chaplin}, {Cegla}, {Watson}, {Davies}, \&
  {Ball}}]{Chaplin2019}
{Chaplin}, W.~J., {Cegla}, H.~M., {Watson}, C.~A., {Davies}, G.~R., \& {Ball},
  W.~H. 2019, \aj, 157, 163, \dodoi{10.3847/1538-3881/ab0c01}

\bibitem[{{Collier Cameron} {et~al.}(2019){Collier Cameron}, {Mortier},
  {Phillips}, {Dumusque}, {Haywood}, {Langellier}, {Watson}, {Cegla}, {Costes},
  {Charbonneau}, {Coffinet}, {Latham}, {Lopez-Morales}, {Malavolta},
  {Maldonado}, {Micela}, {Milbourne}, {Molinari}, {Saar}, {Thompson},
  {Buchschacher}, {Cecconi}, {Cosentino}, {Ghedina}, {Glenday}, {Gonzalez},
  {Li}, {Lodi}, {Lovis}, {Pepe}, {Poretti}, {Rice}, {Sasselov}, {Sozzetti},
  {Szentgyorgyi}, {Udry}, \& {Walsworth}}]{CollierCameron2019}
{Collier Cameron}, A., {Mortier}, A., {Phillips}, D., {et~al.} 2019, \mnras,
  487, 1082, \dodoi{10.1093/mnras/stz1215}

\bibitem[{{Crass} {et~al.}(2021){Crass}, {Gaudi}, {Leifer}, {Beichman},
  {Bender}, {Blackwood}, {Burt}, {Callas}, {Cegla}, {Diddams}, {Dumusque},
  {Eastman}, {Ford}, {Fulton}, {Gibson}, {Halverson}, {Haywood}, {Hearty},
  {Howard}, {Latham}, {L{\"o}hner-B{\"o}ttcher}, {Mamajek}, {Mortier},
  {Newman}, {Plavchan}, {Quirrenbach}, {Reiners}, {Robertson}, {Roy}, {Schwab},
  {Seifahrt}, {Szentgyorgyi}, {Terrien}, {Teske}, {Thompson}, \&
  {Vasisht}}]{Crass2021}
{Crass}, J., {Gaudi}, B.~S., {Leifer}, S., {et~al.} 2021, arXiv e-prints,
  arXiv:2107.14291, \dodoi{10.48550/arXiv.2107.14291}

\bibitem[{{Cretignier} {et~al.}(2024){Cretignier}, {Pietrow}, \&
  {Aigrain}}]{Cretignier2024}
{Cretignier}, M., {Pietrow}, A.~G.~M., \& {Aigrain}, S. 2024, \mnras, 527,
  2940, \dodoi{10.1093/mnras/stad3292}

\bibitem[{{Dalal} {et~al.}(2023){Dalal}, {Haywood}, {Mortier}, {Chaplin}, \&
  {Meunier}}]{Dalal2023}
{Dalal}, S., {Haywood}, R.~D., {Mortier}, A., {Chaplin}, W.~J., \& {Meunier},
  N. 2023, \mnras, 525, 3344, \dodoi{10.1093/mnras/stad2393}

\bibitem[{{Dall} {et~al.}(2006){Dall}, {Santos}, {Arentoft}, {Bedding}, \&
  {Kjeldsen}}]{Dall2006}
{Dall}, T.~H., {Santos}, N.~C., {Arentoft}, T., {Bedding}, T.~R., \&
  {Kjeldsen}, H. 2006, \aap, 454, 341, \dodoi{10.1051/0004-6361:20065021}

\bibitem[{{de Beurs} {et~al.}(2022){de Beurs}, {Vanderburg}, {Shallue},
  {Dumusque}, {Cameron}, {Leet}, {Buchhave}, {Cosentino}, {Ghedina}, {Haywood},
  {Langellier}, {Latham}, {L{\'o}pez-Morales}, {Mayor}, {Micela}, {Milbourne},
  {Mortier}, {Molinari}, {Pepe}, {Phillips}, {Pinamonti}, {Piotto}, {Rice},
  {Sasselov}, {Sozzetti}, {Udry}, \& {Watson}}]{deBeurs2022}
{de Beurs}, Z.~L., {Vanderburg}, A., {Shallue}, C.~J., {et~al.} 2022, \aj, 164,
  49, \dodoi{10.3847/1538-3881/ac738e}

\bibitem[{{Doyle} {et~al.}(2001){Doyle}, {Jevremovi{\'c}}, {Short},
  {Hauschildt}, {Livingston}, \& {Vince}}]{Doyle2001}
{Doyle}, J.~G., {Jevremovi{\'c}}, D., {Short}, C.~I., {et~al.} 2001, \aap, 369,
  L13, \dodoi{10.1051/0004-6361:20010223}

\bibitem[{{Dravins} \& {Ludwig}(2023)}]{Dravins2023}
{Dravins}, D., \& {Ludwig}, H.-G. 2023, \aap, 679, A3,
  \dodoi{10.1051/0004-6361/202347142}

\bibitem[{{Dravins} {et~al.}(2017{\natexlab{a}}){Dravins}, {Ludwig},
  {Dahl{\'e}n}, \& {Pazira}}]{Dravins2017a}
{Dravins}, D., {Ludwig}, H.-G., {Dahl{\'e}n}, E., \& {Pazira}, H.
  2017{\natexlab{a}}, \aap, 605, A91, \dodoi{10.1051/0004-6361/201730901}

\bibitem[{{Dravins} {et~al.}(2017{\natexlab{b}}){Dravins}, {Ludwig},
  {Dahl{\'e}n}, \& {Pazira}}]{Dravins2017b}
---. 2017{\natexlab{b}}, \aap, 605, A90, \dodoi{10.1051/0004-6361/201730900}

\bibitem[{{Dumusque}(2018)}]{Dumusque2018}
{Dumusque}, X. 2018, \aap, 620, A47, \dodoi{10.1051/0004-6361/201833795}

\bibitem[{{Dumusque} {et~al.}(2011){Dumusque}, {Udry}, {Lovis}, {Santos}, \&
  {Monteiro}}]{Dumusque2011}
{Dumusque}, X., {Udry}, S., {Lovis}, C., {Santos}, N.~C., \& {Monteiro},
  M.~J.~P.~F.~G. 2011, \aap, 525, A140, \dodoi{10.1051/0004-6361/201014097}

\bibitem[{{Ellwarth} {et~al.}(2023{\natexlab{a}}){Ellwarth}, {Ehmann},
  {Sch{\"a}fer}, \& {Reiners}}]{Ellwarth2023a}
{Ellwarth}, M., {Ehmann}, B., {Sch{\"a}fer}, S., \& {Reiners}, A.
  2023{\natexlab{a}}, \aap, 680, A62, \dodoi{10.1051/0004-6361/202347615}

\bibitem[{{Ellwarth} {et~al.}(2023{\natexlab{b}}){Ellwarth}, {Sch{\"a}fer},
  {Reiners}, \& {Zechmeister}}]{Ellwarth2023b}
{Ellwarth}, M., {Sch{\"a}fer}, S., {Reiners}, A., \& {Zechmeister}, M.
  2023{\natexlab{b}}, \aap, 673, A19, \dodoi{10.1051/0004-6361/202245612}

\bibitem[{{Elsworth} {et~al.}(1994){Elsworth}, {Howe}, {Isaak}, {McLeod},
  {Miller}, {New}, {Speake}, \& {Wheeler}}]{Elsworth1994}
{Elsworth}, Y., {Howe}, R., {Isaak}, G.~R., {et~al.} 1994, \mnras, 269, 529,
  \dodoi{10.1093/mnras/269.3.529}

\bibitem[{Ford {et~al.}(2021)Ford, Wise, \& Palumbo}]{EchelleCCFs}
Ford, E., Wise, A., \& Palumbo, M. 2021, RvSpectML/EchelleCCFs.jl: v0.1.11,
  v0.1.11,  Zenodo, \dodoi{10.5281/zenodo.4593963}

\bibitem[{{Gibson} {et~al.}(2020){Gibson}, {Howard}, {Rider}, {Roy},
  {Edelstein}, {Kassis}, {Grillo}, {Halverson}, {Sirk}, {Smith}, {Allen},
  {Baker}, {Beichman}, {Berriman}, {Brown}, {Casey}, {Chin}, {Coutts},
  {Cowley}, {Deich}, {Feger}, {Fulton}, {Gers}, {Gurevich}, {Ishikawa},
  {James}, {Jelinsky}, {Kaye}, {Lanclos}, {Li}, {Lilley}, {McCarney}, {Miller},
  {Milner}, {O'Hanlon}, {Pember}, {Raffanti}, {Rockosi}, {Rubenzahl}, {Rumph},
  {Sandford}, {Savage}, {Schwab}, {Seifahrt}, {Shaum}, {Smith}, {Stuermer},
  {Thorne}, {Vandenberg}, {Von Boeckmann}, {Wang}, {Wang}, {Weisfeiler},
  {Wilcox}, {Wishnow}, {Wizinowich}, {Wold}, \& {Wolfenberger}}]{Gibson2020}
{Gibson}, S.~R., {Howard}, A.~W., {Rider}, K., {et~al.} 2020, in Society of
  Photo-Optical Instrumentation Engineers (SPIE) Conference Series, Vol. 11447,
  Ground-based and Airborne Instrumentation for Astronomy VIII, ed. C.~J.
  {Evans}, J.~J. {Bryant}, \& K.~{Motohara}, 1144742,
  \dodoi{10.1117/12.2561783}

\bibitem[{{Gilbertson} {et~al.}(2020){Gilbertson}, {Ford}, \&
  {Dumusque}}]{Gilbertson2020}
{Gilbertson}, C., {Ford}, E.~B., \& {Dumusque}, X. 2020, Research Notes of the
  American Astronomical Society, 4, 59, \dodoi{10.3847/2515-5172/ab8d44}

\bibitem[{{Gray}(1988)}]{Gray1988}
{Gray}, D.~F. 1988, {Lectures on spectral-line analysis: F,G, and K stars}

\bibitem[{{Gray}(2008)}]{Gray2008}
---. 2008, {The Observation and Analysis of Stellar Photospheres}

\bibitem[{{Gray}(2009)}]{Gray2009}
---. 2009, \apj, 697, 1032, \dodoi{10.1088/0004-637X/697/2/1032}

\bibitem[{{Gupta} {et~al.}(2021){Gupta}, {Wright}, {Robertson}, {Halverson},
  {Luhn}, {Roy}, {Mahadevan}, {Ford}, {Bender}, {Blake}, {Hearty}, {Kanodia},
  {Logsdon}, {McElwain}, {Monson}, {Ninan}, {Schwab}, {Stef{\'a}nsson}, \&
  {Terrien}}]{Gupta2021}
{Gupta}, A.~F., {Wright}, J.~T., {Robertson}, P., {et~al.} 2021, \aj, 161, 130,
  \dodoi{10.3847/1538-3881/abd79e}

\bibitem[{{Gupta} {et~al.}(2022){Gupta}, {Luhn}, {Wright}, {Mahadevan}, {Ford},
  {Stef{\'a}nsson}, {Bender}, {Blake}, {Halverson}, {Hearty}, {Kanodia},
  {Logsdon}, {McElwain}, {Ninan}, {Robertson}, {Roy}, {Schwab}, \&
  {Terrien}}]{Gupta2022}
{Gupta}, A.~F., {Luhn}, J., {Wright}, J.~T., {et~al.} 2022, \aj, 164, 254,
  \dodoi{10.3847/1538-3881/ac96f3}

\bibitem[{{Hathaway}(2015)}]{Hathaway2015}
{Hathaway}, D.~H. 2015, Living Reviews in Solar Physics, 12, 4,
  \dodoi{10.1007/lrsp-2015-4}

\bibitem[{{Haywood} {et~al.}(2014){Haywood}, {Collier Cameron}, {Queloz},
  {Barros}, {Deleuil}, {Fares}, {Gillon}, {Lanza}, {Lovis}, {Moutou}, {Pepe},
  {Pollacco}, {Santerne}, {S{\'e}gransan}, \& {Unruh}}]{Haywood2014}
{Haywood}, R.~D., {Collier Cameron}, A., {Queloz}, D., {et~al.} 2014, \mnras,
  443, 2517, \dodoi{10.1093/mnras/stu1320}

\bibitem[{{Hirzberger} {et~al.}(1999){Hirzberger}, {Bonet}, {V{\'a}zquez}, \&
  {Hanslmeier}}]{Hirzberger1999}
{Hirzberger}, J., {Bonet}, J.~A., {V{\'a}zquez}, M., \& {Hanslmeier}, A. 1999,
  \apj, 515, 441, \dodoi{10.1086/307018}

\bibitem[{{Hunter}(2007)}]{Hunter2007}
{Hunter}, J.~D. 2007, Computing in Science and Engineering, 9, 90,
  \dodoi{10.1109/MCSE.2007.55}

\bibitem[{{Jurgenson} {et~al.}(2016){Jurgenson}, {Fischer}, {McCracken},
  {Sawyer}, {Szymkowiak}, {Davis}, {Muller}, \& {Santoro}}]{Jurgenson2016}
{Jurgenson}, C., {Fischer}, D., {McCracken}, T., {et~al.} 2016, in Society of
  Photo-Optical Instrumentation Engineers (SPIE) Conference Series, Vol. 9908,
  Ground-based and Airborne Instrumentation for Astronomy VI, ed. C.~J.
  {Evans}, L.~{Simard}, \& H.~{Takami}, 99086T, \dodoi{10.1117/12.2233002}

\bibitem[{Kramida {et~al.}(2020)Kramida, {Yu.~Ralchenko}, Reader, \& {and NIST
  ASD Team}}]{NISTASD}
Kramida, A., {Yu.~Ralchenko}, Reader, J., \& {and NIST ASD Team}. 2020, {NIST
  Atomic Spectra Database (ver. 5.8), [Online]. Available:
  {\tt{https://physics.nist.gov/asd}} [2021, April 13]. National Institute of
  Standards and Technology, Gaithersburg, MD.}

\bibitem[{{Lakeland} {et~al.}(2024){Lakeland}, {Naylor}, {Haywood}, {Meunier},
  {Rescigno}, {Dalal}, {Mortier}, {Thompson}, {Cameron}, {Dumusque},
  {L{\'o}pez-Morales}, {Pepe}, {Rice}, {Sozzetti}, {Udry}, {Ford}, {Ghedina},
  \& {Lodi}}]{Lakeland2024}
{Lakeland}, B.~S., {Naylor}, T., {Haywood}, R.~D., {et~al.} 2024, \mnras, 527,
  7681, \dodoi{10.1093/mnras/stad3723}

\bibitem[{{Liang} {et~al.}(2024){Liang}, {Winn}, \& {Melchior}}]{Liang2024}
{Liang}, Y., {Winn}, J.~N., \& {Melchior}, P. 2024, \aj, 167, 23,
  \dodoi{10.3847/1538-3881/ad0e01}

\bibitem[{{Liebing} {et~al.}(2021){Liebing}, {Jeffers}, {Reiners}, \&
  {Zechmeister}}]{Liebing2021}
{Liebing}, F., {Jeffers}, S.~V., {Reiners}, A., \& {Zechmeister}, M. 2021,
  \aap, 654, A168, \dodoi{10.1051/0004-6361/202039607}

\bibitem[{{Livingston} {et~al.}(1999){Livingston}, {Wallace}, {Huang}, \&
  {Moise}}]{Livingston1999}
{Livingston}, W., {Wallace}, L., {Huang}, Y., \& {Moise}, E. 1999, in
  Astronomical Society of the Pacific Conference Series, Vol. 183, High
  Resolution Solar Physics: Theory, Observations, and Techniques, ed. T.~R.
  {Rimmele}, K.~S. {Balasubramaniam}, \& R.~R. {Radick}, 494

\bibitem[{{Livingston} {et~al.}(2007){Livingston}, {Wallace}, {White}, \&
  {Giampapa}}]{Livingston2007}
{Livingston}, W., {Wallace}, L., {White}, O.~R., \& {Giampapa}, M.~S. 2007,
  \apj, 657, 1137, \dodoi{10.1086/511127}

\bibitem[{{L{\"o}hner-B{\"o}ttcher} {et~al.}(2017){L{\"o}hner-B{\"o}ttcher},
  {Schmidt}, {Doerr}, {Kentischer}, {Steinmetz}, {Probst}, \&
  {Holzwarth}}]{Lohner-Bottcher2017}
{L{\"o}hner-B{\"o}ttcher}, J., {Schmidt}, W., {Doerr}, H.~P., {et~al.} 2017,
  \aap, 607, A12, \dodoi{10.1051/0004-6361/201731164}

\bibitem[{{L{\"o}hner-B{\"o}ttcher} {et~al.}(2019){L{\"o}hner-B{\"o}ttcher},
  {Schmidt}, {Schlichenmaier}, {Steinmetz}, \&
  {Holzwarth}}]{Lohner-Bottcher2019}
{L{\"o}hner-B{\"o}ttcher}, J., {Schmidt}, W., {Schlichenmaier}, R.,
  {Steinmetz}, T., \& {Holzwarth}, R. 2019, \aap, 624, A57,
  \dodoi{10.1051/0004-6361/201834925}

\bibitem[{{L{\"o}hner-B{\"o}ttcher} {et~al.}(2018){L{\"o}hner-B{\"o}ttcher},
  {Schmidt}, {Stief}, {Steinmetz}, \& {Holzwarth}}]{Lohner-Bottcher2018}
{L{\"o}hner-B{\"o}ttcher}, J., {Schmidt}, W., {Stief}, F., {Steinmetz}, T., \&
  {Holzwarth}, R. 2018, \aap, 611, A4, \dodoi{10.1051/0004-6361/201732107}

\bibitem[{{Meunier} {et~al.}(2017){Meunier}, {Lagrange}, \&
  {Borgniet}}]{Meunier2017}
{Meunier}, N., {Lagrange}, A.~M., \& {Borgniet}, S. 2017, \aap, 607, A6,
  \dodoi{10.1051/0004-6361/201630328}

\bibitem[{{Meunier} {et~al.}(2015){Meunier}, {Lagrange}, {Borgniet}, \&
  {Rieutord}}]{Meunier2015}
{Meunier}, N., {Lagrange}, A.~M., {Borgniet}, S., \& {Rieutord}, M. 2015, \aap,
  583, A118, \dodoi{10.1051/0004-6361/201525721}

\bibitem[{{Meunier} {et~al.}(2023){Meunier}, {Pous}, {Sulis}, {Mary}, \&
  {Lagrange}}]{Meunier2023}
{Meunier}, N., {Pous}, R., {Sulis}, S., {Mary}, D., \& {Lagrange}, A.~M. 2023,
  \aap, 676, A82, \dodoi{10.1051/0004-6361/202346218}

\bibitem[{{Murakawa}(1955)}]{Murakawa1955}
{Murakawa}, K. 1955, Journal of the Physical Society of Japan, 10, 336,
  \dodoi{10.1143/JPSJ.10.336}

\bibitem[{{Pall{\'e}} {et~al.}(1999){Pall{\'e}}, {Roca Cort{\'e}s},
  {Jim{\'e}nez}, {GOLF Team}, \& {Virgo Team}}]{Palle1999}
{Pall{\'e}}, P.~L., {Roca Cort{\'e}s}, T., {Jim{\'e}nez}, A., {GOLF Team}, \&
  {Virgo Team}. 1999, in Astronomical Society of the Pacific Conference Series,
  Vol. 173, Stellar Structure: Theory and Test of Connective Energy Transport,
  ed. A.~{Gimenez}, E.~F. {Guinan}, \& B.~{Montesinos}, 297

\bibitem[{{Palumbo} {et~al.}(2022){Palumbo}, {Ford}, {Wright}, {Mahadevan},
  {Wise}, \& {L{\"o}hner-B{\"o}ttcher}}]{Palumbo2022}
{Palumbo}, Michael~L., I., {Ford}, E.~B., {Wright}, J.~T., {et~al.} 2022, \aj,
  163, 11, \dodoi{10.3847/1538-3881/ac32c2}

\bibitem[{Palumbo {et~al.}(2023{\natexlab{a}})Palumbo, Ford, Wright, Mahadevan,
  Wise, \& Lohner-Bottcher}]{palumbo_iii_2023_8271530}
Palumbo, M.~L., Ford, E.~B., Wright, J.~T., {et~al.} 2023{\natexlab{a}}, GRASS,
   Zenodo, \dodoi{10.5281/zenodo.5512774}

\bibitem[{Palumbo {et~al.}(2023{\natexlab{b}})Palumbo, Ford, Wright, Mahadevan,
  Wise, \& Löhner-Böttcher}]{palumbo_iii_michael_l_2023_8271417}
---. 2023{\natexlab{b}}, GRASS Input Data, 1.1,  Zenodo,
  \dodoi{10.5281/zenodo.8271417}

\bibitem[{{Pepe} {et~al.}(2002){Pepe}, {Mayor}, {Galland}, {Naef}, {Queloz},
  {Santos}, {Udry}, \& {Burnet}}]{Pepe2002}
{Pepe}, F., {Mayor}, M., {Galland}, F., {et~al.} 2002, \aap, 388, 632,
  \dodoi{10.1051/0004-6361:20020433}

\bibitem[{{Pepe} {et~al.}(2021){Pepe}, {Cristiani}, {Rebolo}, {Santos},
  {Dekker}, {Cabral}, {Di Marcantonio}, {Figueira}, {Lo Curto}, {Lovis},
  {Mayor}, {M{\'e}gevand}, {Molaro}, {Riva}, {Zapatero Osorio}, {Amate},
  {Manescau}, {Pasquini}, {Zerbi}, {Adibekyan}, {Abreu}, {Affolter}, {Alibert},
  {Aliverti}, {Allart}, {Allende Prieto}, {{\'A}lvarez}, {Alves}, {Avila},
  {Baldini}, {Bandy}, {Barros}, {Benz}, {Bianco}, {Borsa}, {Bourrier},
  {Bouchy}, {Broeg}, {Calderone}, {Cirami}, {Coelho}, {Conconi}, {Coretti},
  {Cumani}, {Cupani}, {D'Odorico}, {Damasso}, {Deiries}, {Delabre},
  {Demangeon}, {Dumusque}, {Ehrenreich}, {Faria}, {Fragoso}, {Genolet},
  {Genoni}, {G{\'e}nova Santos}, {Gonz{\'a}lez Hern{\'a}ndez}, {Hughes},
  {Iwert}, {Kerber}, {Knudstrup}, {Landoni}, {Lavie}, {Lillo-Box}, {Lizon},
  {Maire}, {Martins}, {Mehner}, {Micela}, {Modigliani}, {Monteiro}, {Monteiro},
  {Moschetti}, {Murphy}, {Nunes}, {Oggioni}, {Oliveira}, {Oshagh}, {Pall{\'e}},
  {Pariani}, {Poretti}, {Rasilla}, {Rebord{\~a}o}, {Redaelli}, {Santana
  Tschudi}, {Santin}, {Santos}, {S{\'e}gransan}, {Schmidt}, {Segovia},
  {Sosnowska}, {Sozzetti}, {Sousa}, {Span{\`o}}, {Su{\'a}rez Mascare{\~n}o},
  {Tabernero}, {Tenegi}, {Udry}, \& {Zanutta}}]{Pepe2021}
{Pepe}, F., {Cristiani}, S., {Rebolo}, R., {et~al.} 2021, \aap, 645, A96,
  \dodoi{10.1051/0004-6361/202038306}

\bibitem[{{Petersburg} {et~al.}(2020){Petersburg}, {Ong}, {Zhao}, {Blackman},
  {Brewer}, {Buchhave}, {Cabot}, {Davis}, {Jurgenson}, {Leet}, {McCracken},
  {Sawyer}, {Sharov}, {Tronsgaard}, {Szymkowiak}, \&
  {Fischer}}]{Petersburg2020}
{Petersburg}, R.~R., {Ong}, J.~M.~J., {Zhao}, L.~L., {et~al.} 2020, \aj, 159,
  187, \dodoi{10.3847/1538-3881/ab7e31}

\bibitem[{{Piskunov} \& {Kochukhov}(2002)}]{Piskunov2002}
{Piskunov}, N., \& {Kochukhov}, O. 2002, \aap, 381, 736,
  \dodoi{10.1051/0004-6361:20011517}

\bibitem[{{Povich} {et~al.}(2001){Povich}, {Giampapa}, {Valenti}, {Tilleman},
  {Barden}, {Deming}, {Livingston}, \& {Pilachowski}}]{Povich2001}
{Povich}, M.~S., {Giampapa}, M.~S., {Valenti}, J.~A., {et~al.} 2001, \aj, 121,
  1136, \dodoi{10.1086/318745}

\bibitem[{{Queloz} {et~al.}(2001){Queloz}, {Henry}, {Sivan}, {Baliunas},
  {Beuzit}, {Donahue}, {Mayor}, {Naef}, {Perrier}, \& {Udry}}]{Queloz2001}
{Queloz}, D., {Henry}, G.~W., {Sivan}, J.~P., {et~al.} 2001, \aap, 379, 279,
  \dodoi{10.1051/0004-6361:20011308}

\bibitem[{{Rajpaul} {et~al.}(2015){Rajpaul}, {Aigrain}, {Osborne}, {Reece}, \&
  {Roberts}}]{Rajpaul2015}
{Rajpaul}, V., {Aigrain}, S., {Osborne}, M.~A., {Reece}, S., \& {Roberts}, S.
  2015, \mnras, 452, 2269, \dodoi{10.1093/mnras/stv1428}

\bibitem[{{Rajpaul} {et~al.}(2016){Rajpaul}, {Aigrain}, \&
  {Roberts}}]{Rajpaul2016}
{Rajpaul}, V., {Aigrain}, S., \& {Roberts}, S. 2016, \mnras, 456, L6,
  \dodoi{10.1093/mnrasl/slv164}

\bibitem[{{Reiners} {et~al.}(2016{\natexlab{a}}){Reiners}, {Lemke}, {Bauer},
  {Beeck}, \& {Huke}}]{Reiners2016a}
{Reiners}, A., {Lemke}, U., {Bauer}, F., {Beeck}, B., \& {Huke}, P.
  2016{\natexlab{a}}, \aap, 595, A26, \dodoi{10.1051/0004-6361/201629088}

\bibitem[{{Reiners} {et~al.}(2016{\natexlab{b}}){Reiners}, {Mrotzek}, {Lemke},
  {Hinrichs}, \& {Reinsch}}]{Reiners2016b}
{Reiners}, A., {Mrotzek}, N., {Lemke}, U., {Hinrichs}, J., \& {Reinsch}, K.
  2016{\natexlab{b}}, \aap, 587, A65, \dodoi{10.1051/0004-6361/201527530}

\bibitem[{{Rieutord} \& {Rincon}(2010)}]{Rieutord2010}
{Rieutord}, M., \& {Rincon}, F. 2010, Living Reviews in Solar Physics, 7, 2,
  \dodoi{10.12942/lrsp-2010-2}

\bibitem[{{Scherrer} {et~al.}(2012){Scherrer}, {Schou}, {Bush}, {Kosovichev},
  {Bogart}, {Hoeksema}, {Liu}, {Duvall}, {Zhao}, {Title}, {Schrijver},
  {Tarbell}, \& {Tomczyk}}]{Scherrer2012}
{Scherrer}, P.~H., {Schou}, J., {Bush}, R.~I., {et~al.} 2012, \solphys, 275,
  207, \dodoi{10.1007/s11207-011-9834-2}

\bibitem[{{Schwab} {et~al.}(2016){Schwab}, {Rakich}, {Gong}, {Mahadevan},
  {Halverson}, {Roy}, {Terrien}, {Robertson}, {Hearty}, {Levi}, {Monson},
  {Wright}, {McElwain}, {Bender}, {Blake}, {St{\"u}rmer}, {Gurevich},
  {Chakraborty}, \& {Ramsey}}]{Schwab2016}
{Schwab}, C., {Rakich}, A., {Gong}, Q., {et~al.} 2016, in Society of
  Photo-Optical Instrumentation Engineers (SPIE) Conference Series, Vol. 9908,
  Ground-based and Airborne Instrumentation for Astronomy VI, ed. C.~J.
  {Evans}, L.~{Simard}, \& H.~{Takami}, 99087H, \dodoi{10.1117/12.2234411}

\bibitem[{{Scott} {et~al.}(2015){Scott}, {Asplund}, {Grevesse}, {Bergemann}, \&
  {Sauval}}]{Scott2015}
{Scott}, P., {Asplund}, M., {Grevesse}, N., {Bergemann}, M., \& {Sauval}, A.~J.
  2015, \aap, 573, A26, \dodoi{10.1051/0004-6361/201424110}

\bibitem[{{Smitha} {et~al.}(2020){Smitha}, {Holzreuter}, {van Noort}, \&
  {Solanki}}]{Smitha2020}
{Smitha}, H.~N., {Holzreuter}, R., {van Noort}, M., \& {Solanki}, S.~K. 2020,
  \aap, 633, A157, \dodoi{10.1051/0004-6361/201937041}

\bibitem[{{Stenflo} \& {Lindegren}(1977)}]{Stenflo1977}
{Stenflo}, J.~O., \& {Lindegren}, L. 1977, \aap, 59, 367

\bibitem[{{Stief} {et~al.}(2019){Stief}, {L{\"o}hner-B{\"o}ttcher}, {Schmidt},
  {Steinmetz}, \& {Holzwarth}}]{Stief2019}
{Stief}, F., {L{\"o}hner-B{\"o}ttcher}, J., {Schmidt}, W., {Steinmetz}, T., \&
  {Holzwarth}, R. 2019, \aap, 622, A34, \dodoi{10.1051/0004-6361/201834538}

\bibitem[{{Strassmeier} {et~al.}(2015){Strassmeier}, {Ilyin}, {J{\"a}rvinen},
  {Weber}, {Woche}, {Barnes}, {Bauer}, {Beckert}, {Bittner}, {Bredthauer},
  {Carroll}, {Denker}, {Dionies}, {DiVarano}, {D{\"o}scher}, {Fechner},
  {Feuerstein}, {Granzer}, {Hahn}, {Harnisch}, {Hofmann}, {Lesser}, {Paschke},
  {Pankratow}, {Plank}, {Pl{\"u}schke}, {Popow}, \&
  {Sablowski}}]{Strassmeier2015}
{Strassmeier}, K.~G., {Ilyin}, I., {J{\"a}rvinen}, A., {et~al.} 2015,
  Astronomische Nachrichten, 336, 324, \dodoi{10.1002/asna.201512172}

\bibitem[{{Sulis} {et~al.}(2023){Sulis}, {Lendl}, {Cegla}, {Rodr{\'\i}guez
  D{\'\i}az}, {Bigot}, {Van Grootel}, {Bekkelien}, {Cameron}, {Maxted},
  {Simon}, {Lovis}, {Scandariato}, {Bruno}, {Nardiello}, {Bonfanti},
  {Fridlund}, {Persson}, {Salmon}, {Sousa}, {Wilson}, {Krenn}, {Hoyer},
  {Santerne}, {Ehrenreich}, {Alibert}, {Alonso}, {Anglada}, {B{\'a}rczy},
  {Barrado y Navascues}, {Barros}, {Baumjohann}, {Beck}, {Beck}, {Benz},
  {Billot}, {Bonfils}, {Borsato}, {Brandeker}, {Broeg}, {Cabrera}, {Charnoz},
  {Corral van Damme}, {Csizmadia}, {Davies}, {Deleuil}, {Deline}, {Delrez},
  {Demangeon}, {Demory}, {Erikson}, {Fortier}, {Fossati}, {Gandolfi}, {Gillon},
  {G{\"u}del}, {Heng}, {Isaak}, {Kiss}, {Laskar}, {Lecavelier des Etangs},
  {Magrin}, {Munari}, {Nascimbeni}, {Olofsson}, {Ottensamer}, {Pagano},
  {Pall{\'e}}, {Peter}, {Piotto}, {Pollacco}, {Queloz}, {Ragazzoni}, {Rando},
  {Rauer}, {Ribas}, {Rieder}, {Santos}, {S{\'e}gransan}, {Smith},
  {Steinberger}, {Steller}, {Szab{\'o}}, {Thomas}, {Udry}, {Walton}, \&
  {Wolter}}]{Sulis2023}
{Sulis}, S., {Lendl}, M., {Cegla}, H.~M., {et~al.} 2023, \aap, 670, A24,
  \dodoi{10.1051/0004-6361/202244223}

\bibitem[{{Vitas} {et~al.}(2009){Vitas}, {Viticchi{\`e}}, {Rutten}, \&
  {V{\"o}gler}}]{Vitas2009}
{Vitas}, N., {Viticchi{\`e}}, B., {Rutten}, R.~J., \& {V{\"o}gler}, A. 2009,
  \aap, 499, 301, \dodoi{10.1051/0004-6361/200810600}

\bibitem[{{Vogt} {et~al.}(1987){Vogt}, {Penrod}, \& {Hatzes}}]{Vogt1987}
{Vogt}, S.~S., {Penrod}, G.~D., \& {Hatzes}, A.~P. 1987, \apj, 321, 496,
  \dodoi{10.1086/165647}

\bibitem[{{Wise} {et~al.}(2022){Wise}, {Plavchan}, {Dumusque}, {Cegla}, \&
  {Wright}}]{Wise2022}
{Wise}, A., {Plavchan}, P., {Dumusque}, X., {Cegla}, H., \& {Wright}, D. 2022,
  \apj, 930, 121, \dodoi{10.3847/1538-4357/ac649b}

\bibitem[{{Zhao} \& {Tinney}(2020)}]{Zhao2020}
{Zhao}, J., \& {Tinney}, C.~G. 2020, \mnras, 491, 4131,
  \dodoi{10.1093/mnras/stz3254}

\bibitem[{{Zhao} \& {Dumusque}(2023)}]{Zhao2023}
{Zhao}, Y., \& {Dumusque}, X. 2023, \aap, 671, A11,
  \dodoi{10.1051/0004-6361/202244568}

\end{thebibliography}
\bibliographystyle{aasjournal}
          
\clearpage
\appendix

\vspace{-3mm}
\section{Specific \grass\ Implementation Details} \label{app:implementation}

\grass\ uses disk- and time-resolved solar spectra to construct synthetic time-series, disk-integrated spectra. Several changes to the software implementation of \grass\ have been made since \citet{Palumbo2022}. In addition to the overview of the methods provided in \S\ref{methods}, we describe specific aspects of the input data treatment and software implementation in the following subsections. \par 

\subsection{Data Pre-processing} \label{sub:preprocess}

\begin{figure*}[!htb]
    \epsscale{1.15}
    \plotone{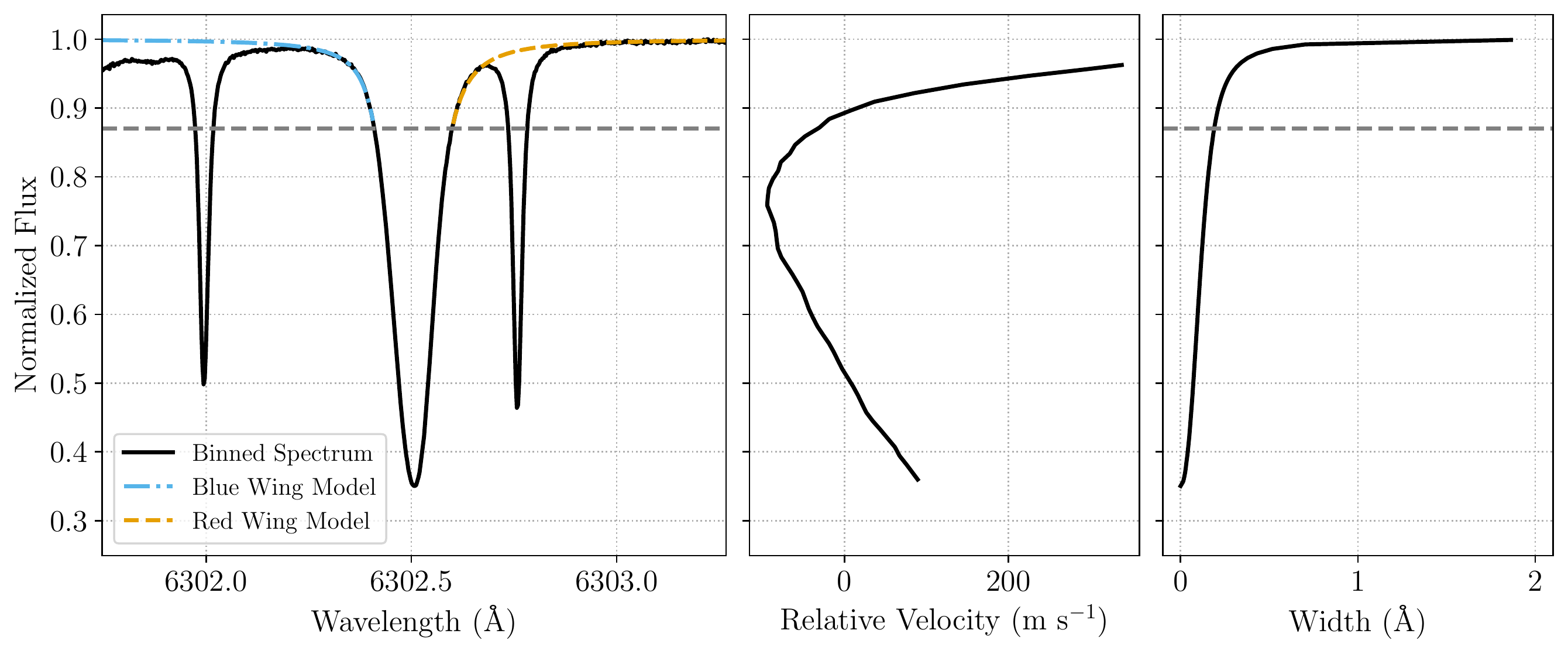}
    \caption{Example Voigt fits to observed line profile (left-hand subplot) and resulting input data. The bisector (middle subplot) is measured directly from the observed line profile, whereas the width as a function of intensity (right-hand subplot) is measured from the best-fit Voigt wings above $\sim$88\% of the continuum flux in this example.}
    \label{fig:input_fit_example}  
\end{figure*}

To reconstruct and model line profiles (see \S\ref{synthesis_procedure}), we first measure time-resolved line bisectors and line widths as a function of intensity from the LARS solar observations. As explained in \S\ref{data} and in \citet{Palumbo2022}, we refer to these measurements as the ``input data.'' To measure these input data, we follow a procedure similar to that described in \S3.2.1 of \citet{Palumbo2022}. In brief: we bin the reduced and wavelength-calibrated solar spectra to a 15-second cadence, isolate each solar line, measure line bisectors, and then fit separate Voigt profiles to each line wing in order to measure line width as a function of continuum-normalized intensity within the line. In the following paragraphs, we describe each of these steps in detail. Example line-wing fits and input data are shown in Figure~\ref{fig:input_fit_example}. \par 

As the first step in binning the reduced spectra, we resample each spectrum onto a common wavelength grid, implementing the flux-preserving resampling algorithm presented in \citet{Carnall2017}. Since nearly all of the observations conducted for \citet{Lohner-Bottcher2018}, \citet{Stief2019}, and \citet{Lohner-Bottcher2019} were performed with exposure and read-out times permitting binning to a 15-second cadence, the time-series line profiles produced by \grass\ have a 15-second time step. The minority of spectra that were observed with a cadence that cannot be binned to 15 seconds are excluded from pre-processing. To carry out the 15-second binning of the observed solar spectra, we perform a weighted average of the flux in each wavelength element, with weights computed as the inverse square of the reduction-pipeline-reported flux uncertainty (details of the raw data reduction are given in \citealt{Lohner-Bottcher2017} and \citealt{Lohner-Bottcher2018}). \par

With the spectra binned, we perform an approximate continuum normalization and then isolate each line of interest in order to measure line bisectors and widths as a function of depth within the line. Together, these two quantities can be thought of as a lossless compression of each line profile shape that will be reconstructed in the synthesis stage of \grass\ (see \S\ref{synthesis_procedure}). To measure line bisectors and widths at each depth, we perform a simple linear interpolation in normalized flux, consistent with the methods used by \citet{Gray1988} and \citet{Dall2006}. As in \citet{Palumbo2022}, we fit separate Voigt profiles to the red and blue wings of each line in order to measure a smooth line width into the continuum (because the deep solar lines of interest are often increasingly blended in their wings). We use a Voigt profile, rather than e.g., a Gaussian, since the different lines are shaped quite differently in the wings and core (see Figure~\ref{fig:spectra}). Voigt profiles are able to capture this diversity in shape more flexibly and accurately. The effects of this approach are discussed specifically for the Fe \textsc{I} 5434 \AA\ line in \S4.1 of \citet{Palumbo2022} and generally in \S\ref{validation}. \par 

Like the width measurements, the bisector measurements become fraught near the continuum, owing both to blends and measurement uncertainty which is inversely proportional to the first derivative of the spectrum at the intensity of interest \citep{Gray2008}. To compensate for this, we fit a polynomial to the lower 80\% of each bisector and model the remaining top 20\% as the extrapolation of the best-fit polynomial. Similarly to the fitting procedure employed by \citet{Zhao2023}, we use a third-order polynomial for bisectors at $\mu > 0.4$, and a first-order polynomial for bisectors at $\mu \leq 0.4$. Example disk-resolved spectra with model fits and input data with extrapolations are shown in Figure~\ref{fig:input_fit_example}. \par 

Rather than writing the pre-processed data for each limb position to separate FITS files, as in \grass\ v1.0 \citep{Palumbo2022}, we instead write all data for a given line to an HDF5 file, in order to reduce the storage size of the input data and excess I/O latency during synthesis. The input data are available in a Zenodo record \citep{palumbo_iii_michael_l_2023_8271417}, which \grass\ automatically downloads upon installation. \grass\ includes additional functions used to read and manipulate the data stored in these files. These functions are described in the package documentation available on GitHub. \par   

\subsection{Changes to Model Grid} \label{subsec:model_grid}

\begin{figure*}[!htb]
    \epsscale{1.1}
    \plotone{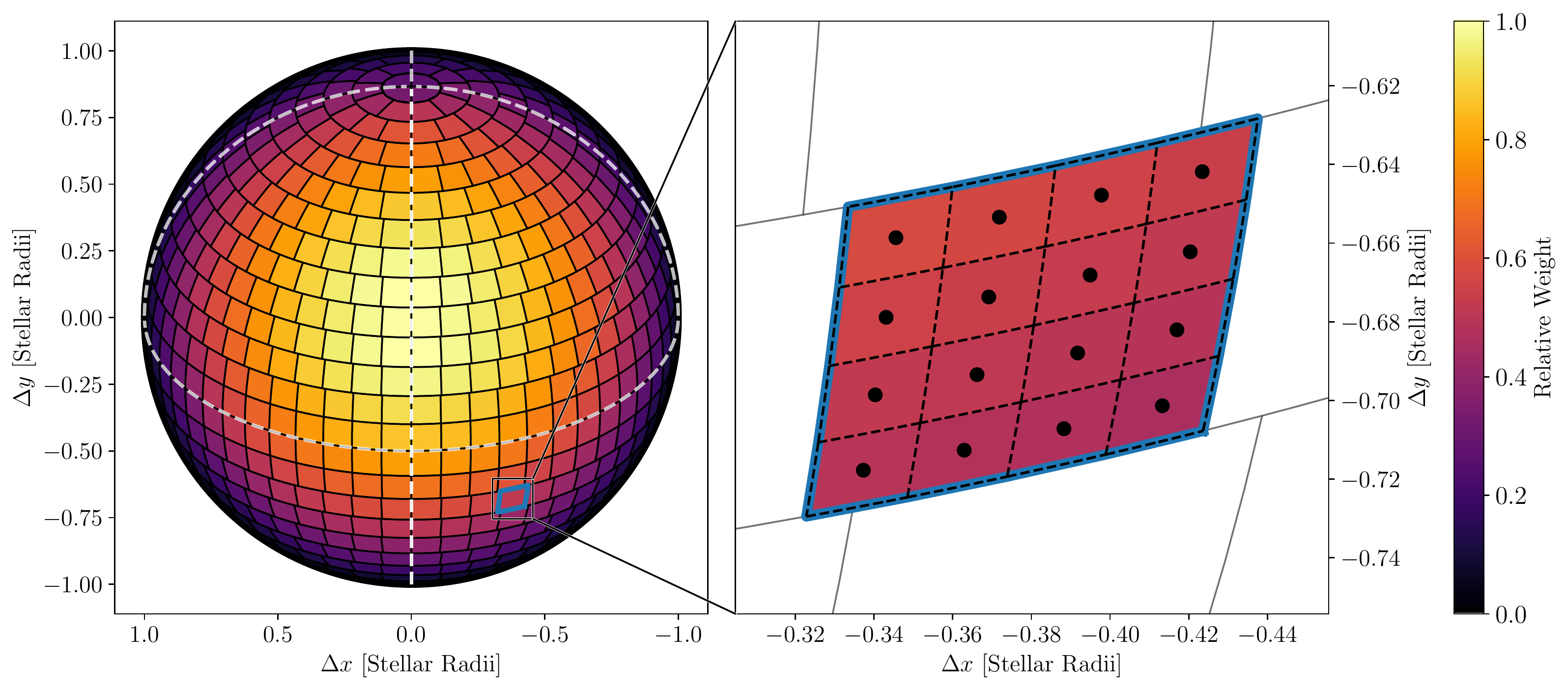}
    \caption{Tiling (left-hand plot) and sub-tiling (right-hand plot) of an example stellar surface viewed at slight inclination. In both plots, the tiles and sub-tiles are drawn artificially large for illustrative clarity. The coloring indicates the fractional weight of each tile normalized to the largest weight value; individual weights are given by the product of the limb darkening and projected area of the tile. The stellar equator and the zero- and ninety-degree meridians are overplotted as white dashed lines. \textit{Left:} The stellar surface is divided into tiles according to the prescription developed by \citet{Vogt1987}. The number of latitudinal slices (and consequently the total number of tiles) is set such that the average tile area corresponds to the average angular size of the input observations, as described in \S\ref{subsec:model_grid}. \textit{Right:} Zoom-in on the tile highlighted by the blue border in the left-hand plot, with example sub-tiling. Because the grid resolution dictated by the angular size of the input observations would normally introduce discretization errors in the rotation velocity, limb darkening, and projected tile area computations, these quantities are calculated at the centers of sub-tiles within each larger tile (indicated by the black dots). As in \cite{Cegla2019a}, the sub-tiles are spaced evenly in stellar latitude and longitude within each larger tile.} 
    \label{fig:model_grid}
\end{figure*}

As explained in \S3.2.3 and \S3.2.4 of \citet{Palumbo2022}, v1.0 of \grass\ synthesized disk-integrated line profiles by integrating over an evenly sampled 2D stellar grid projected on the sky plane. To enable more complex modeling of stellar geometries and to be consistent with other Sun-as-a-star simulations, v2.0 of \grass\ now tiles the surface of a sphere in stellar longitude and latitude, following from the procedures used in \citet{Vogt1987} and \citet{Piskunov2002}. In brief, we first divide the star into $N_{\phi}$ latitude slices. The number of longitudes sampled in each latitude slice is then given by:

\begin{equation}
    N_{\theta}(\phi) = \frac{2\pi \cos(\phi)}{\Delta\phi},
\end{equation}

\noindent where $\phi$ is the latitude at the center of a given latitudinal slice and $\Delta\phi$ the latitude span of the slices. In each slice, $N_{\theta}$ is rounded up to the nearest whole number. As noted in \citet{Vogt1987}, this tiling scheme produces tiles whose areas varies minimally with latitude, except near the poles where the tiles become triangular. Tiles that are located entirely on the invisible hemisphere of the star do not contribute to the integrated flux. Because the tiles are no longer of equivalent projected area (as in \texttt{GRASS} v1.0, \citealt{Palumbo2022}), we now weight the contribution of each tile to the disk-integrated flux by the projected area of each tile, in addition to limb darkening. An example grid with color-coded weights is shown at left in Figure~\ref{fig:model_grid}. \par 

The number of longitudinal elements is set by the latitude increment $\Delta\phi$, which is equal to $180^{\circ} / N_\phi$; consequently, we parameterize the resolution of the spatial grid in terms of only $N_{\phi}$. In previous works that have implemented the \citet{Vogt1987} tiling scheme (e.g., \citealt{Reiners2016a}), $N_\phi$ is customarily set to some very large number in order to minimize errors introduced by the discretization of the stellar surface elements. However, as explained \S4.2 of \citet{Palumbo2022}, the resolution of the spatial grid must correspond to the (average) angular size of observed patches in \citet{Lohner-Bottcher2018}  and \citet{Lohner-Bottcher2019}; i.e., there is an optimal $N_\phi$ that cannot be freely chosen. With our modified tiling procedure, we find that $N_\phi = 197$ yields tile sizes corresponding to the intensity-weighted average area of the observed patches. As in \S4.2 and Figure~4 of \citet{Palumbo2022}, we have verified that this resolution produces an RMS RV broadly consistent with those observed by \citet{Elsworth1994} and \citet{Palle1999}, with the exact RMS RV depending on the input data used to synthesize the spectra (see \S\ref{subsec:line_by_line}). \par 

Generally, at $N_\phi = 197$ the star is not tiled densely enough and appreciable ($\sim$$\ms$) errors in the rotational velocity are introduced by the sparse tessellation. To circumvent this problem, we follow the procedure developed by \cite{Cegla2019a} and compute the limb darkening, projected rotational velocity, and projected areas in each tile as the (weighted) average of those values computed on a 40-by-40 grid of sub-tiles. Within each larger tile, the sub-tiles are evenly spaced in stellar latitude and longitude. The limb-darkened intensity assigned to each large tile is then given as the mean intensity across the corresponding sub-tiles and the rotational velocity as the weighted mean of the sub-tile velocities. This sub-tiling scheme is illustrated at right in Figure~\ref{fig:model_grid}. \par 

To calibrate the necessary number of sub-tiles, we computed the disk-summed rotational velocity and projected tile area as a function of the number of sub-tiles. Ideally, as the number of sub-tiles is increased, the sum of the tile rotational velocities should tend to zero, and the sum of the projected areas should approach $\pi$. With a 40-by-40 grid of sub-tiles in each larger stellar tile, we find that the disk-summed rotational velocity error is $\sim$$3.7 \times 10^{-9}$ $\ms$; the error in the total projected area of the disk is $\sim$$5.9 \times 10^{-9}$. To assess the impact of these errors on the synthesized line profiles, we compared line profiles generated with an arbitrarily high density of sub-tiles (1600-by-1600 grid) and with the standard 40-by-40 grid. Taking the high-resolution line profile as the fiducial, the flux errors did not exceed $\sim$$8.9 \times 10^{-10}$; likewise, the line velocity errors were sufficiently small at $8.9 \times 10^{-7}$ $\ms$. \par 

\subsection{Treatment of Convective Blueshift} \label{sec:cbs}

Stellar absorption lines are known to exhibit convective blueshift that varies with both limb angle \citep{Lohner-Bottcher2018, Lohner-Bottcher2019} and depth \citep{Reiners2016b, Ellwarth2023b}. By default, \grass\ uses only the convective blueshifts in the solar observations in its construction of line profiles (as is the case for the simulations and analysis presented in this work). However, if \grass\ is used to model a line of arbitrary depth (following the procedure described in \S3.2.2 of \citealt{Palumbo2022}), it may also be desirable to artificially prescribe the disk-integrated convective blueshift of the synthetic line profile. To enable this modeling, \grass\ now optionally includes an additional convective blueshift term in order to reproduce the solar scaling relation given in \citet{Reiners2016b}. Specifically, we use Equation 2 of \citet{Reiners2016b} which gives $\Delta v_{\rm conv}$ as a function of disk-integrated, continuum-normalized line depth $d$ as:

\begin{equation}
    \Delta v_{\rm conv} = -504.891-43.7963d-145.560 d^2+884.308 d^3,
\end{equation}

\noindent where $\Delta v_{\rm conv}$ is measured in units of $\ms$. In implementation, \grass\ draws the appropriate convective blueshift from a look-up table tabulated from this polynomial relation, rather than evaluating the polynomial for each synthetic line. \par

\subsection{GPU Implementation} \label{subsec:gpu}

\begin{figure}[!htb]
\gridline{\fig{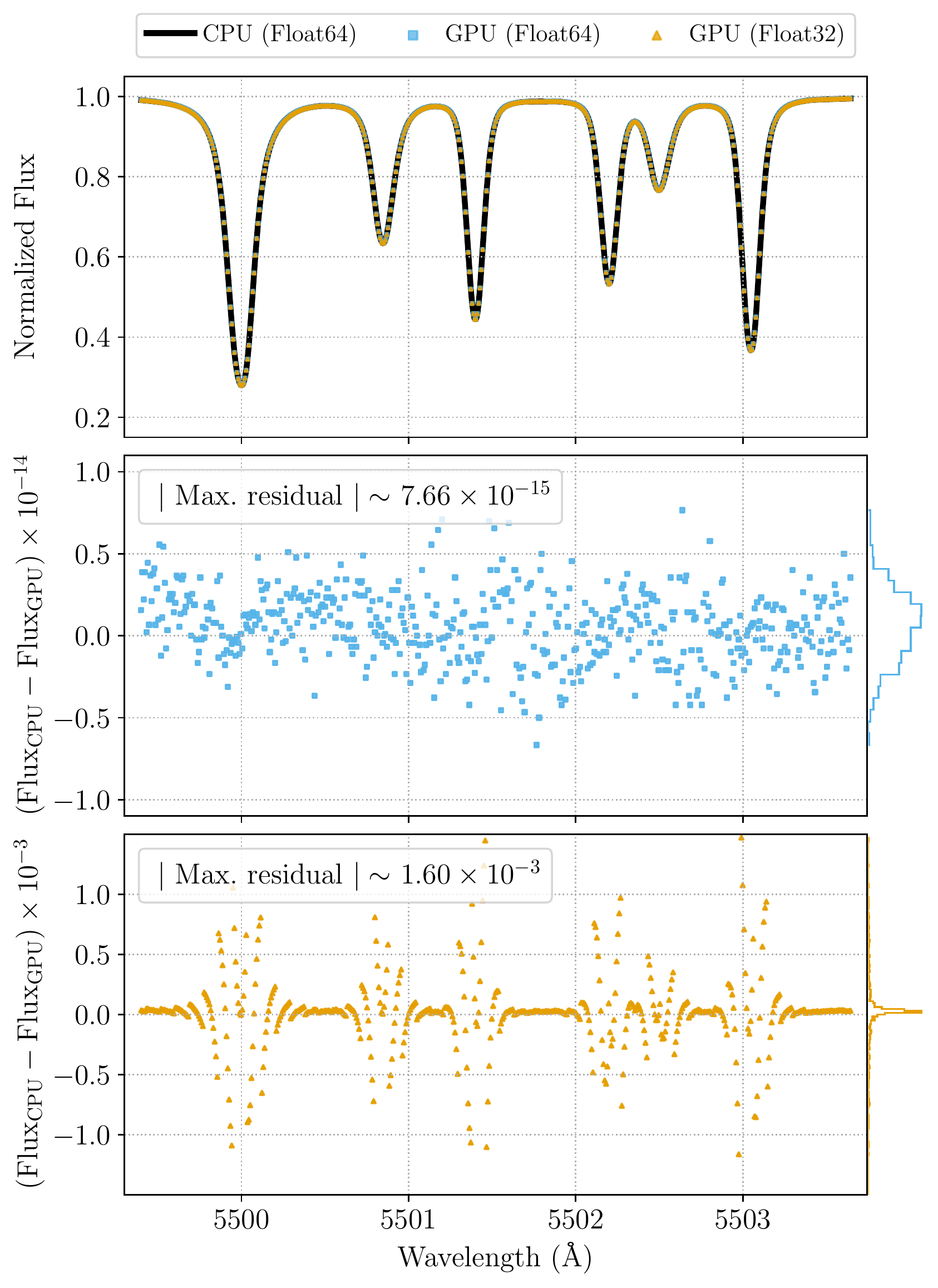}{0.455\textwidth}{(a)}
          \fig{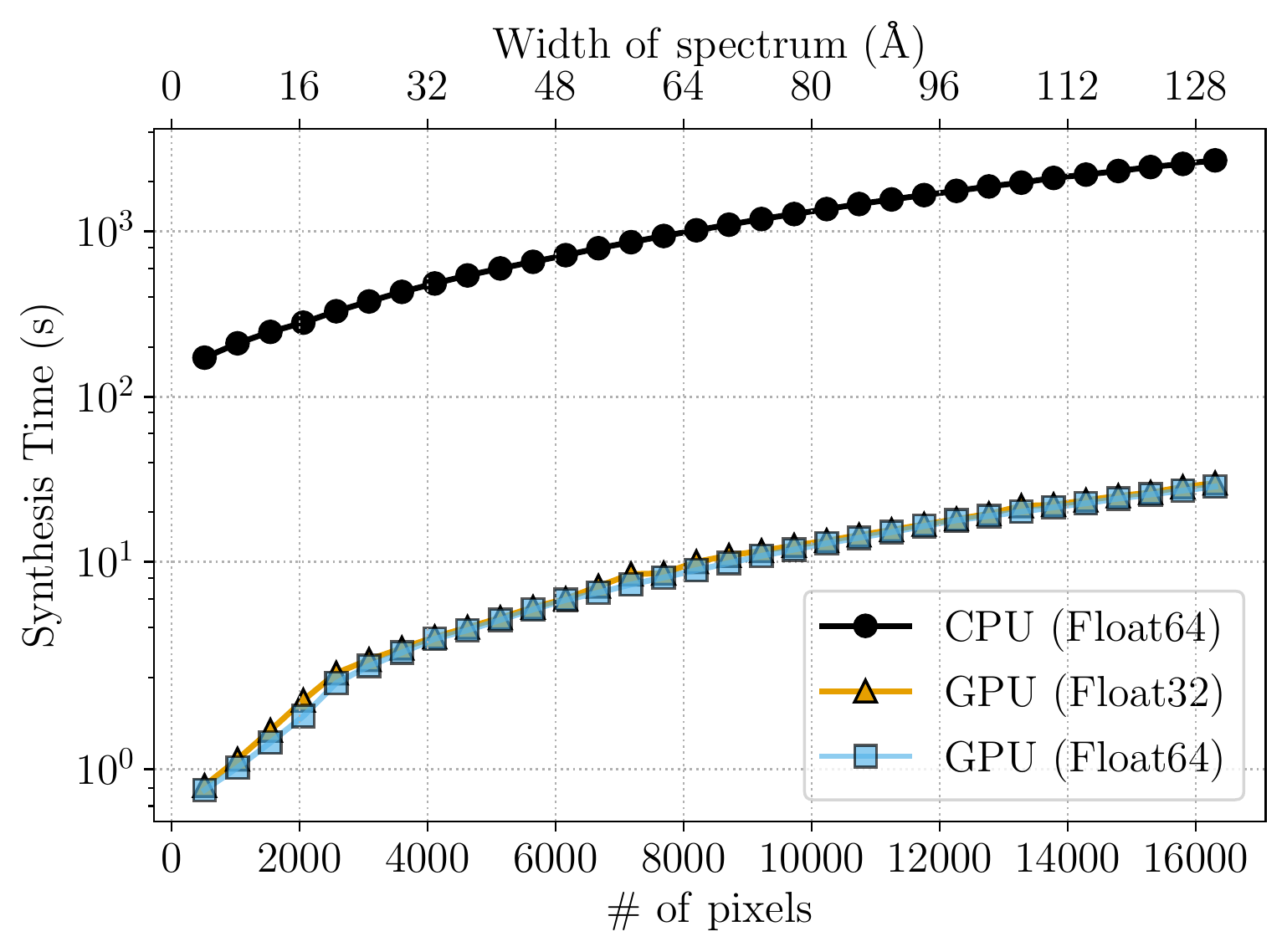}{0.45\textwidth}{(b)}}
    \caption{\textit{Panel (a):} Synthesis accuracy for the \grass\ GPU implementation assessed against the CPU implementation fiducial. Spectra synthesized at double precision on a CPU and at both single and double precision on a GPU are shown in the top panel. Residuals for each of the GPU precisions are shown in the middle and bottom panels. Marginal distributions of the residuals are shown at the right of the middle and bottom panels. Note the differences in $y$-axis scales between each panel. Compared to the double-precision synthesis, which is accurate to several parts in $10^{15}$, single-precision synthesis is only accurate to a few parts in $10^{3}$ as the result of catastrophic cancellation in repeated interpolation calculations. This effect is most pronounced closer to the line core, where many repeated interpolations and in-place multiplications are performed, compared to the line wings and continuum. Because flux errors at the level of $1\times10^{-3}$ propagate to velocity errors of order $\sim$several$\cms$, we strongly advise against the use of the single-precision GPU implementation of \grass. \textit{Panel (b):} \grass\ performance by implementation architecture. GPU benchmarks were performed with an NVIDIA Tesla V100S GPU. For these benchmarks, the size of the synthesized spectrum (i.e., number of lines) was varied, with the number of time steps fixed at $N_t = 50$, and the number of latitude slices fixed at $N_\phi = 197$. Even for smaller problem sizes, the GPU implementation outperforms the CPU implementation because of the large overhead incurred by computing the necessary geometrical parameters and weights (see \S\ref{subsec:model_grid}) on the CPU. The double-precision GPU implementation achieves a factor of $\gtrsim$100 speed-up for larger problem sizes. Owing the large loss in accuracy in the Float32 implementation (see Figure~\ref{fig:accuracy}), we do not recommend use of single-precision synthesis with \grass.}
    \label{fig:accuracy}
\end{figure}

\citet{Palumbo2022} presented v1.0 of \grass, which synthesized spectra serially on a CPU. Depending on the number of spectral resolution elements and the number of time steps in the synthesis, the required computation time could become quite large. As shown in another recent work \citep{Zhao2023}, thoughtful use of parallelization can greatly increase the performance and usability of stellar modeling codes, particularly owing to the performance boost enabled by performing the necessary interpolations (see \S3.2.3 of \citealt{Palumbo2022}) of spectra \textit{en masse}. In order to compute spectra more efficiently, we have implemented a GPU version of \grass. \par 

The GPU kernels are written in \texttt{Julia} using the \texttt{CUDA.jl} \citep{besard2018juliagpu, besard2019prototyping} package for use with NVIDIA GPUs. Multiple steps of the updated synthesis procedure have been parallelized. Specifically, the spectral synthesis computations and interpolations for each spatial grid cell and wavelength element are performed in parallel, in addition to the computation of weights governed by the viewing geometry, limb darkening, and differential rotation of the model star. The same high-level functions used to generate synthetic spectra are able to interface with the GPU kernels such that no knowledge of GPU computation is needed by users of \grass\ to take advantage of this implementation. As in the CPU implementation, the computation of the disk integrated spectrum is performed in place to avoid excess memory allocation, which could otherwise exceed the available VRAM at moderate problem sizes. \par

As shown in Panel (a) of Figure~\ref{fig:accuracy}, the double-precision GPU implementation of \grass\ produces flux values that are accurate to the CPU implementation within about $\sim$several parts in $10^{15}$ when using the same random number generator (RNG) seed. Due to catastrophic cancellation arising in the interpolation step of the spectral synthesis (described in \S3.2.3 of \citealt{Palumbo2022}), the flux values produced by the single-precision GPU synthesis are accurate to the CPU implementation at only about a couple parts in $10^{3}$. Propagating these errors in flux to velocity, the single-precision flux error produces a velocity error of order $\sim$several$\cms$, whereas the double-precision GPU velocity error amounts to only $\sim$1$\times 10^{-10}$ $\ms$. Given the appreciable velocity error in the single-precision implementation, only the double-precision GPU implementation of \grass\ was used for computations and results presented in this work, and we strongly discourage the use of single-precision GPU implementation for synthesizing spectra with \grass. To prevent unwitting single-precision computations, \grass\ will throw a warning if any GPU allocations are made with single-precision floats. \par

To assess the relative performance of the CPU and GPU implementations, we performed benchmarks of \grass\ with a 2.5 GHz Intel Xeon Gold 6248 CPU and a NVIDIA Tesla V100S GPU running with CUDA Driver Version 550.54.14 and CUDA Toolkit Version 12.4. Because the GPU implementation performs parallel computations over the spatial and spectral grid cells, we evaluated performance as a function of the number of spectral resolution elements (or somewhat equivalently, the width of the spectrum in units of length), because the number of spatial grid cells should be fixed for physical validity (see \S 4.2 of \citealt{Palumbo2022}). The number of time steps in the simulation was held fixed at $N_t = 50$, corresponding to 12.5 minutes of time given the time step of 15 seconds. The results of these benchmarks are shown in Panel (b) of Figure~\ref{fig:accuracy}. For larger problem sizes, the GPU implementation outperforms the CPU implementation by a factor of about 100$\times$. \par

\section{Notes on Specific Lines} \label{app:iag}

\begin{figure*}[h!]
\gridline{\fig{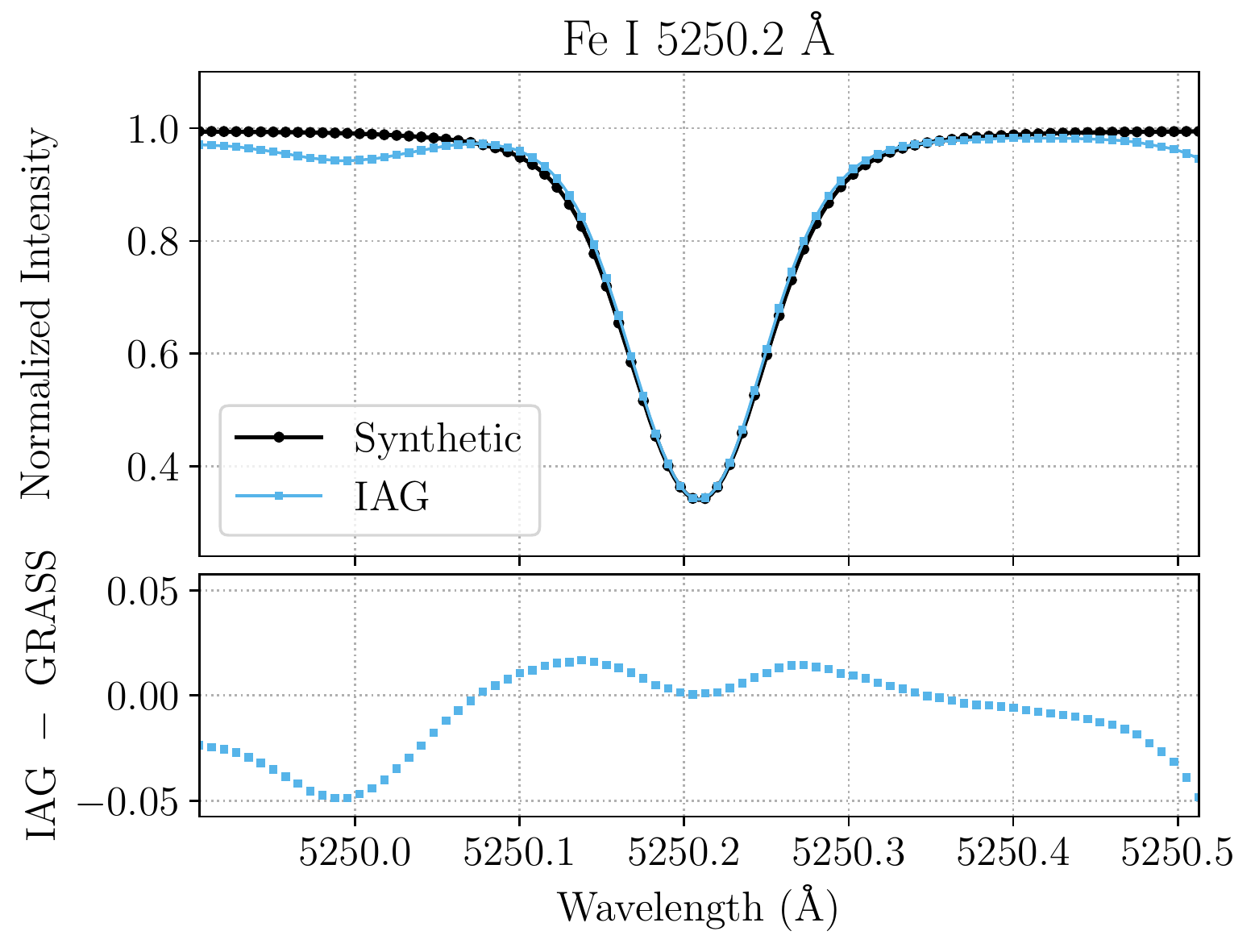}{0.455\textwidth}{}
          \fig{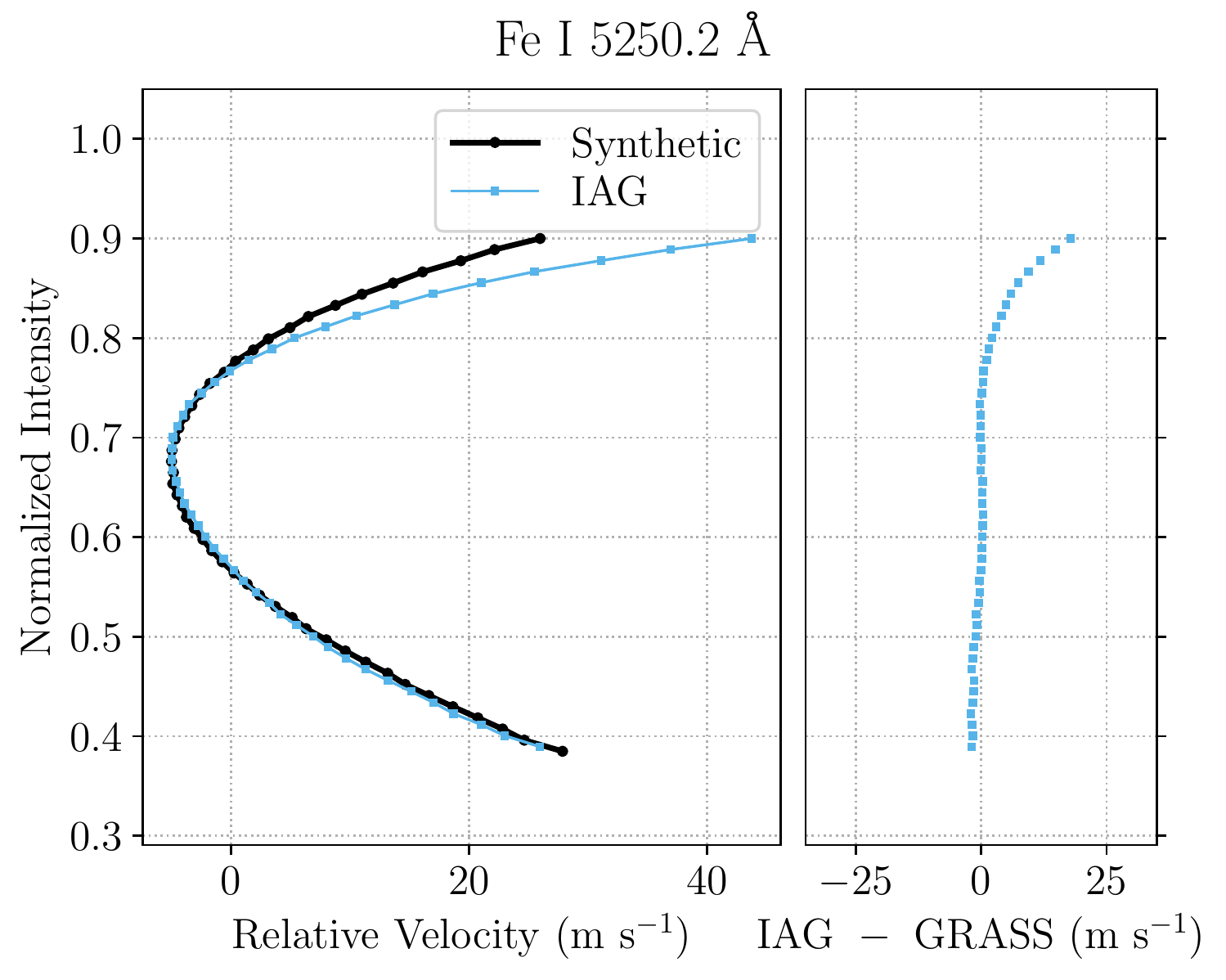}{0.45\textwidth}{}}
\gridline{\fig{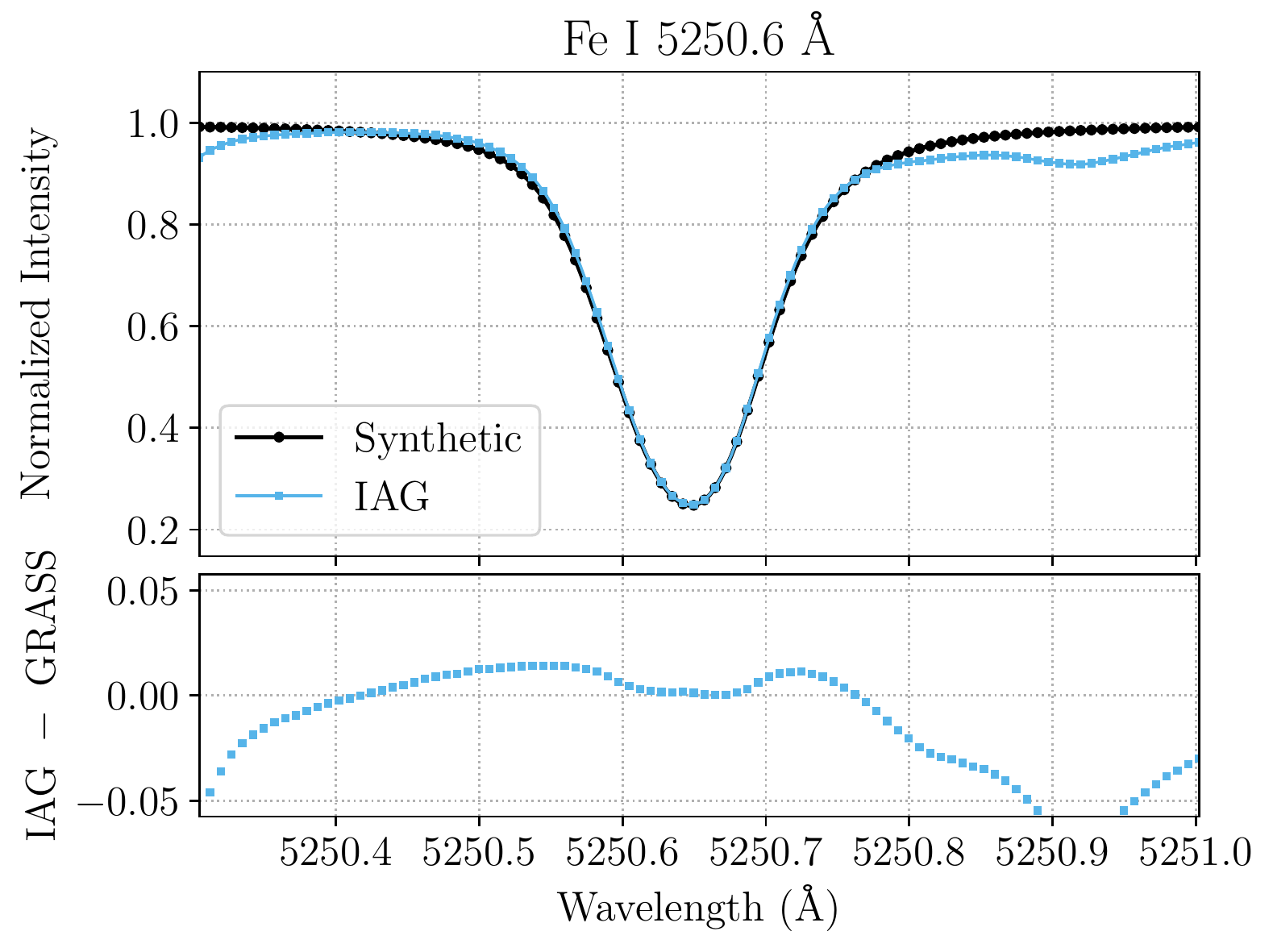}{0.455\textwidth}{}
          \fig{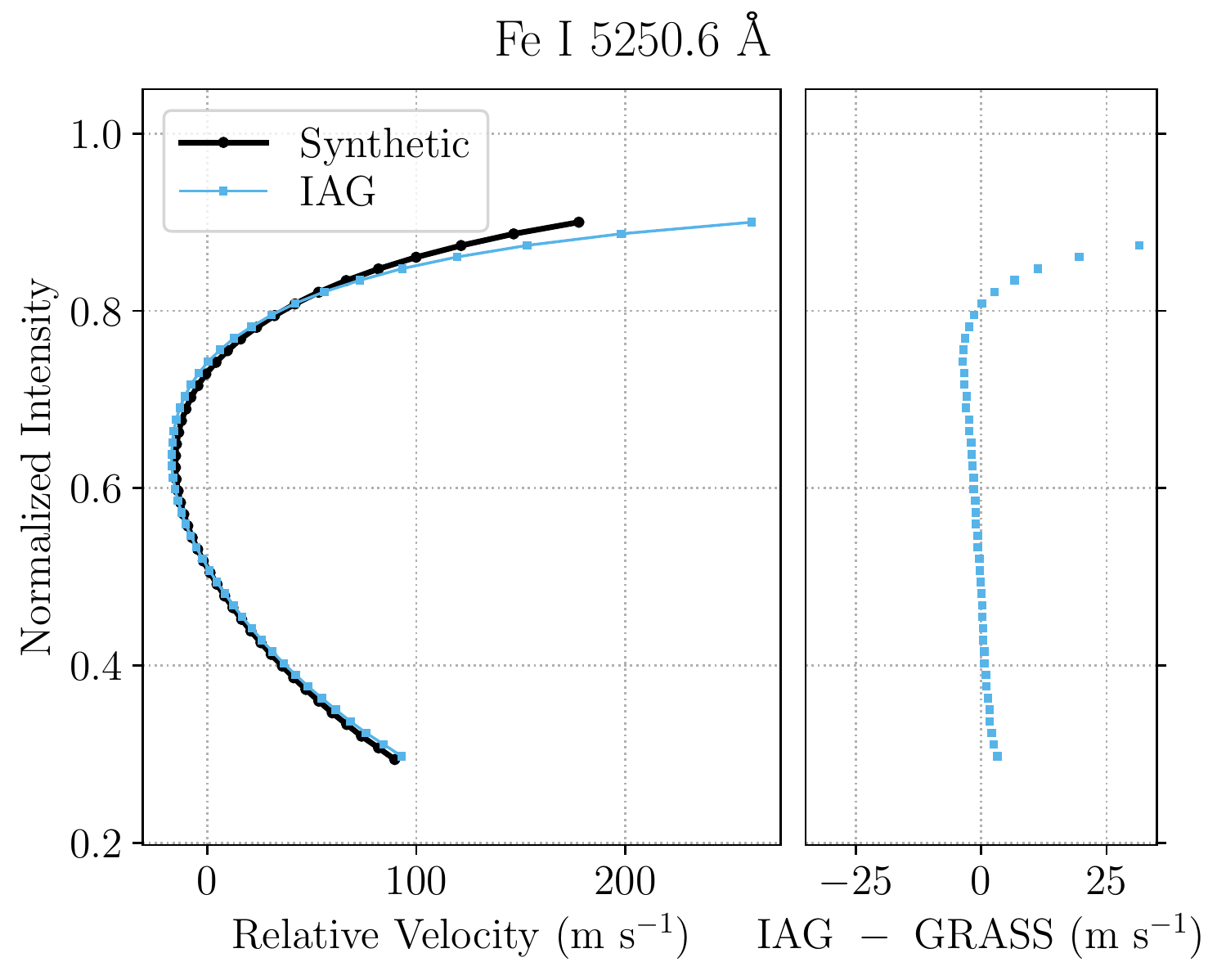}{0.45\textwidth}{}}
\gridline{\fig{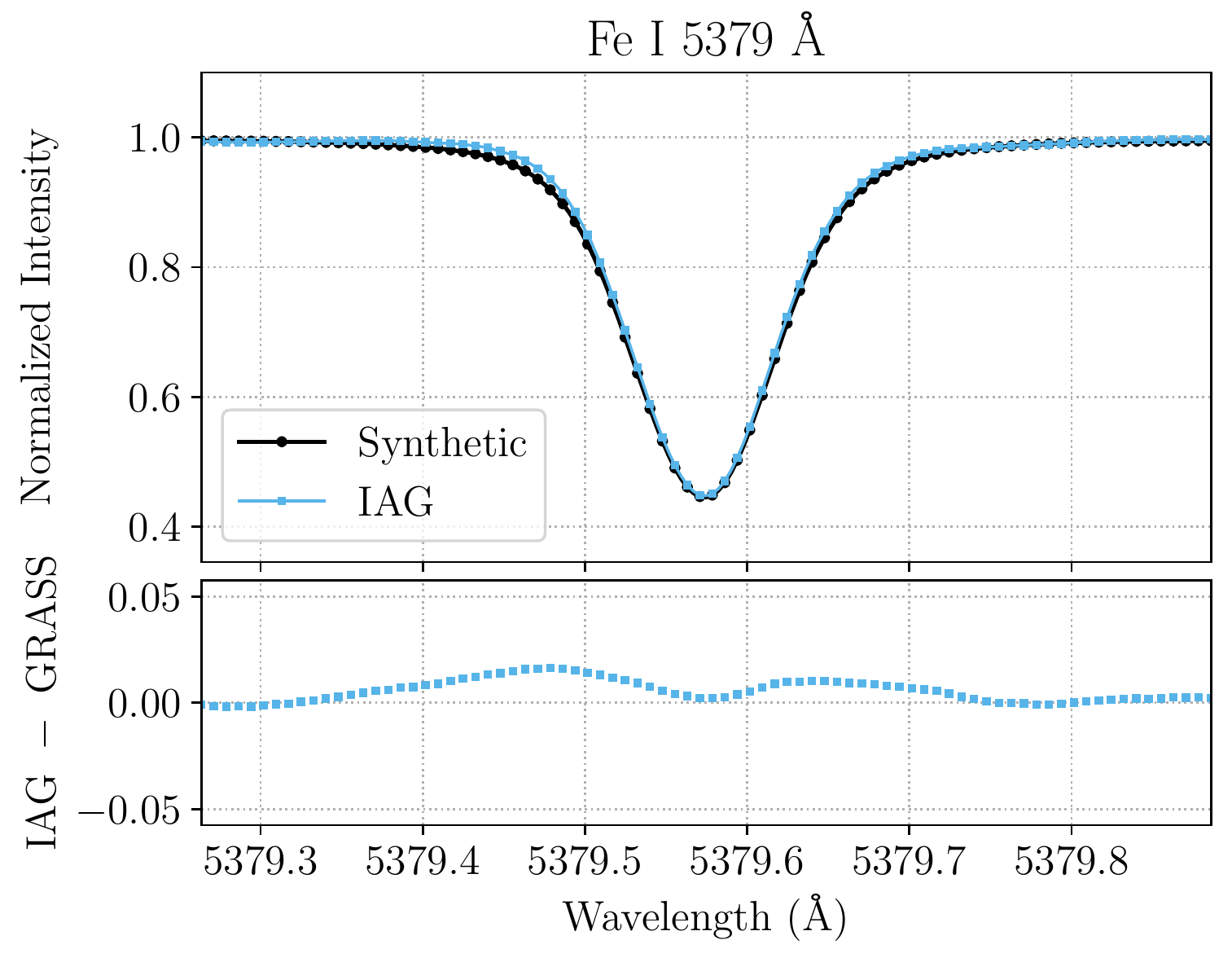}{0.455\textwidth}{}
          \fig{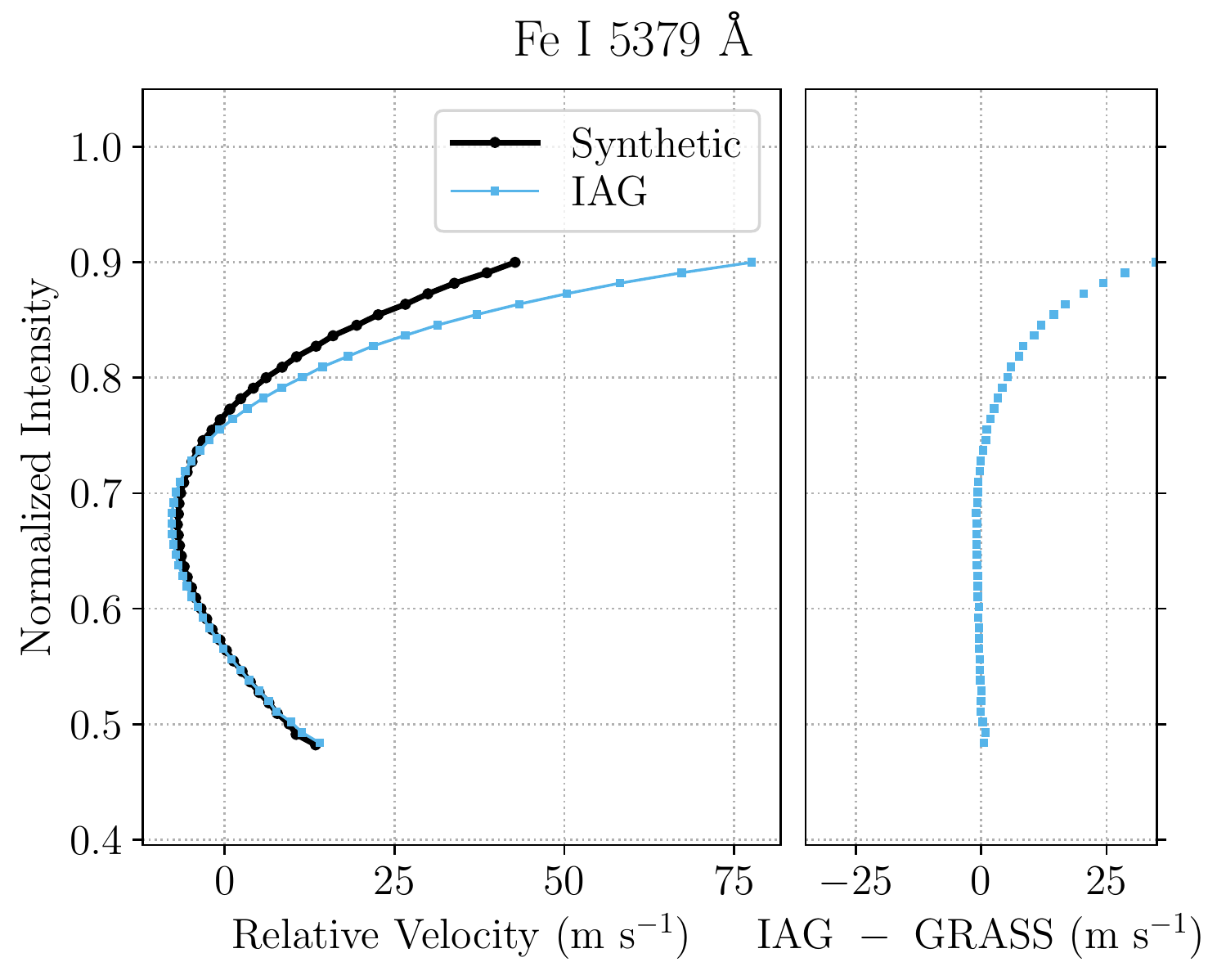}{0.45\textwidth}{}}
\caption{Comparisons between line profiles (left column) and bisectors (right column) for synthetic (black) and observed (blue) disk-integrated lines. The complete figure set (44 figures) is available in the online journal.}
\label{fig:iag_bis_first}  
\end{figure*}

In \citet{Palumbo2022}, we demonstrated that \grass\ accurately reproduced the line profile and bisector of the Fe\textsc{I} 5434 \AA\ line by comparison to the IAG Solar Atlas \citep{Reiners2016b}. In this work, we have performed this semi-quantitative comparison for all 22 lines used in the \grass\ template line library (see \S\ref{validation}). An example set of comparisons is shown in Figure~\ref{fig:iag_bis_first}; line profile and bisector comparison plots are available for all 22 lines in the online journal. 
Below, we provide comments on specific details relevant to each line. For additional discussion of the convective blueshift of these lines, as well as their historical use in the heliophysics literature, refer to \citet{Lohner-Bottcher2018} and \citet{Lohner-Bottcher2019}. \par 

\subsection{Lines around 5251 \AA}

Two lines in this region were studied: Fe \textsc{I} 5250.2 \AA\ and Fe \textsc{I} 5250.6. As noted in \citet{Lohner-Bottcher2019}, other lines in this region were extremely weak or significantly blended in the wings. Both Fe \textsc{I} 5250.2 \AA\ and Fe \textsc{I} 5250.6 exhibit classical, smooth ``C''-shaped bisectors, which are well-modeled by \grass. \par 

\subsection{Lines around 5381 \AA}

Four lines in this region were studied: Fe \textsc{I} 5379 \AA, Ti \textsc{II} 5381 \AA, Fe \textsc{I} 5382 \AA, and Fe \textsc{I} 5383 \AA. A fifth line, C \textsc{I} 5380 \AA, was noted but excluded from the analysis owing to its very modest depth. Two lines, Fe \textsc{I} 5379 \AA and Ti \textsc{II} 5381 \AA, exhibit classical ``C''-shaped bisectors.  Fe \textsc{I} 5383 \AA, on the other hand, has a ``\textbackslash''-shaped bisector which are typically observed for disk-resolved bisectors near the limb. In this case, as noted by \citet{Lohner-Bottcher2019}, the strong blueward trend is caused by blends in the blue wing of the line. Lastly, \textsc{I} 5382 is quite shallow, but its bisector is suggestive of the upper region of a ``C''-shaped bisector. \grass\ models Fe \textsc{I} 5379 \AA\ and Ti \textsc{II} 5381 \AA\ quite well, except for diminished curvature in the top $\sim$20\% of the Fe \textsc{I} 5379 \AA line. \grass\ also reproduces the ``\textbackslash''-shaped bisector of Fe \textsc{I} 5383 \AA\ quite well up to $\sim$60\% of the continuum. However, the shape of the bisector of this region is shaped by the cores of weak lines in the wing of Fe I \textsc{I} 5383 \AA; small changes in the relative depths of these lines (e.g., from activity, as the Sun was more active during the observation of the IAG Atlas; see \citealt{Hathaway2015} and \citealt{Reiners2016b}) could lead to this discrepancy. Lastly, \grass\ underestimates the amplitude of the shallow Fe \textsc{I} 5382 \AA\ bisector. Being the shallowest line that we model with \grass\ (with fractional depth $\sim$0.2), the amplitude underestimate is not entirely surprising. We note that the mismatch between the IAG and \grass\ bisectors in general becomes most extreme above $\sim$80\% continuum flux; the entirety of this line is above this level. \par 

\subsection{Lines around 5434 \AA}

Six lines in this region were studied: Mn \textsc{I} 5432 \AA, Fe \textsc{I} 5432 \AA, Fe \textsc{I} 5434 \AA, Ni \textsc{I} 5435 \AA, Fe \textsc{I} 5436.3 \AA, and Fe \textsc{I} 5436.6 \AA. Fe \textsc{I} 5434 \AA\ was the line modeled in the initial presentation of \grass\ in \citet{Palumbo2022}. Lines with classical ``C''-shaped bisectors included Fe \textsc{I} 5432 and Fe \textsc{I} 5434. Lines whose bisectors visually correspond to the upper half of the classical ``C'' include Mn \textsc{I} 5432 \AA, Fe \textsc{I} 5436.3 \AA, and Fe \textsc{I} 5436.6 \AA. As discussed in \S\ref{subsubsec:manganese}, the Mn \textsc{I} 5432 \AA\ line is particularly broad as the result of the prodigious hyperfine structure of Mn. Ni \textsc{I} 5435 \AA\ is blended in the wing such that it's bisector is ``\textbackslash''-shaped, as seen for Fe \textsc{I} 5383 \AA. \grass\ generally models these lines well (with only very small differences in the curvature of bisectors). Somewhat notably, \grass\ underestimates the curvature in the upper region of the ``C'' for Fe \textsc{I} 5434 \AA, but this discrepancy is due to blends in the red line wing of the line which were modeled out in the pre-processing of the LARS data (as discussed in greater detail in \S4.1 of \citealt{Palumbo2022}).\par 

\vspace{-1mm}
\subsection{Lines around 5578 \AA}

Two lines in this region were studied: Fe \textsc{I} 5576 \AA\ and Ni \textsc{I} 5578 \AA. Both lines exhibit classical ``C''-shaped bisectors which are well-modeled by \grass. However, \grass\ does somewhat underestimate the curvature of  Fe \textsc{I} 5576 \AA\ above $\sim$70\% flux. \par 

\subsection{Lines around 6150 \AA}

Two lines in this region were studied: Fe \textsc{II} 6149 \AA\ and Fe \textsc{I} 6151 \AA. Both lines exhibit bisectors which visually correspond to the upper $\sim$half of the classical ``C'' shape. Considering the modest depths of these lines (especially so in the case of  Fe \textsc{II} 6149 \AA), \grass\ models the bisectors of these lines remarkably well. \par 

\subsection{Lines around 6171 \AA}

Four lines in this region were studied: Ca \textsc{I} 6169.0 \AA, Ca \textsc{I} 6169.5 \AA, Fe \textsc{I} 6170 \AA, and Fe \textsc{I} 6173 \AA. Two lines, Ca \textsc{I} 6169.0 \AA\ and Fe \textsc{I} 6173 \AA, exhibit classical ``C''-shaped bisectors that are well modeled by \grass. Fe \textsc{I} 6173 \AA\ is notably one of the most magnetically sensitive lines in the visible portion of the spectrum, and such has been extensively used to study the solar line-of-sight magnetic field (e.g., the Helioseismic Magnetic Imager of the Solar Dynamics Observatory, \citealt{Scherrer2012}). \grass\ does somewhat underestimate the curvature of this line. The remaining two lines, Ca \textsc{I} 6169.5 \AA\ and Fe \textsc{I} 6170 \AA, are blended such that their bisectors deviate from the classical ``C'' shape. Ca \textsc{I} 6169.5 \AA\ exhibits a ``\textbackslash''-shaped bisector as in Ni \textsc{I} 5435 \AA\ and Fe \textsc{I} 5383 \AA. Fe \textsc{I} 6170 \AA\ exhibits a bisector unlike those of the other blend lines studied in this work; it is shaped like a ``/`` up about $\sim$80\% continuum before hooking back to the blue. \grass\ faithfully reproduces the majority of these two bisectors, but deviates slightly in the uppermost regions. \par 

\subsection{Lines around 6302 \AA}

Two lines were studied in this last region: Fe \textsc{I} 6301 \AA\ and Fe \textsc{I} 6302 \AA. The \citet{Cegla2013, Cegla2018, Cegla2019a} series of papers synthesized the Fe \textsc{I} 6302 \AA, which has also been widely used in the solar physics. Both lines are otherwise widely used in the solar physics literature (see, e.g, the Introduction of \citealt{Smitha2020} for a review). Both lines have classical ``C''-shaped bisectors and are notably flanked by deep telluric oxygen lines. The bisectors of these lines are well-modeled by \grass\ up to $\sim$70\% of the continuum; above this level, \grass\ notably underestimates the redward curvature of the bisector relative to the IAG bisector. It is plausible that these deviations are caused by variations in the strength of the oxygen telluric lines between the respective observing sites of the IAG Atlas (G\"{o}ttingen, Germany; altitude $\sim$150 meters) and the LARS data used as input for \grass\ (Teide Observatory, Canary Islands; altitude $\sim$2400 meters). \par

\begin{figure*}[h!]
\gridline{\fig{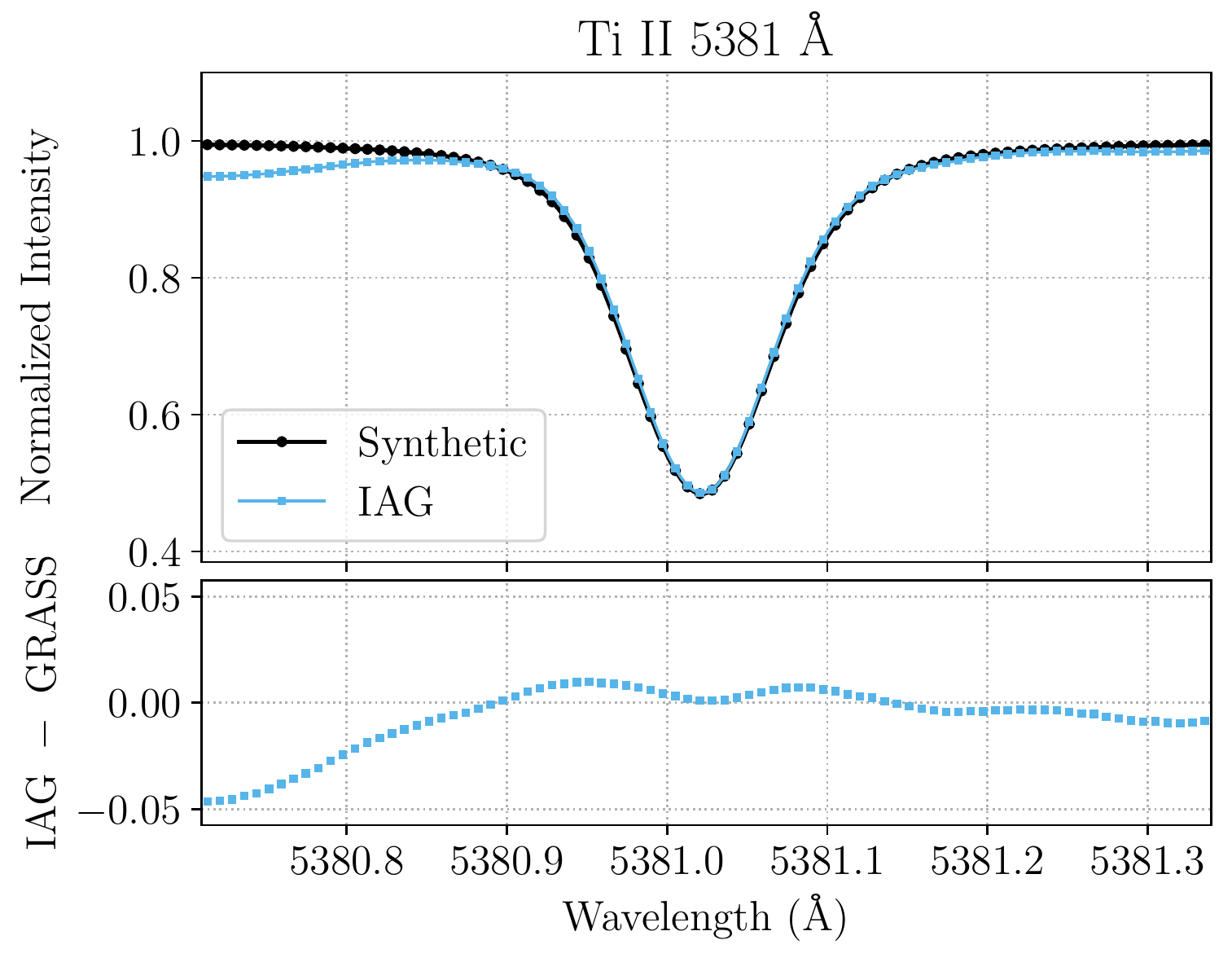}{0.455\textwidth}{}
          \fig{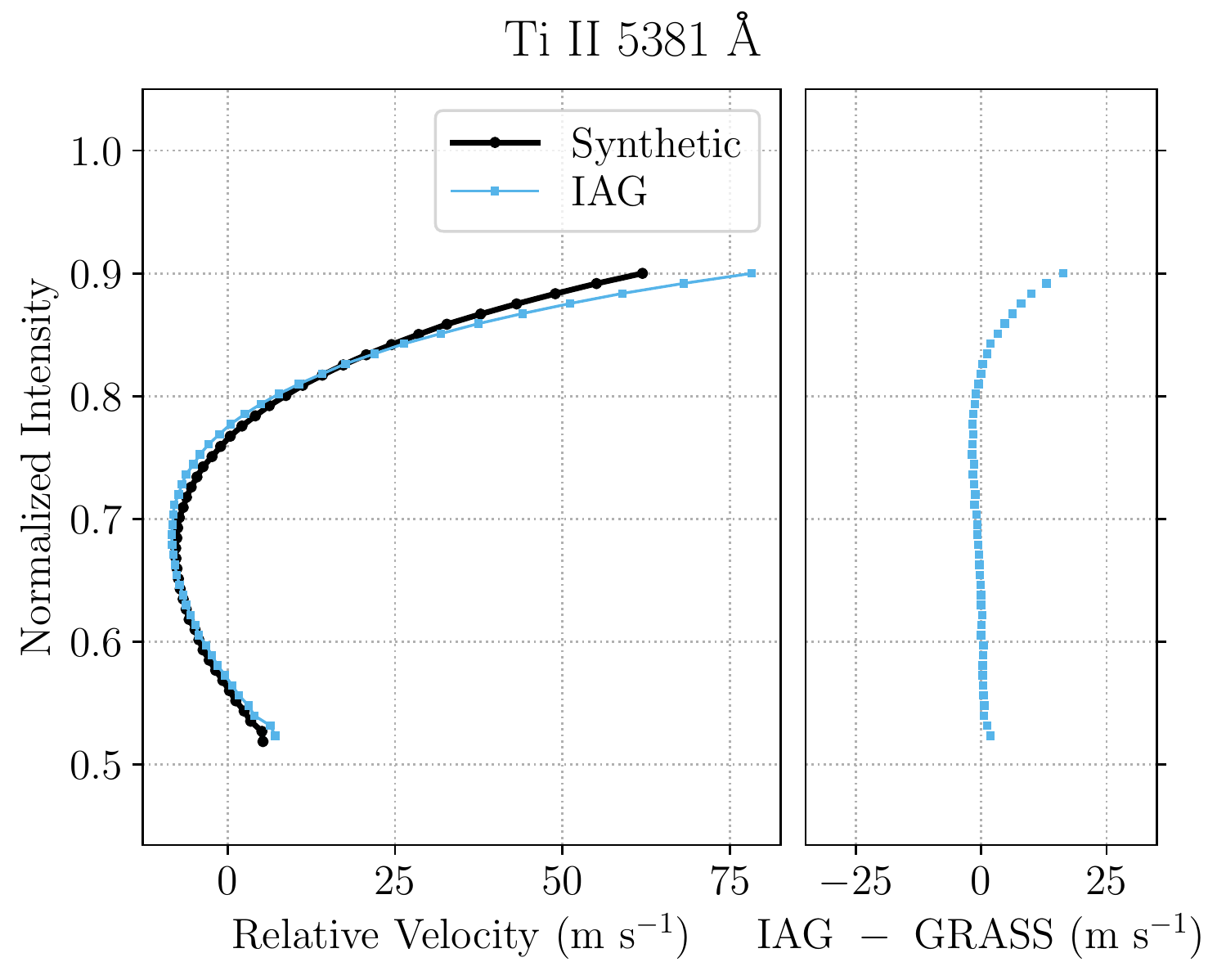}{0.45\textwidth}{}}
\gridline{\fig{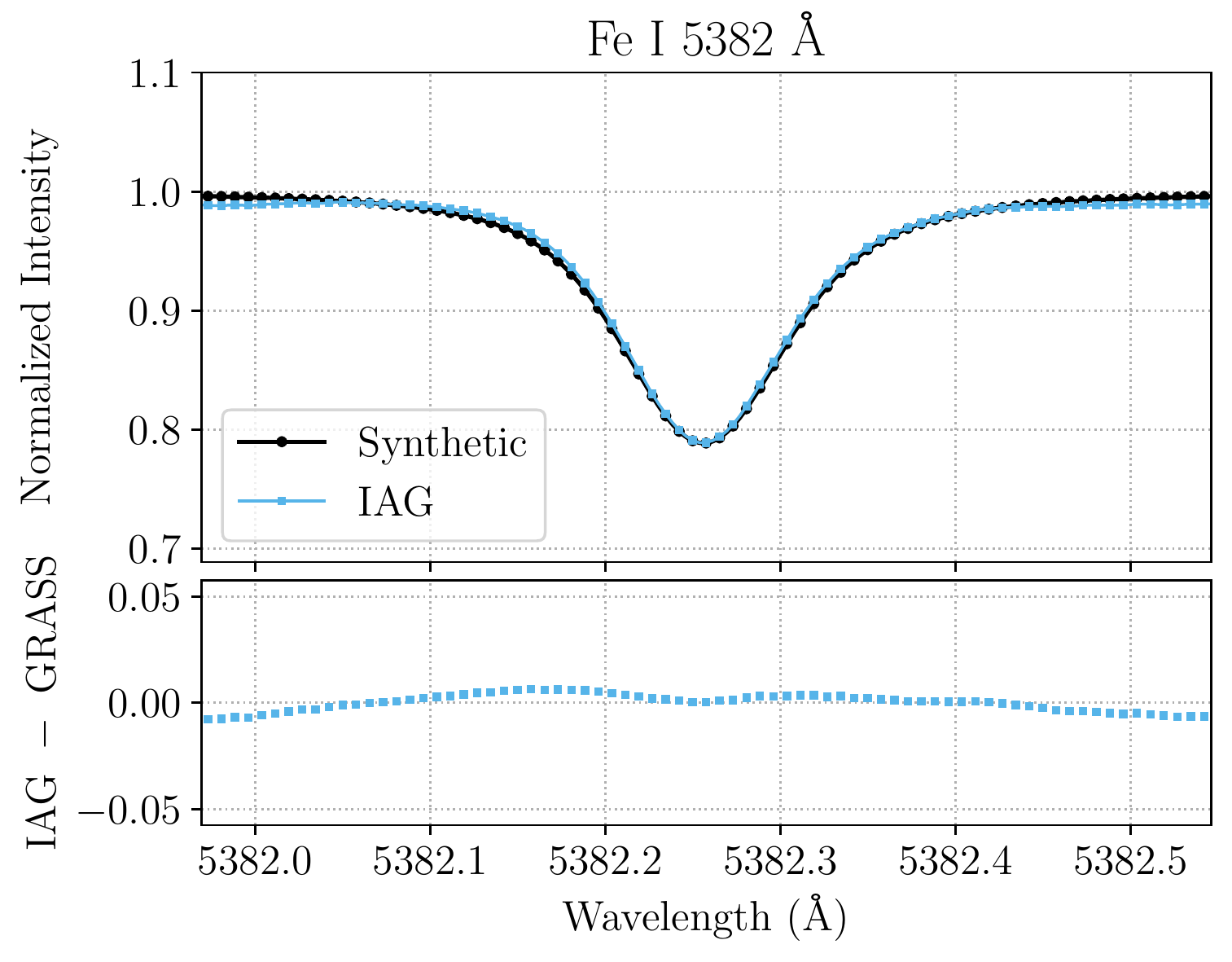}{0.455\textwidth}{}
          \fig{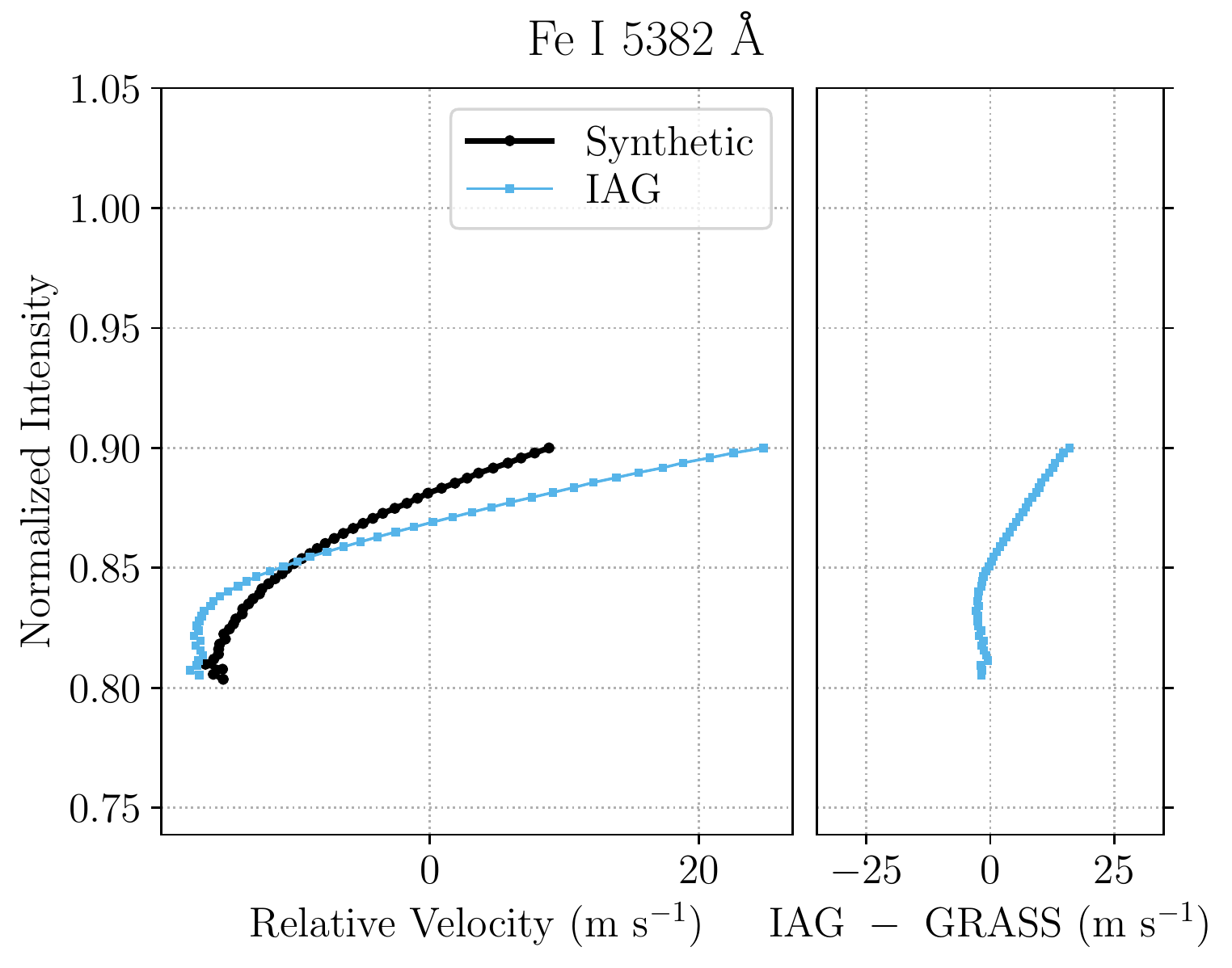}{0.45\textwidth}{}}
\gridline{\fig{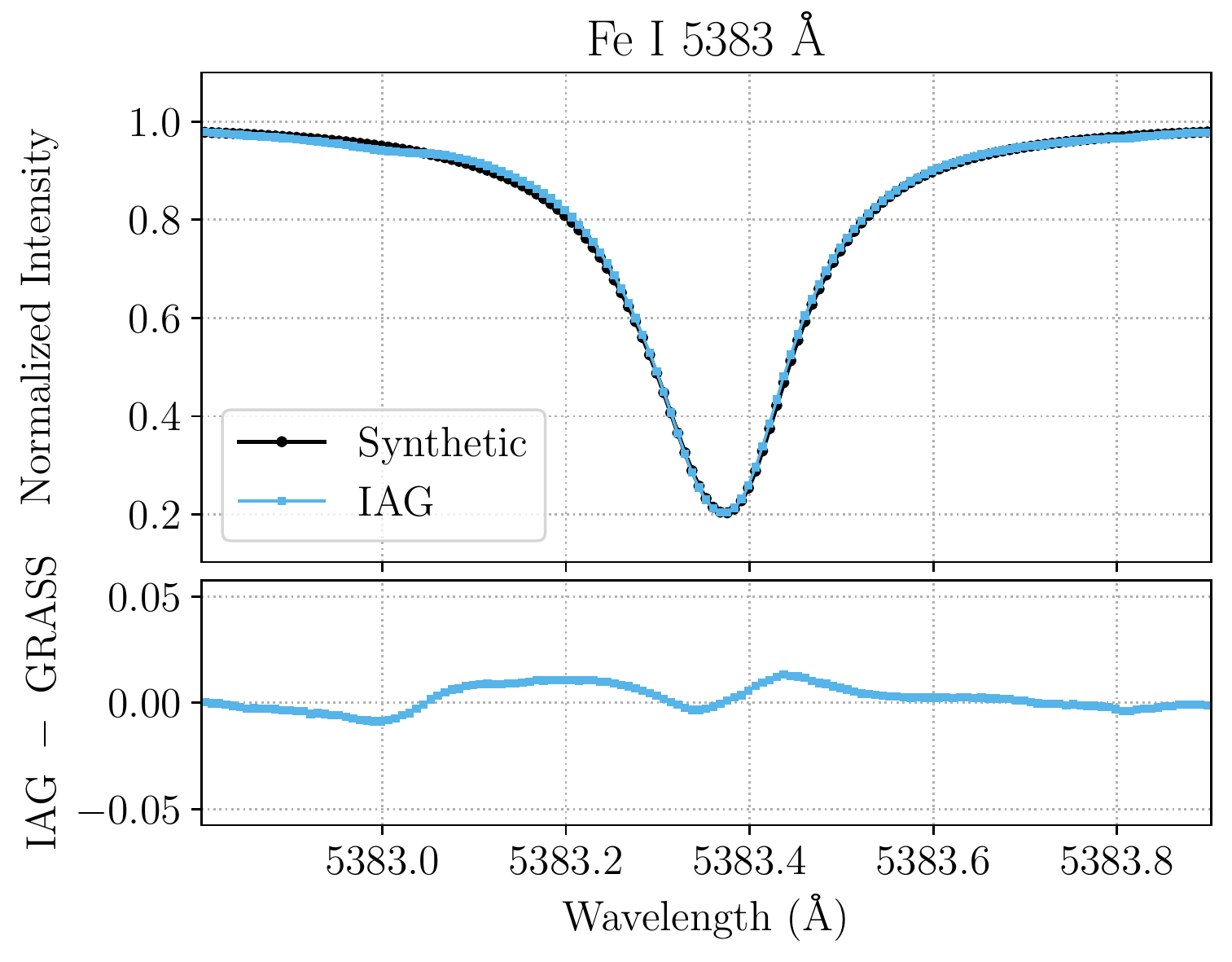}{0.455\textwidth}{}
          \fig{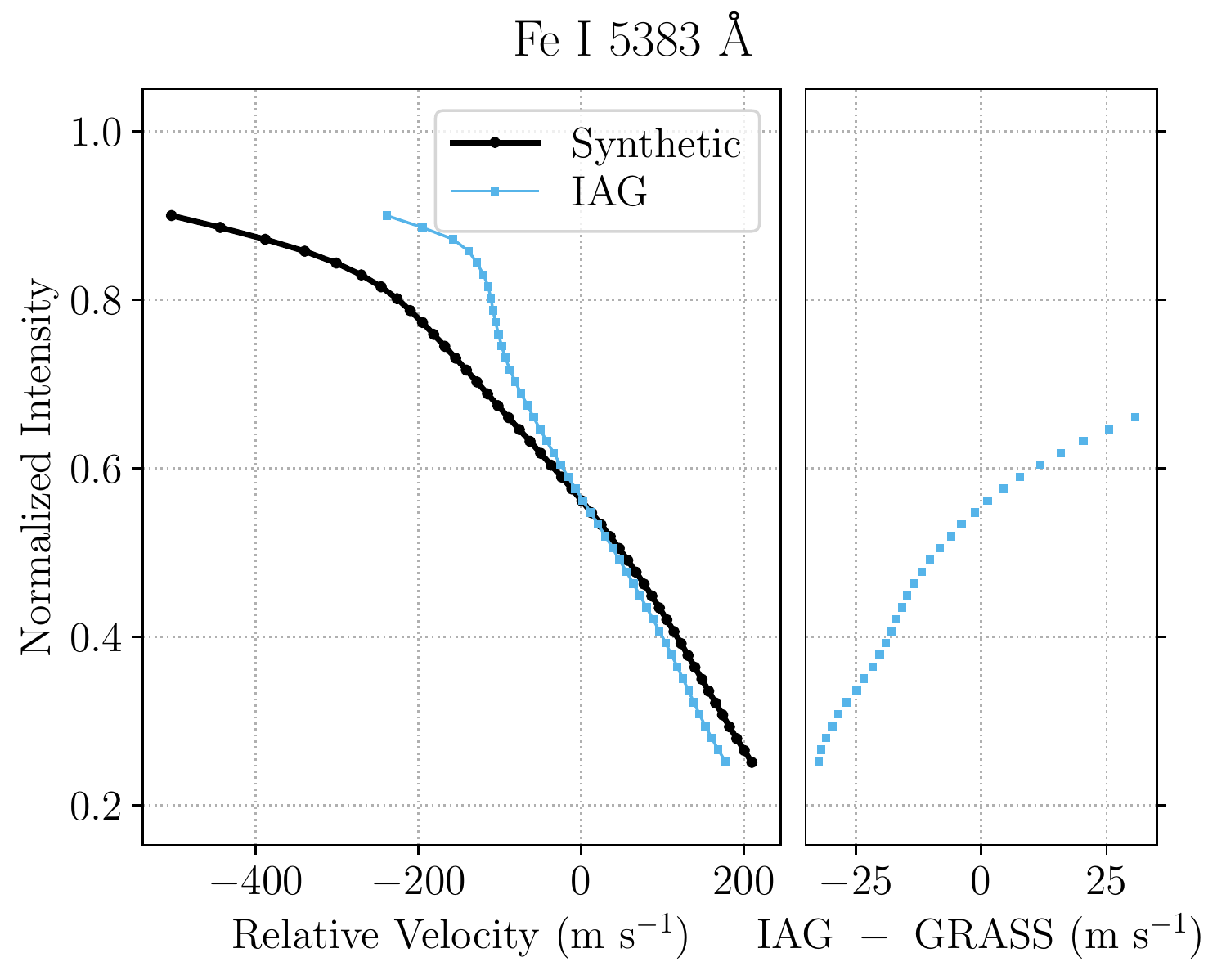}{0.45\textwidth}{}}
\caption{Same as previous figure.}
\label{fig:iag_bis_two}  
\end{figure*}

\begin{figure*}
\gridline{\fig{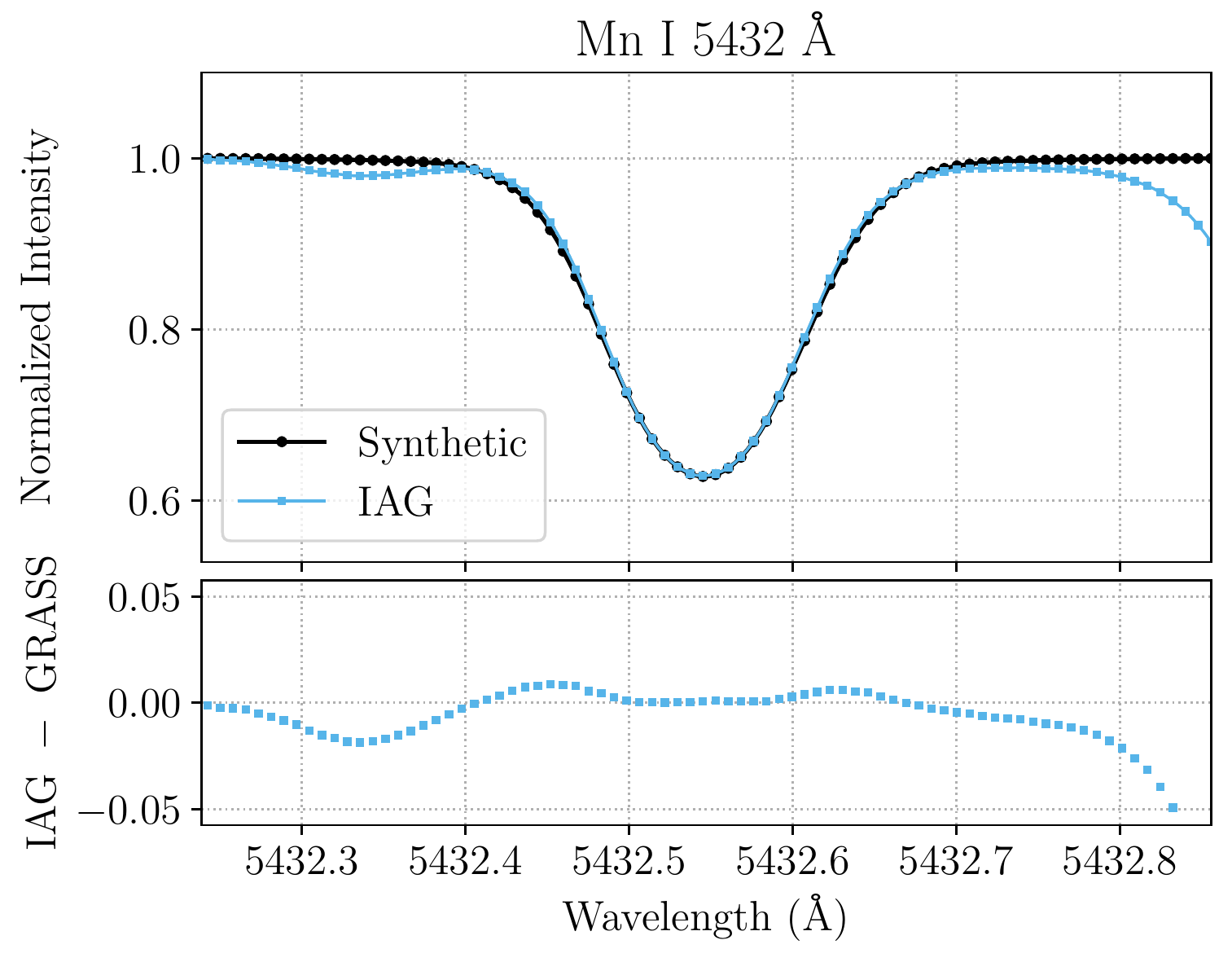}{0.455\textwidth}{}
          \fig{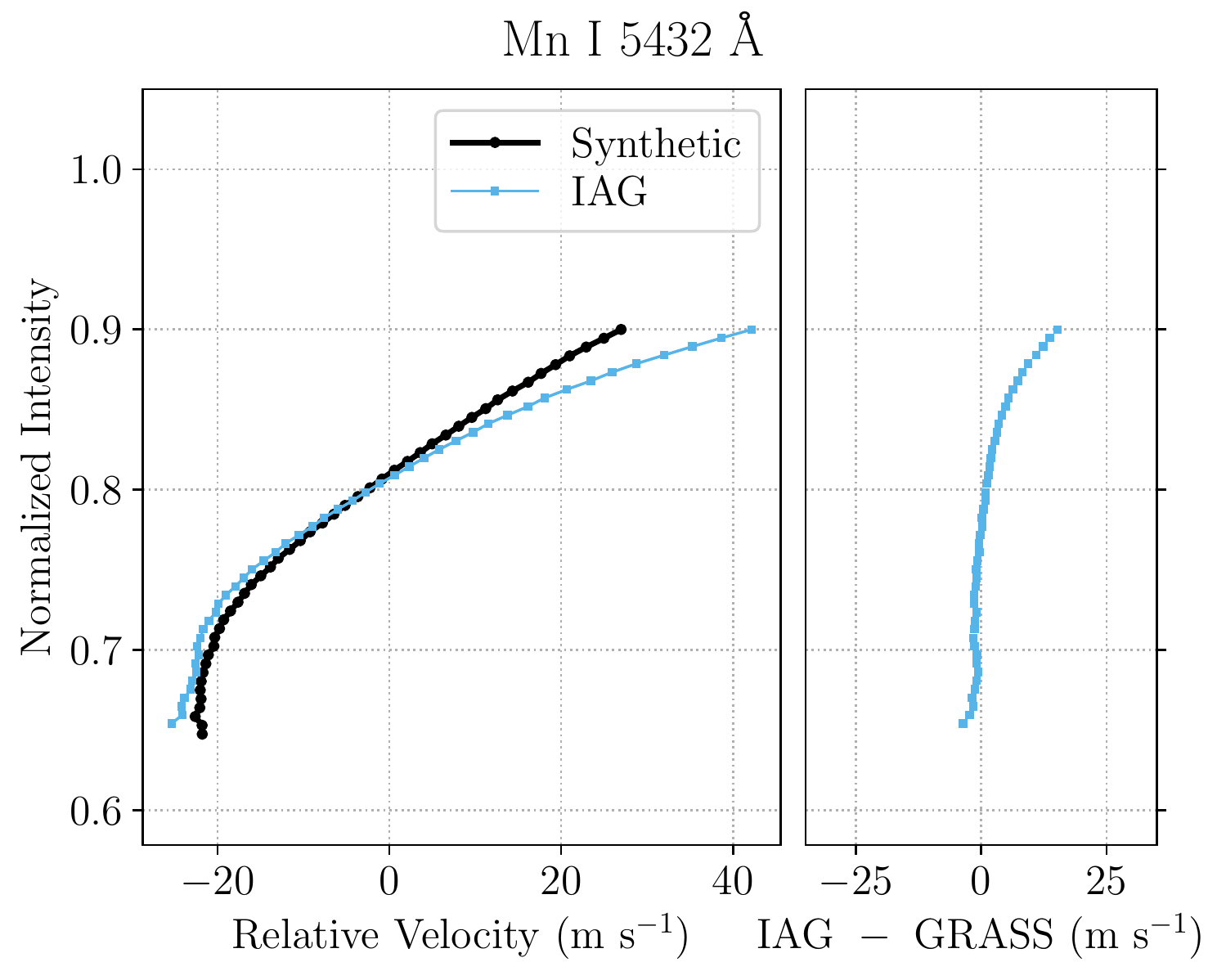}{0.45\textwidth}{}}
\gridline{\fig{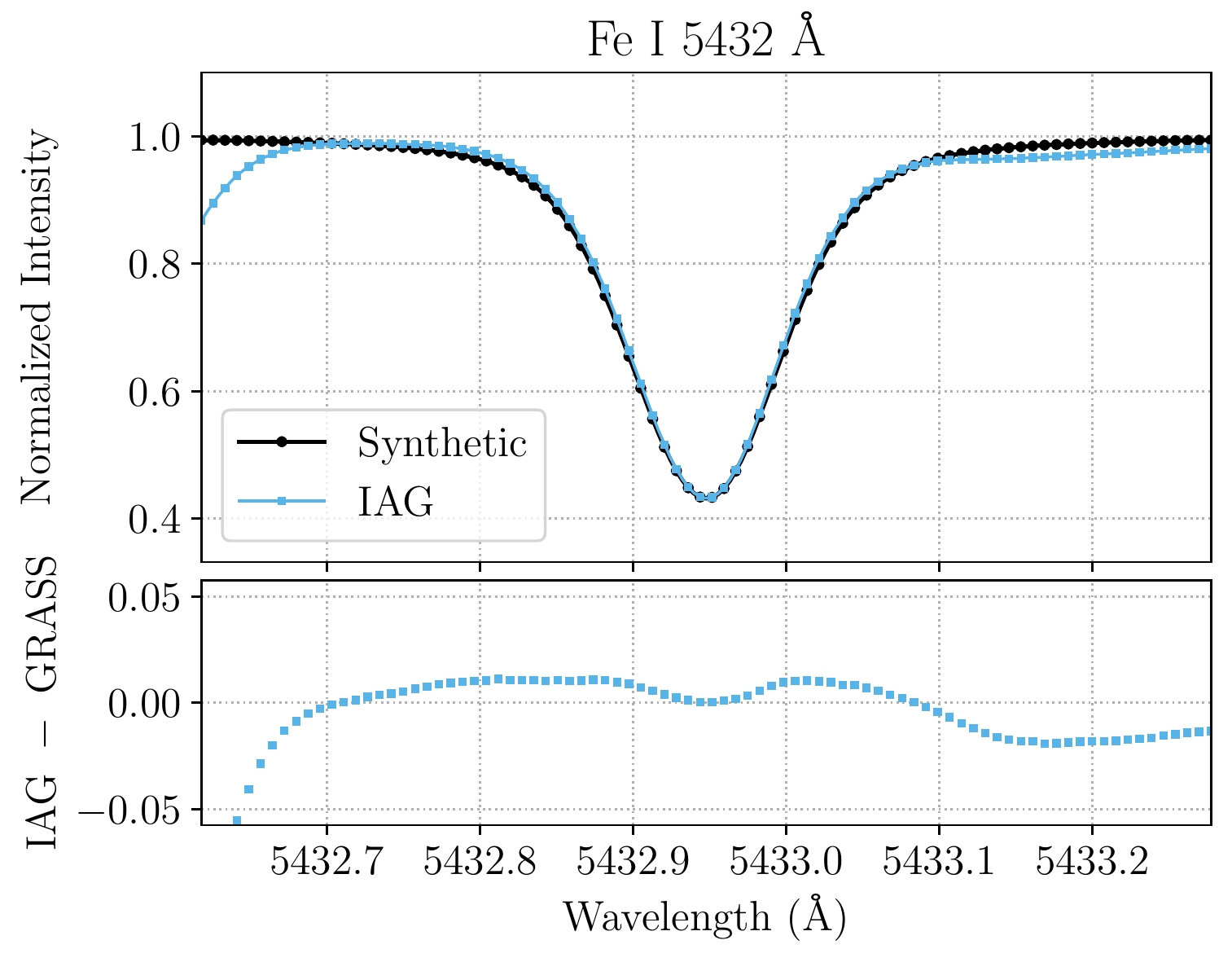}{0.455\textwidth}{}
          \fig{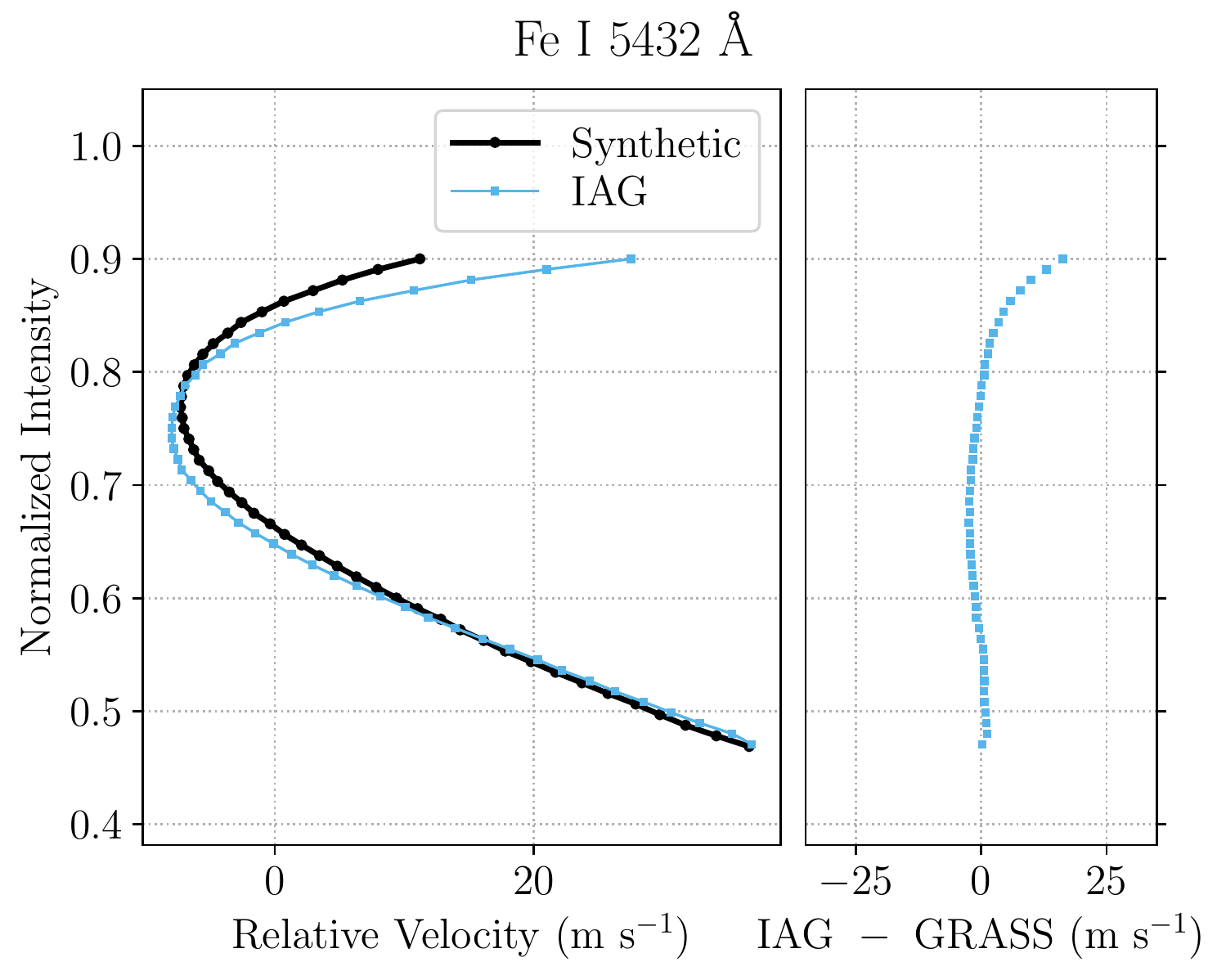}{0.45\textwidth}{}}
\gridline{\fig{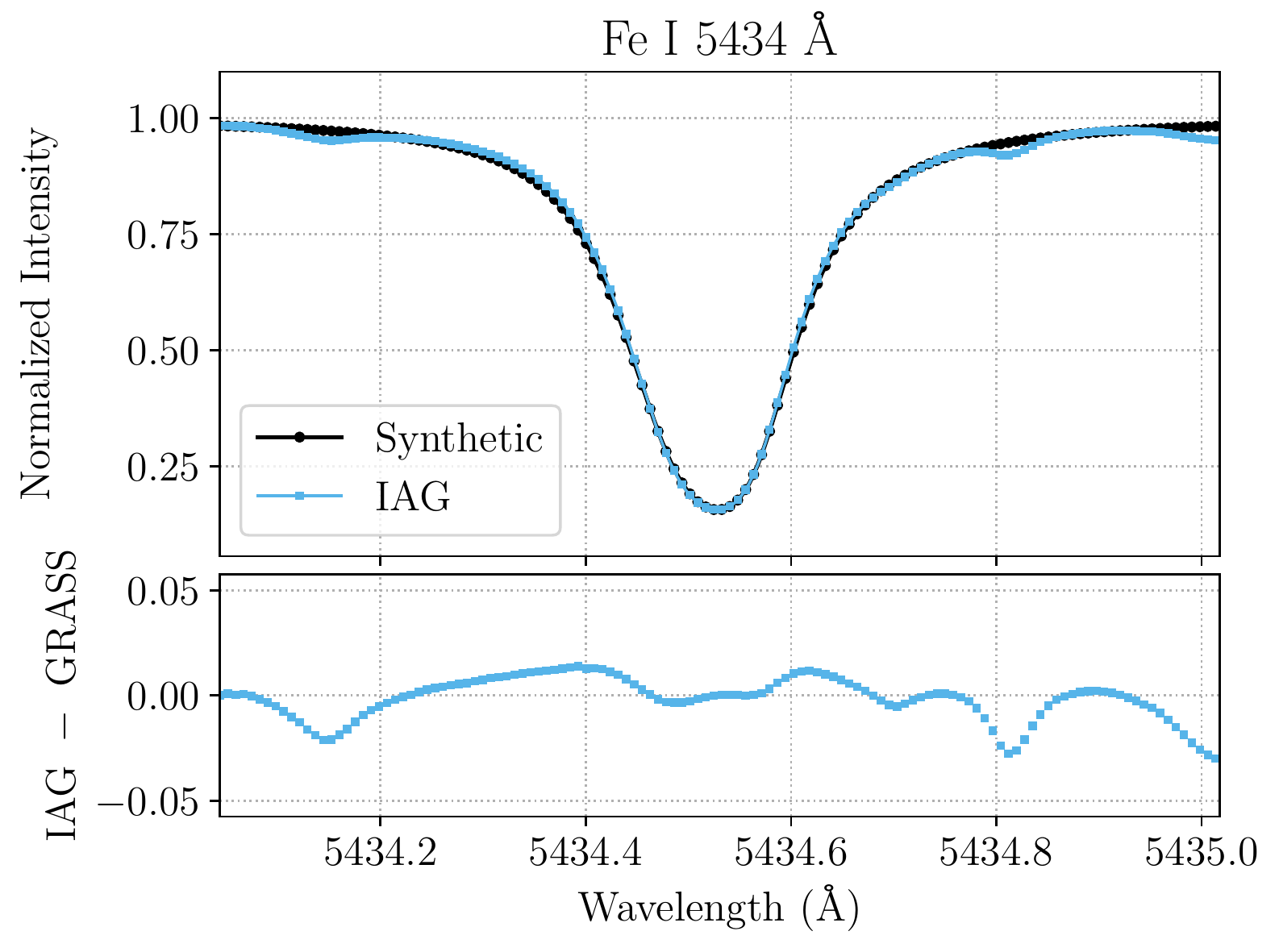}{0.455\textwidth}{}
          \fig{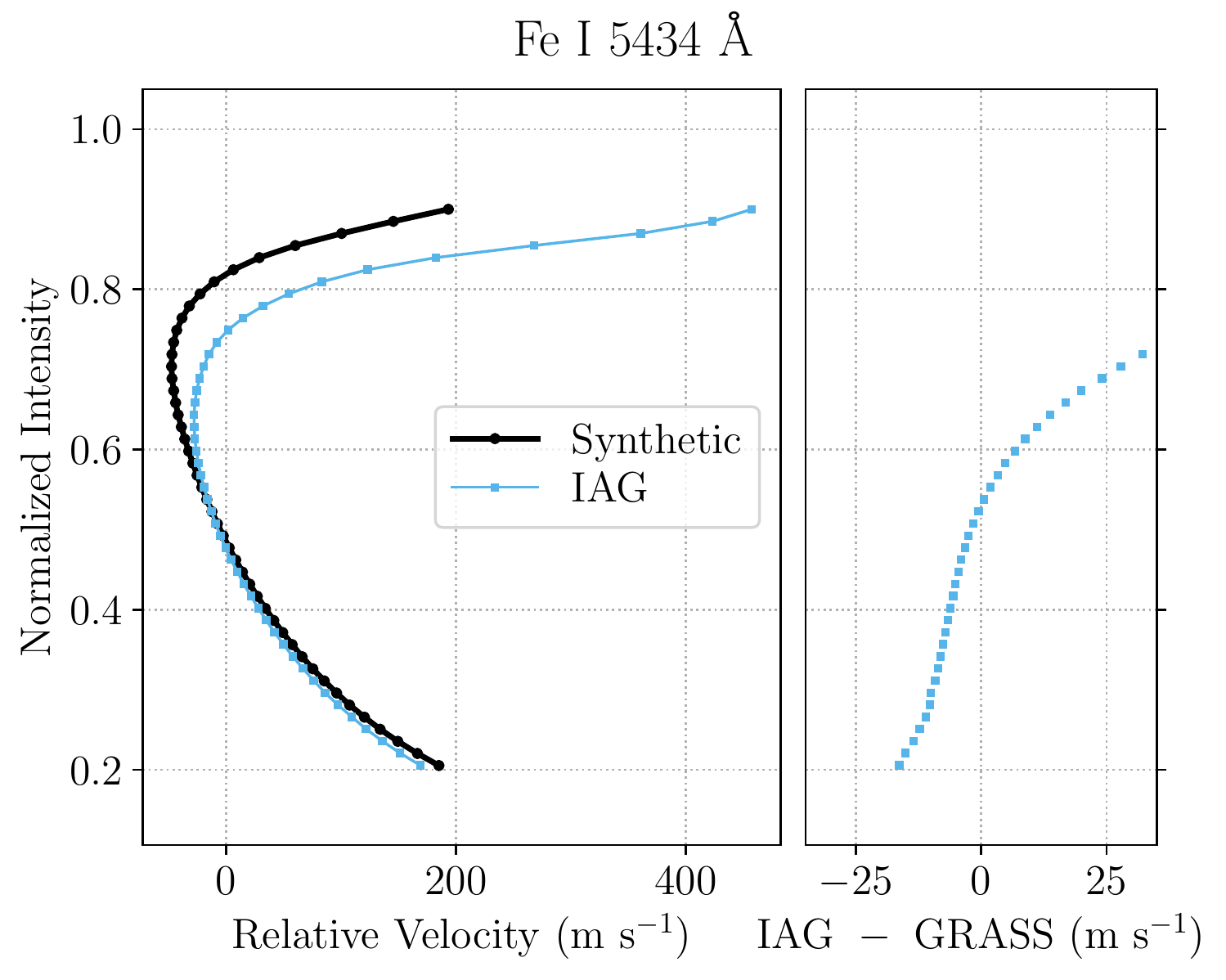}{0.45\textwidth}{}}
\caption{Same as previous figure.}
\label{fig:iag_bis_three}  
\end{figure*}

\begin{figure*}
\gridline{\fig{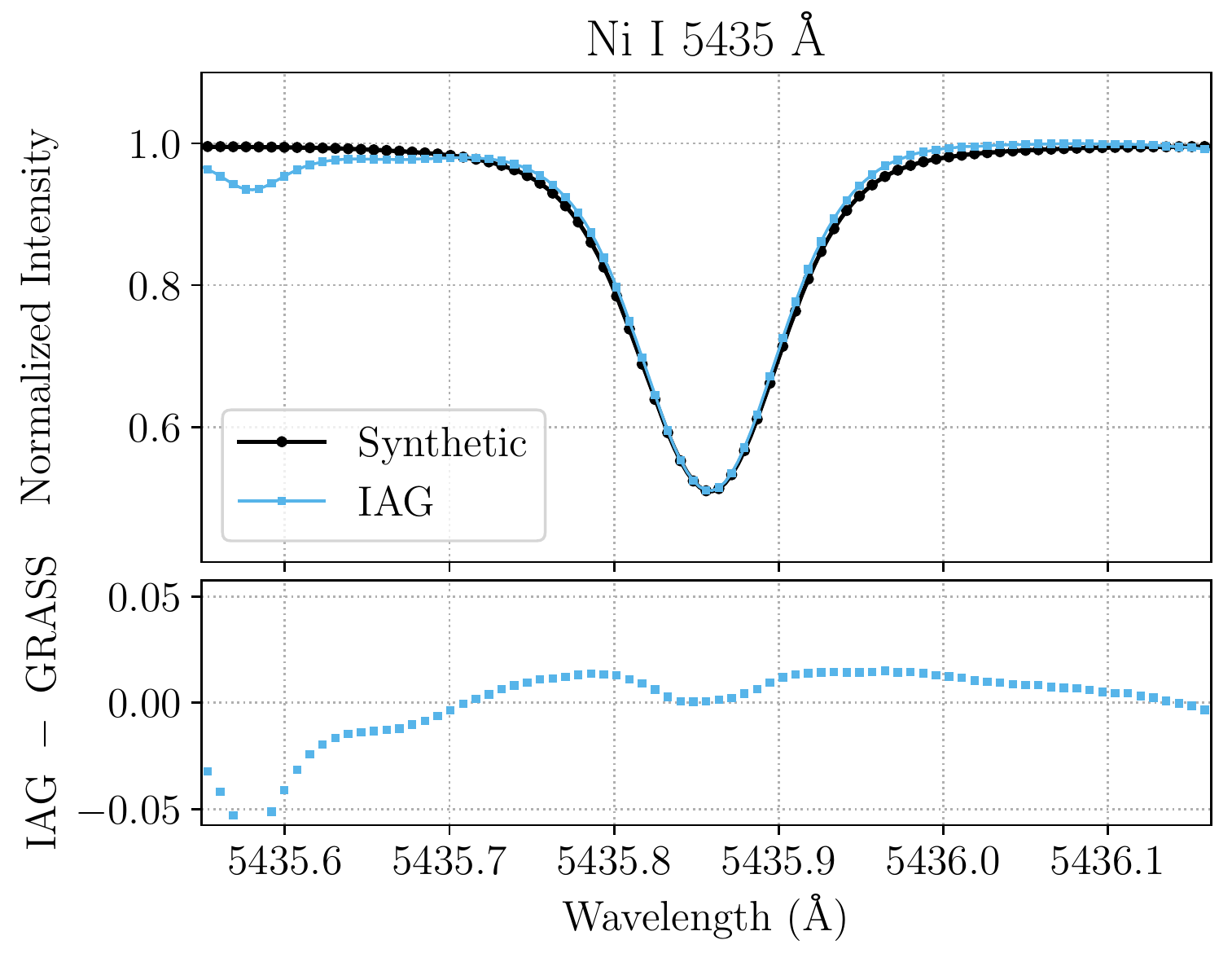}{0.455\textwidth}{}
          \fig{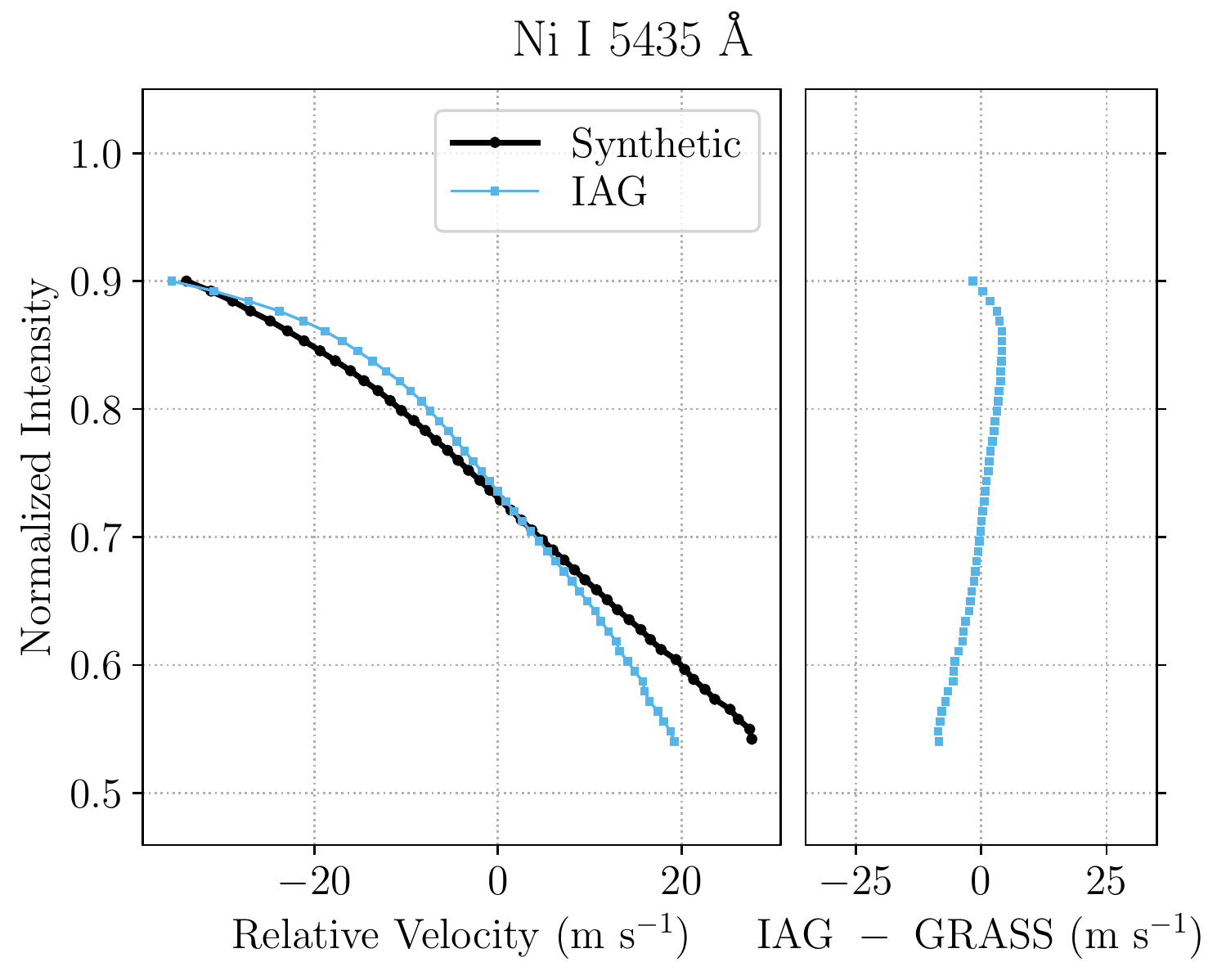}{0.45\textwidth}{}}
\gridline{\fig{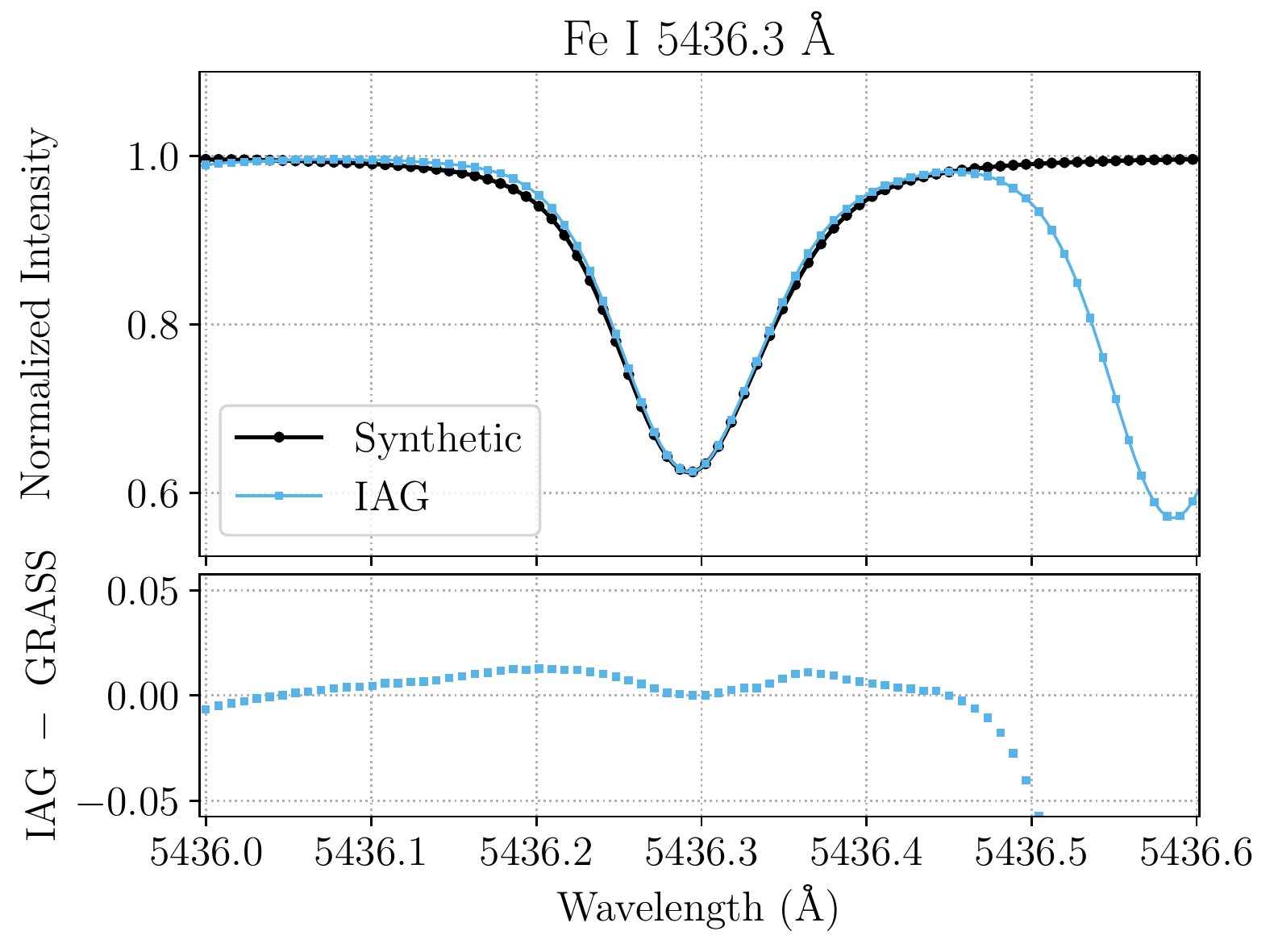}{0.455\textwidth}{}
          \fig{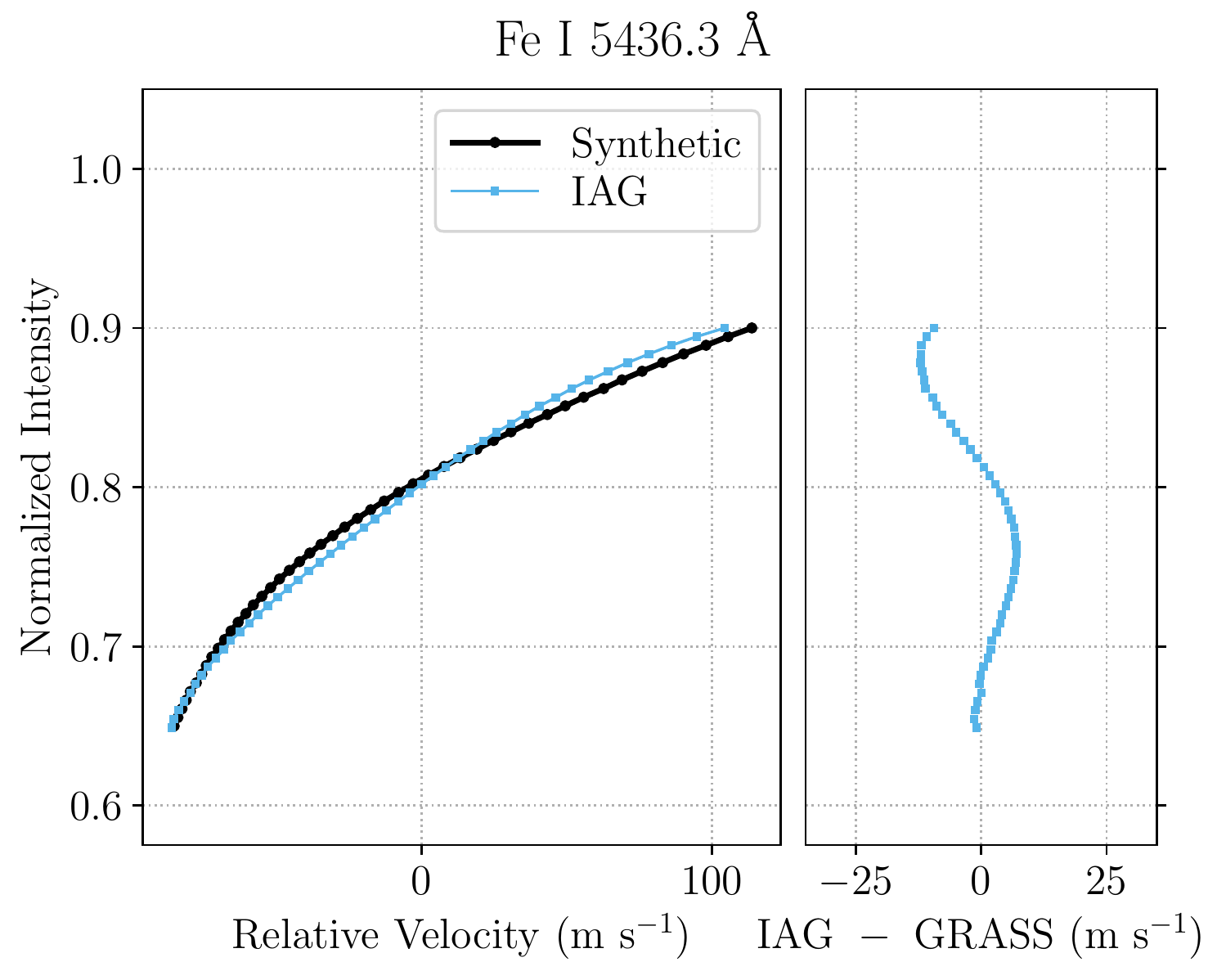}{0.45\textwidth}{}}
\gridline{\fig{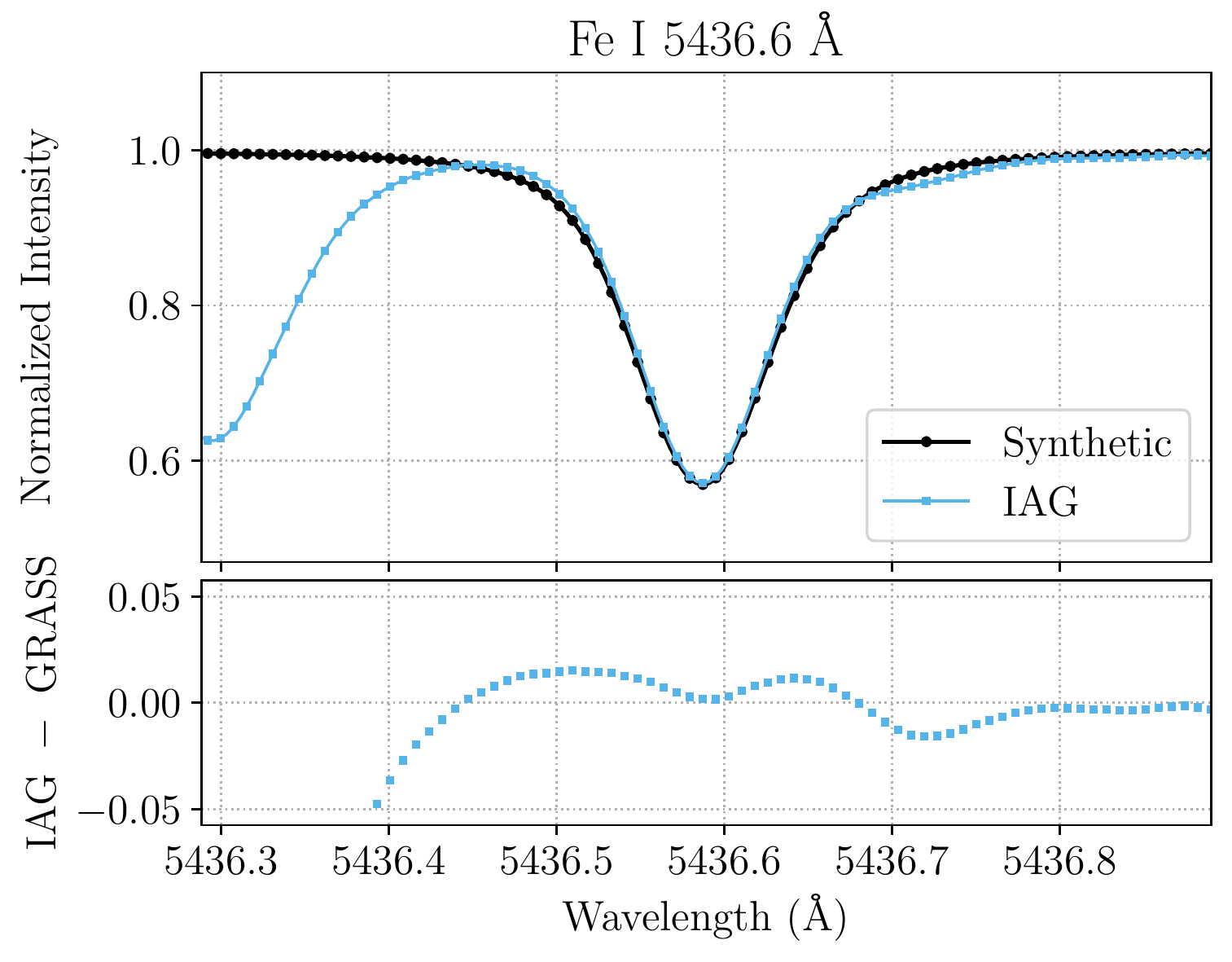}{0.455\textwidth}{}
          \fig{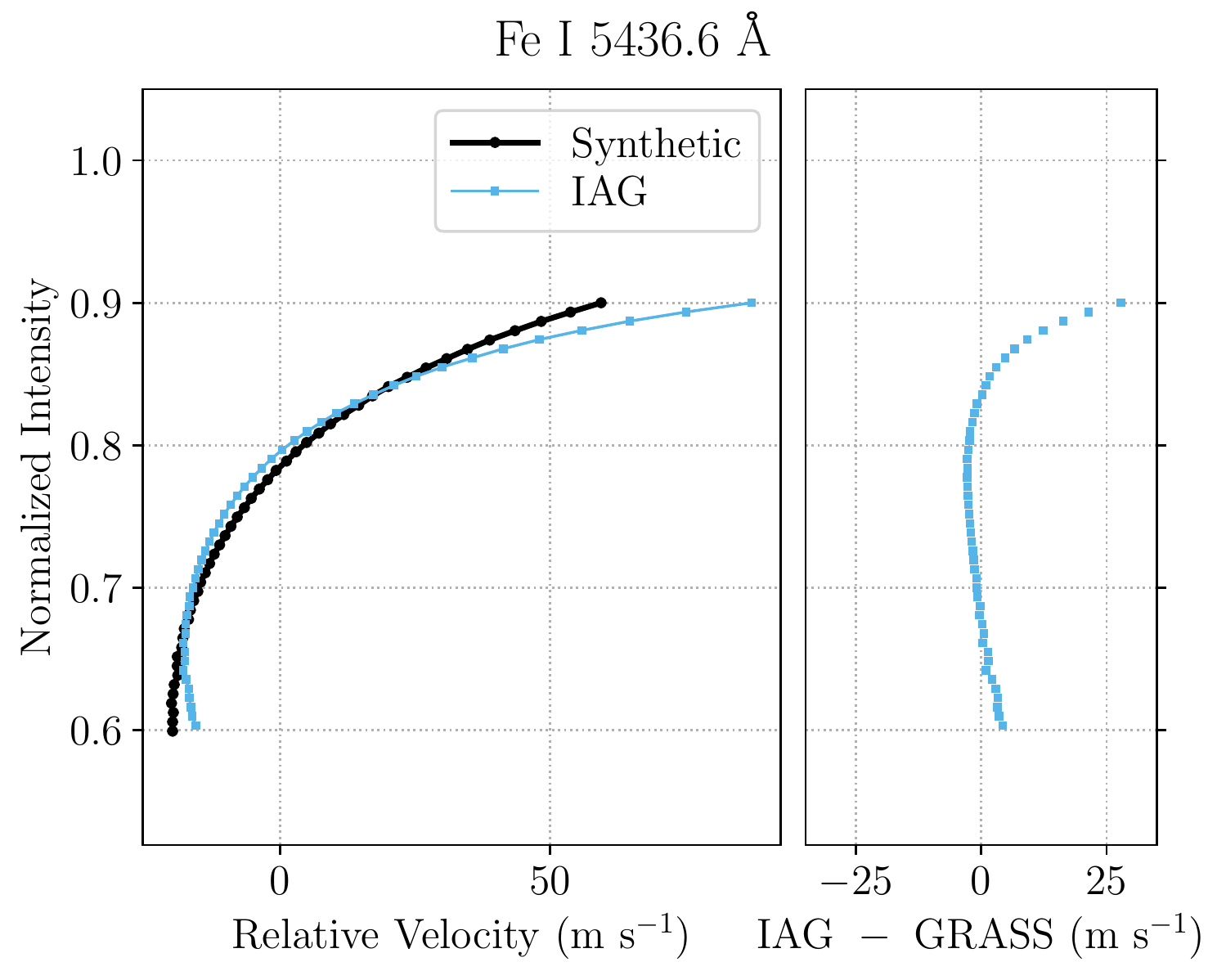}{0.45\textwidth}{}}
\caption{Same as previous figure.}
\label{fig:iag_bis_four}  
\end{figure*}

\begin{figure*}
\gridline{\fig{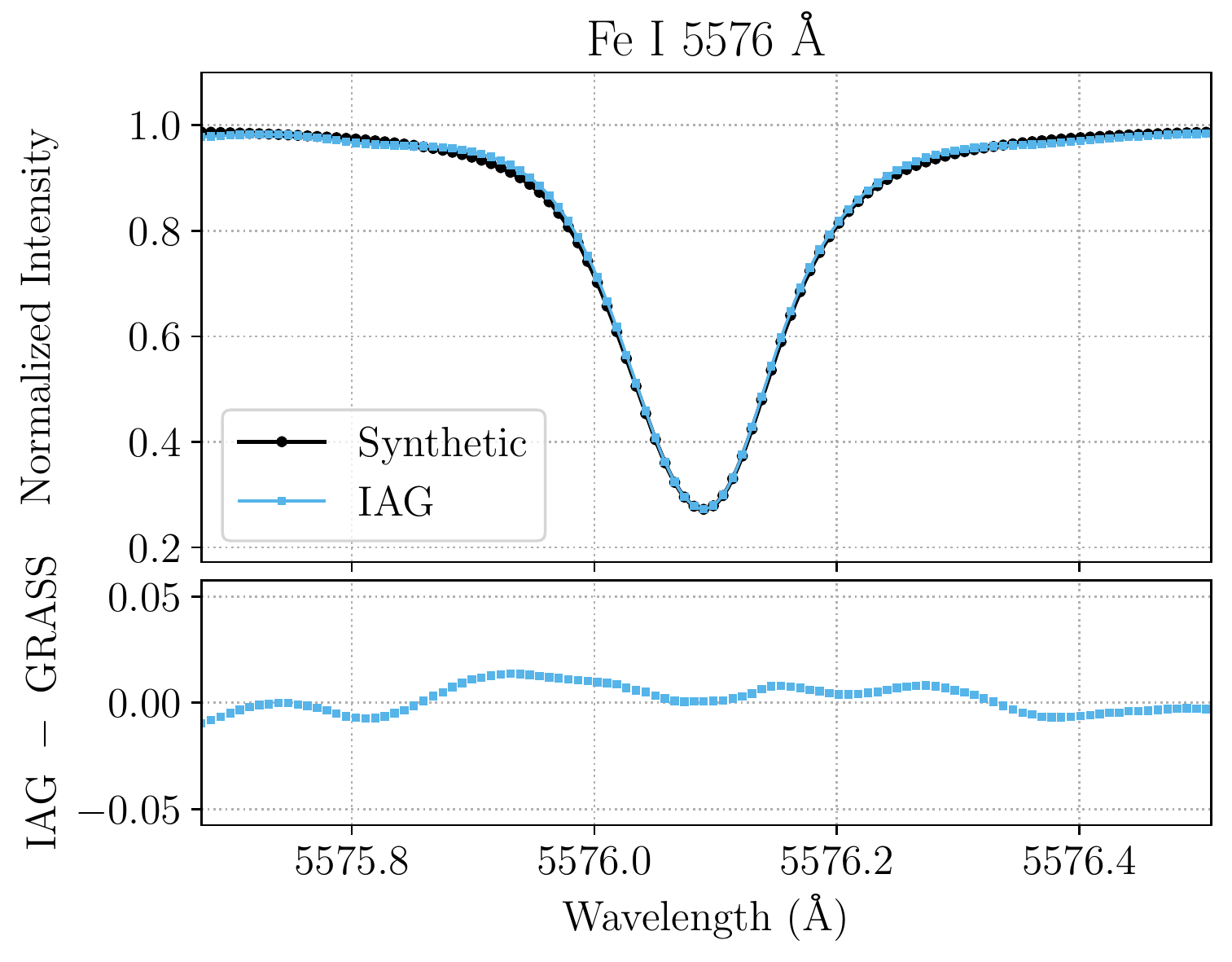}{0.455\textwidth}{}
          \fig{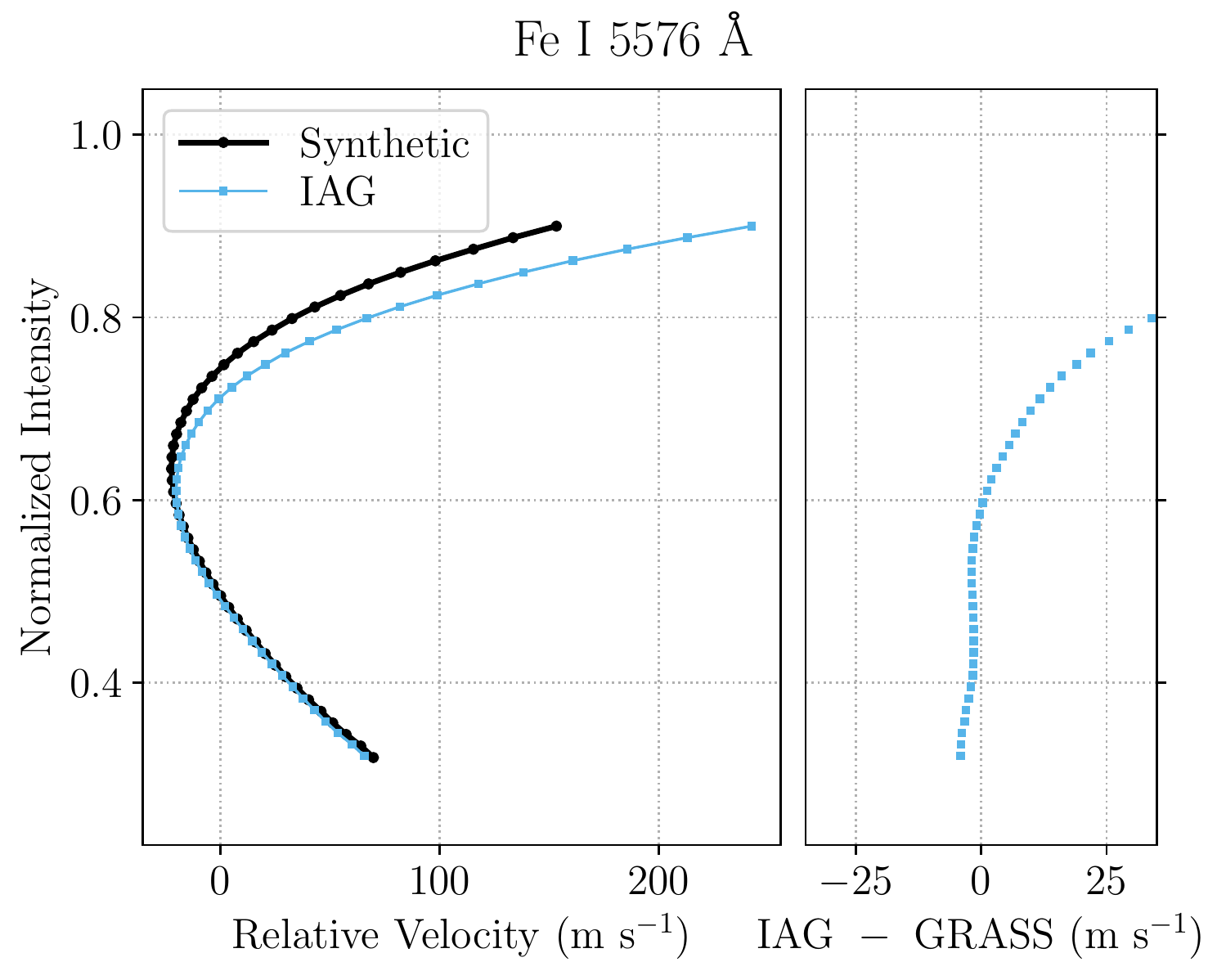}{0.45\textwidth}{}}
\gridline{\fig{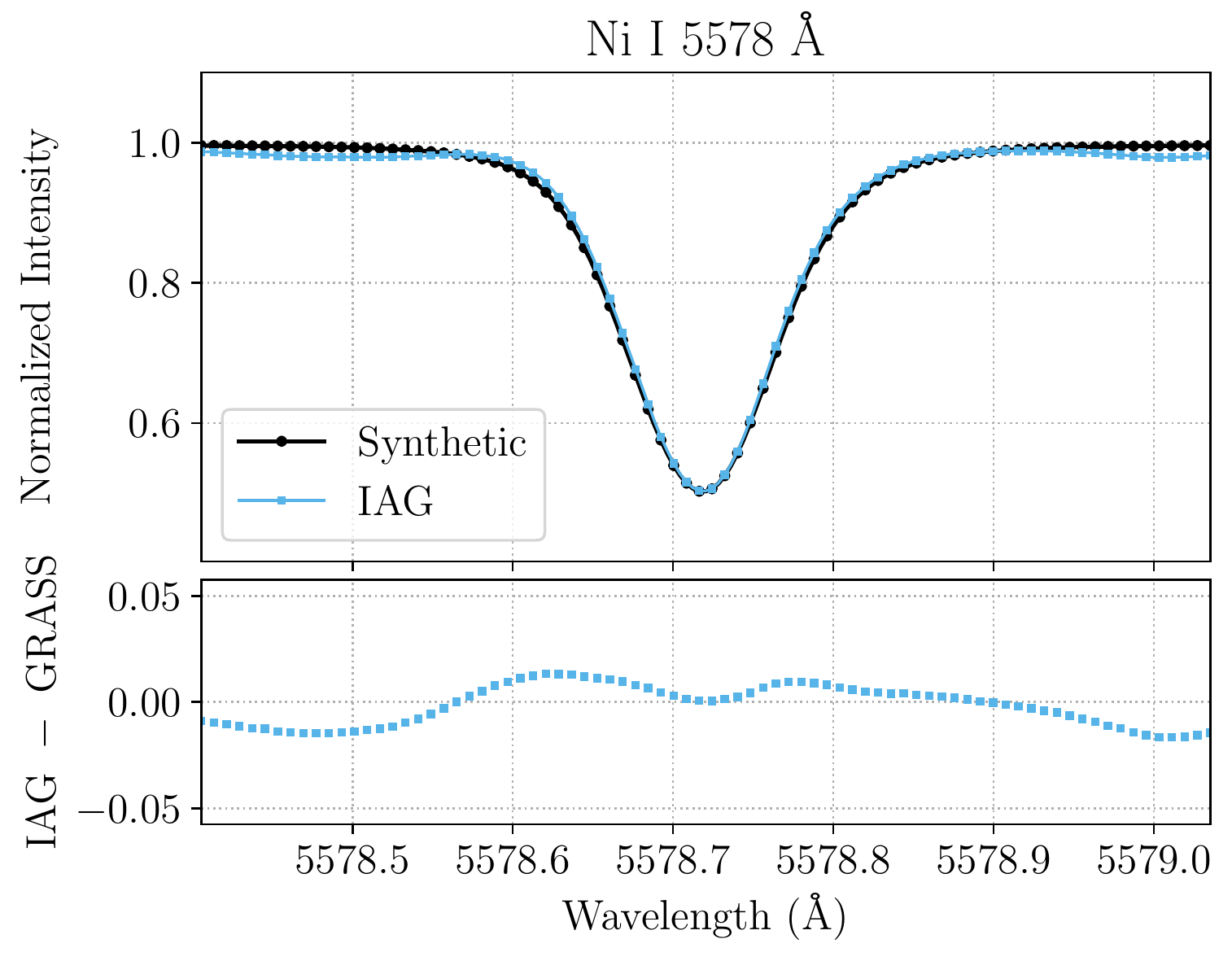}{0.455\textwidth}{}
          \fig{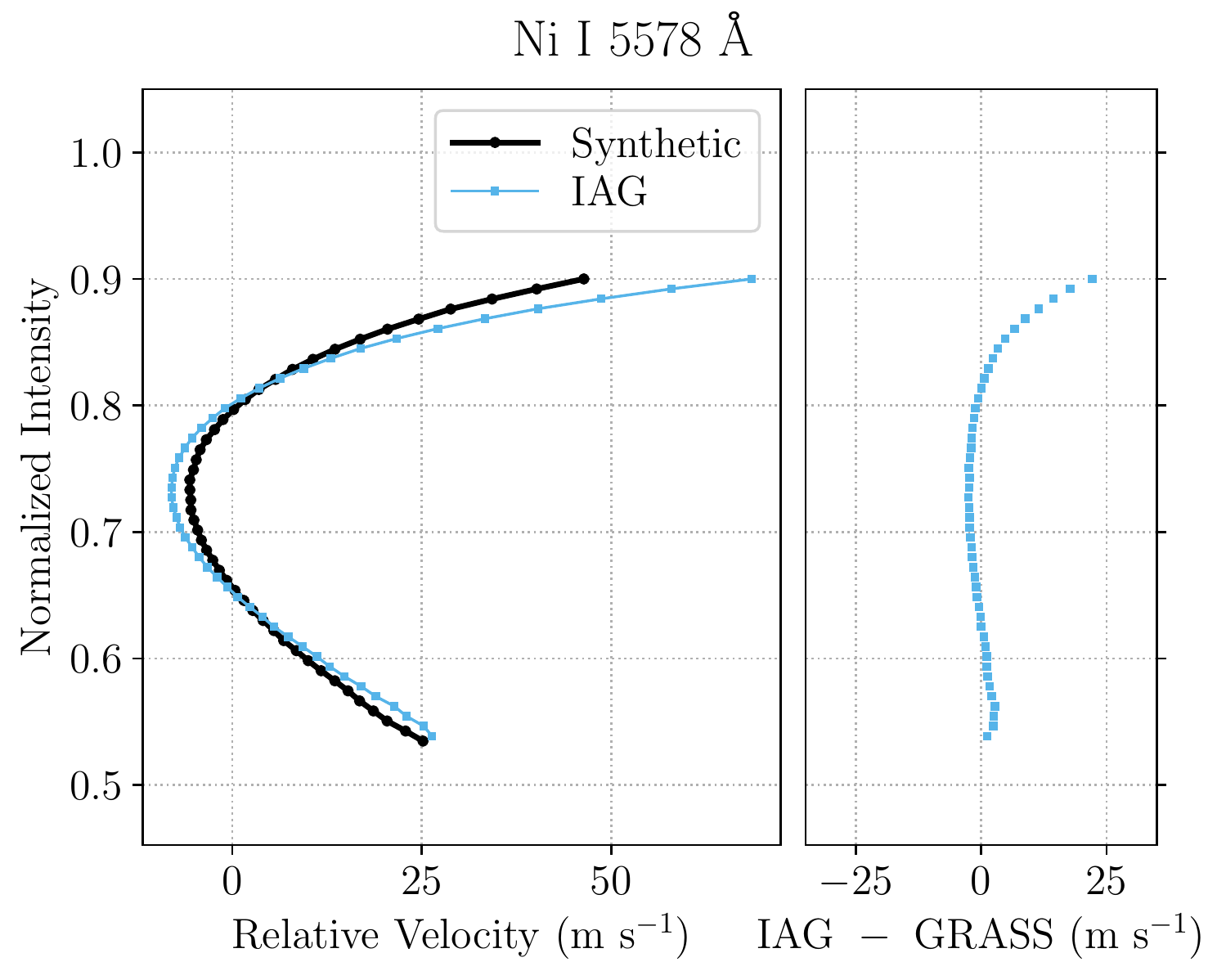}{0.45\textwidth}{}}
\gridline{\fig{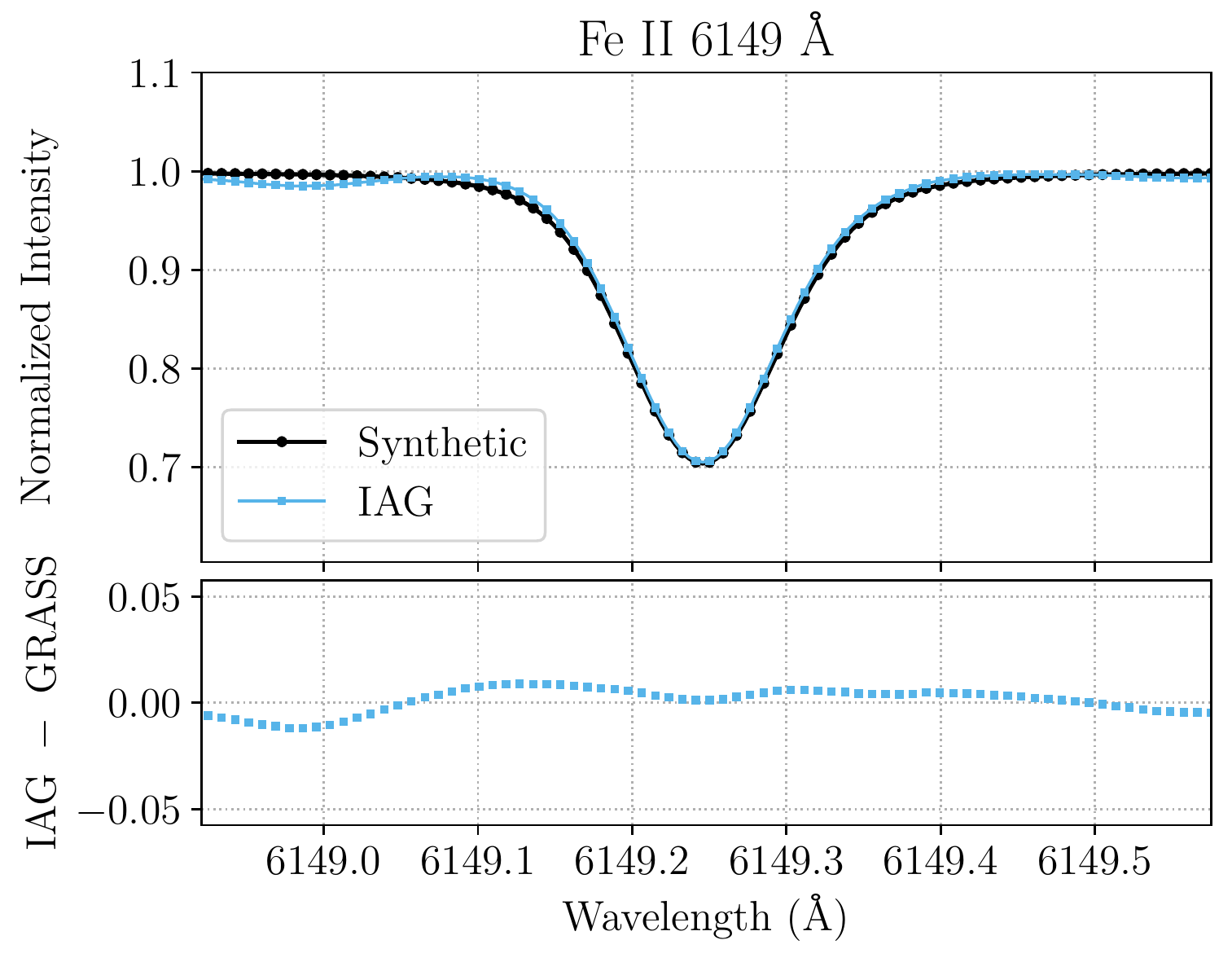}{0.455\textwidth}{}
          \fig{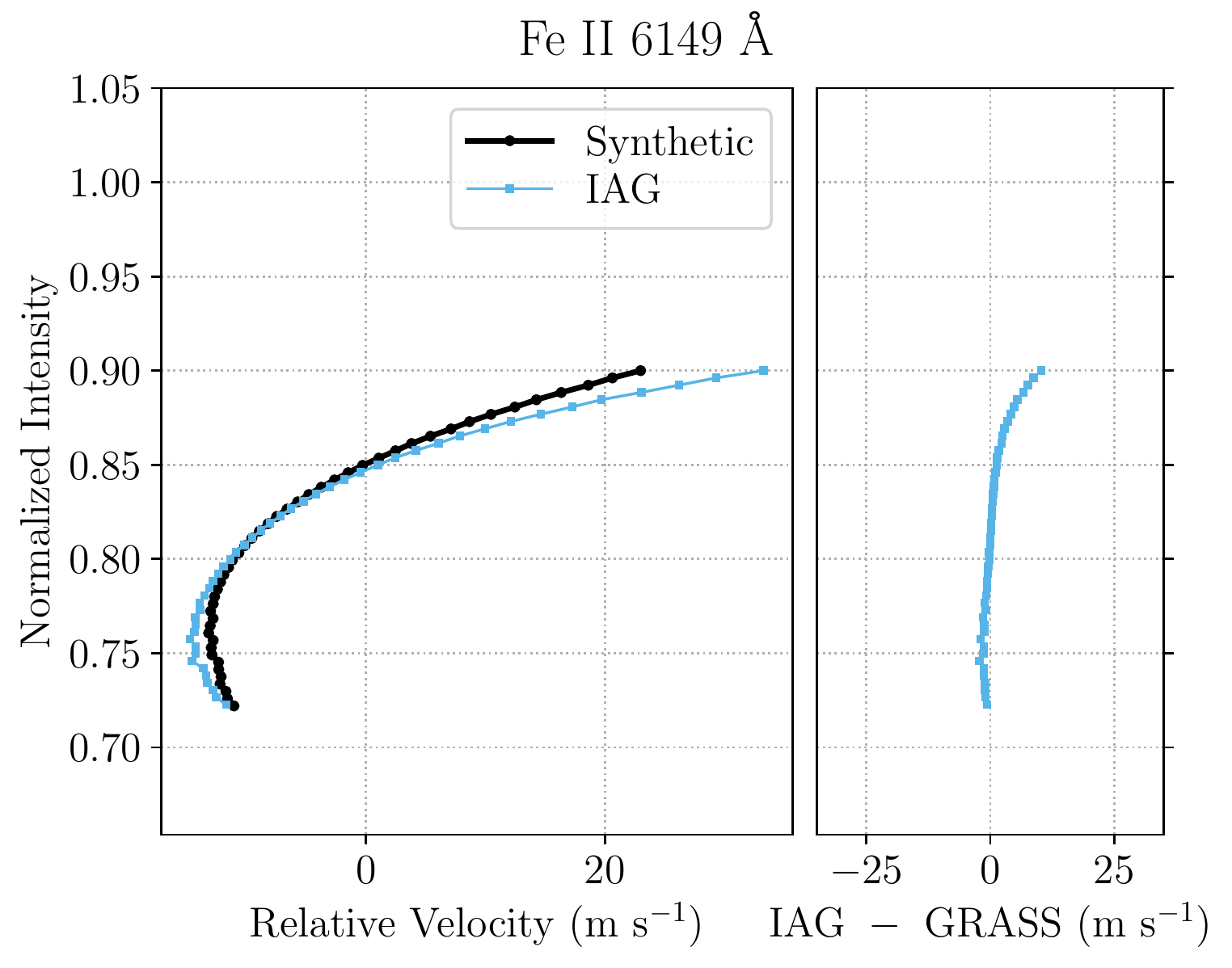}{0.45\textwidth}{}}
\caption{Same as previous figure.}
\label{fig:iag_bis_five}  
\end{figure*}

\begin{figure*}
\gridline{\fig{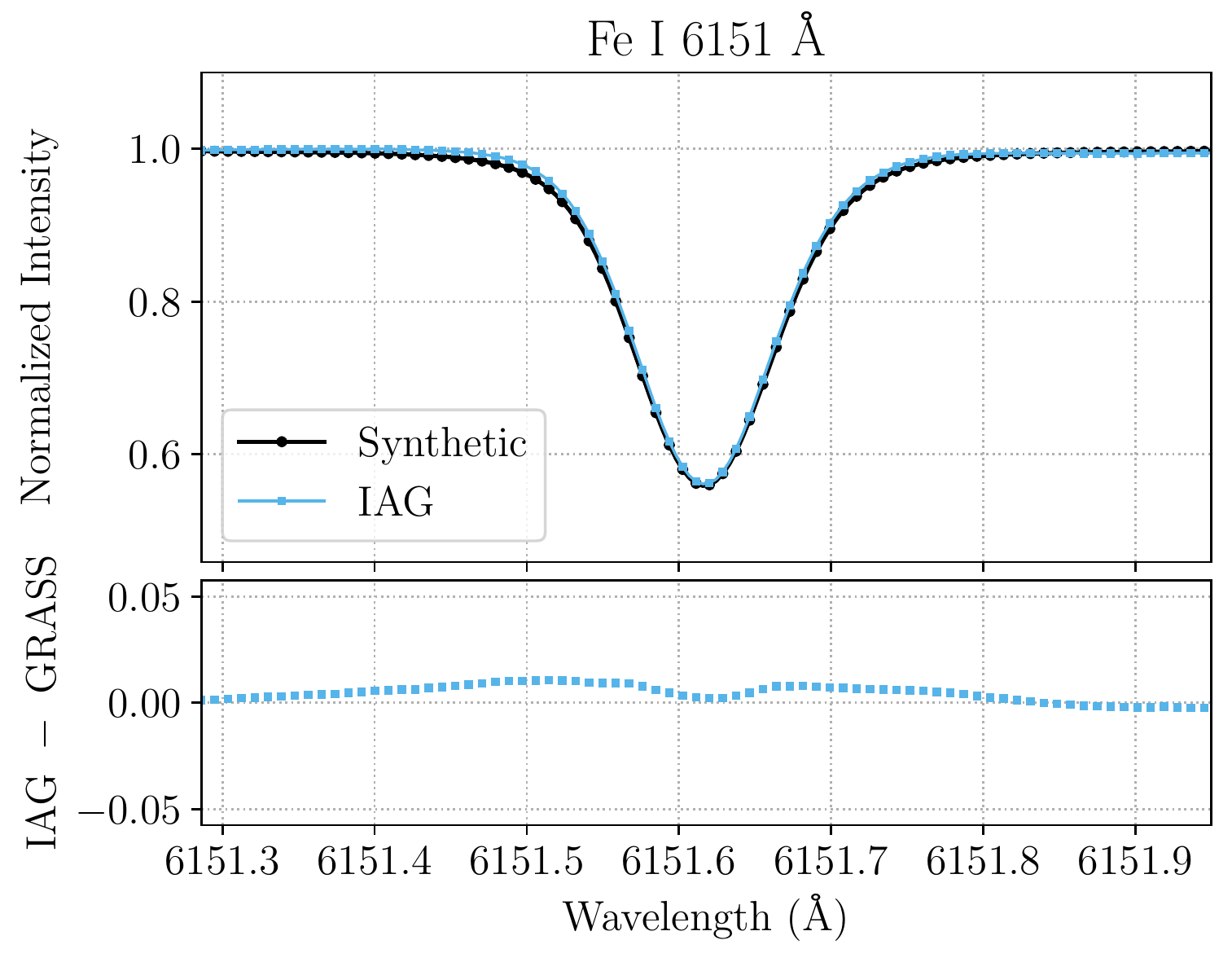}{0.455\textwidth}{}
          \fig{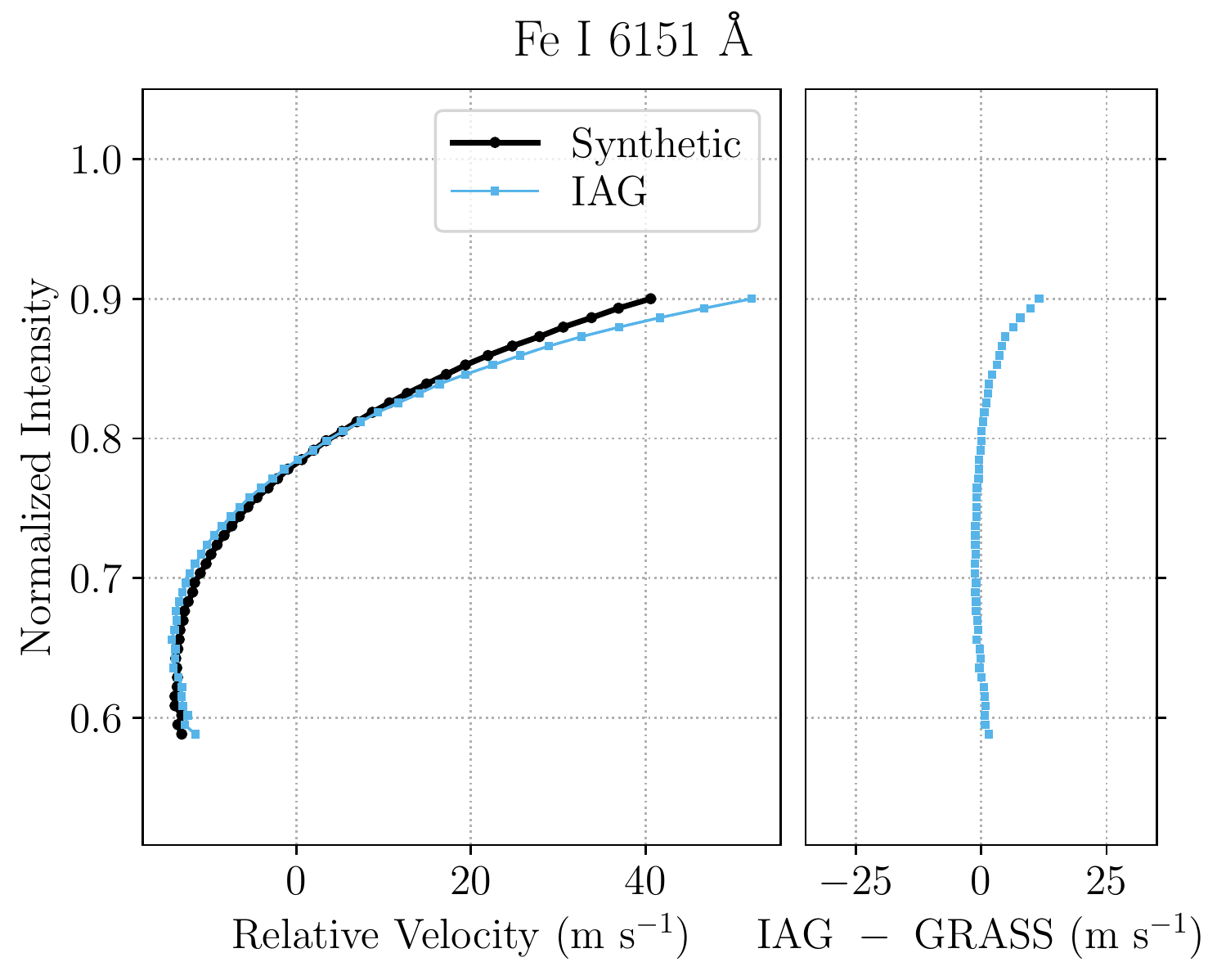}{0.45\textwidth}{}}
\gridline{\fig{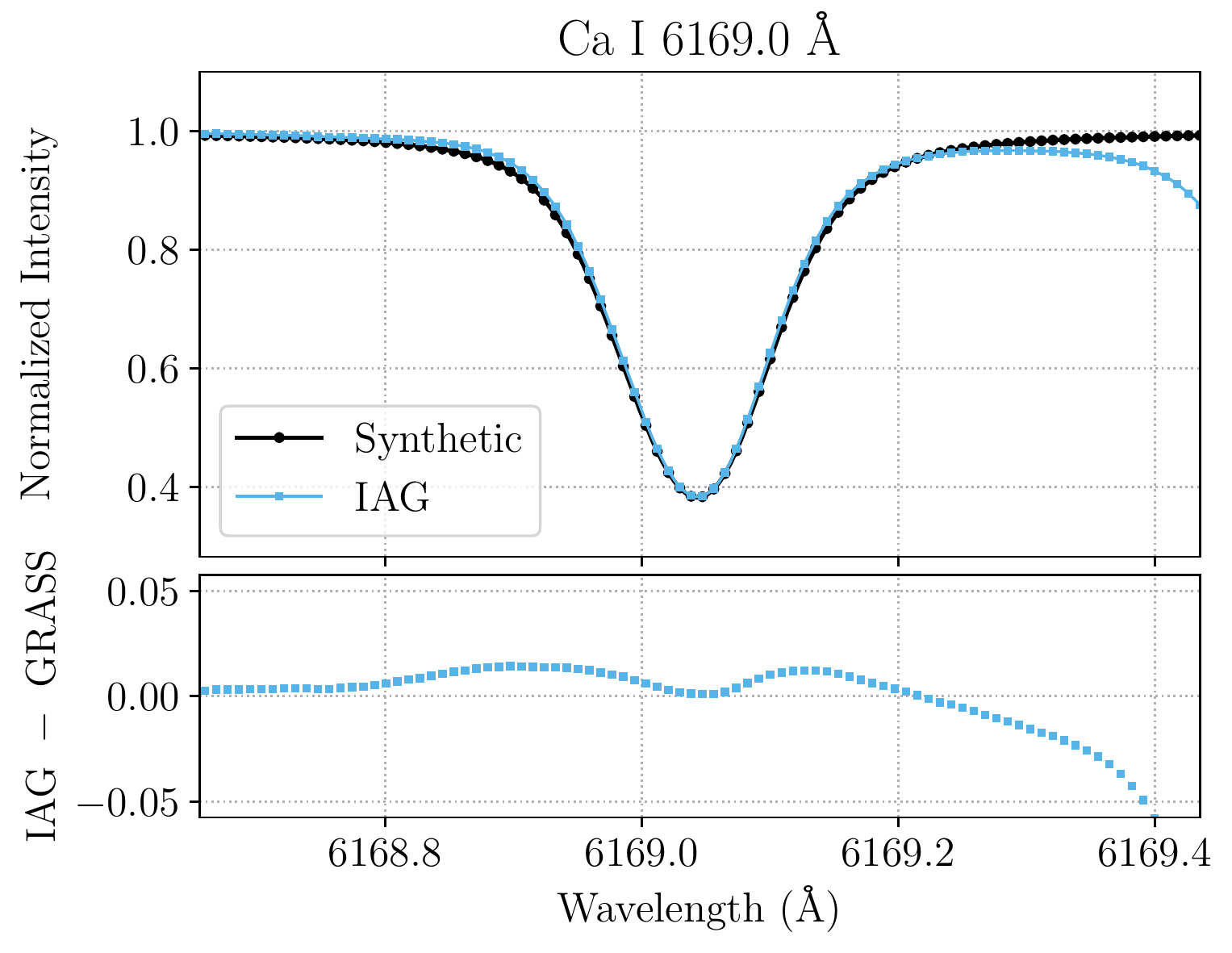}{0.455\textwidth}{}
          \fig{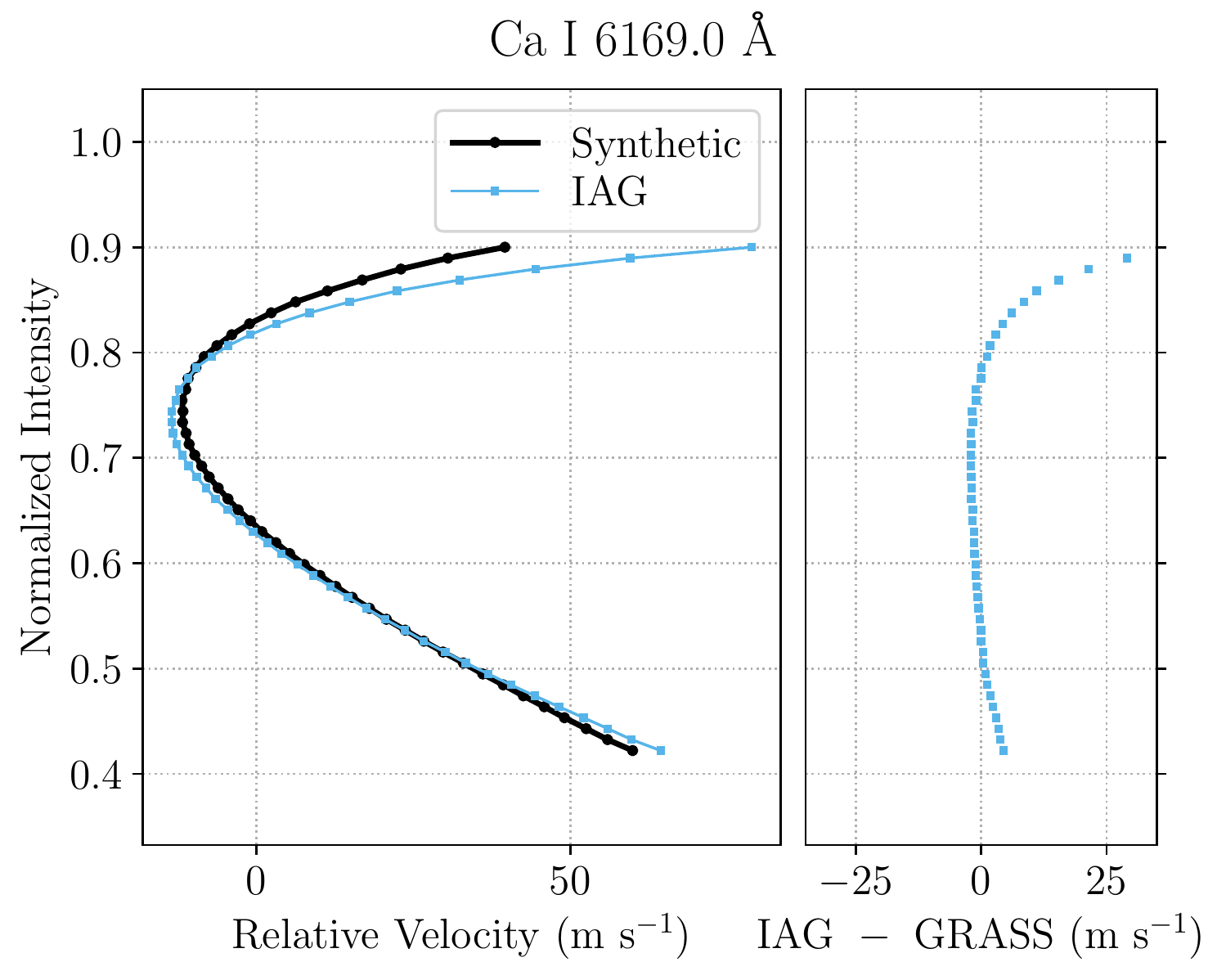}{0.45\textwidth}{}}
\gridline{\fig{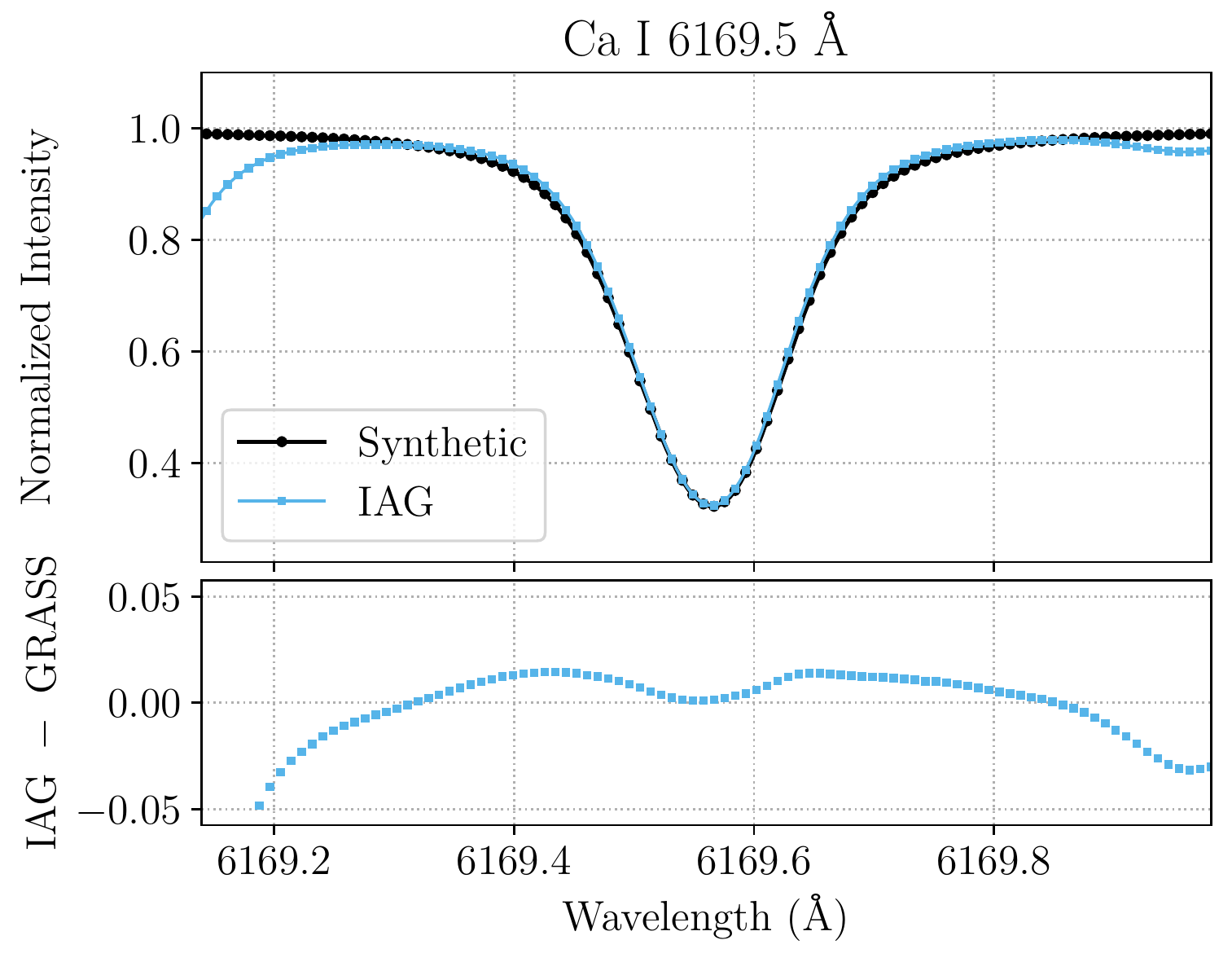}{0.455\textwidth}{}
          \fig{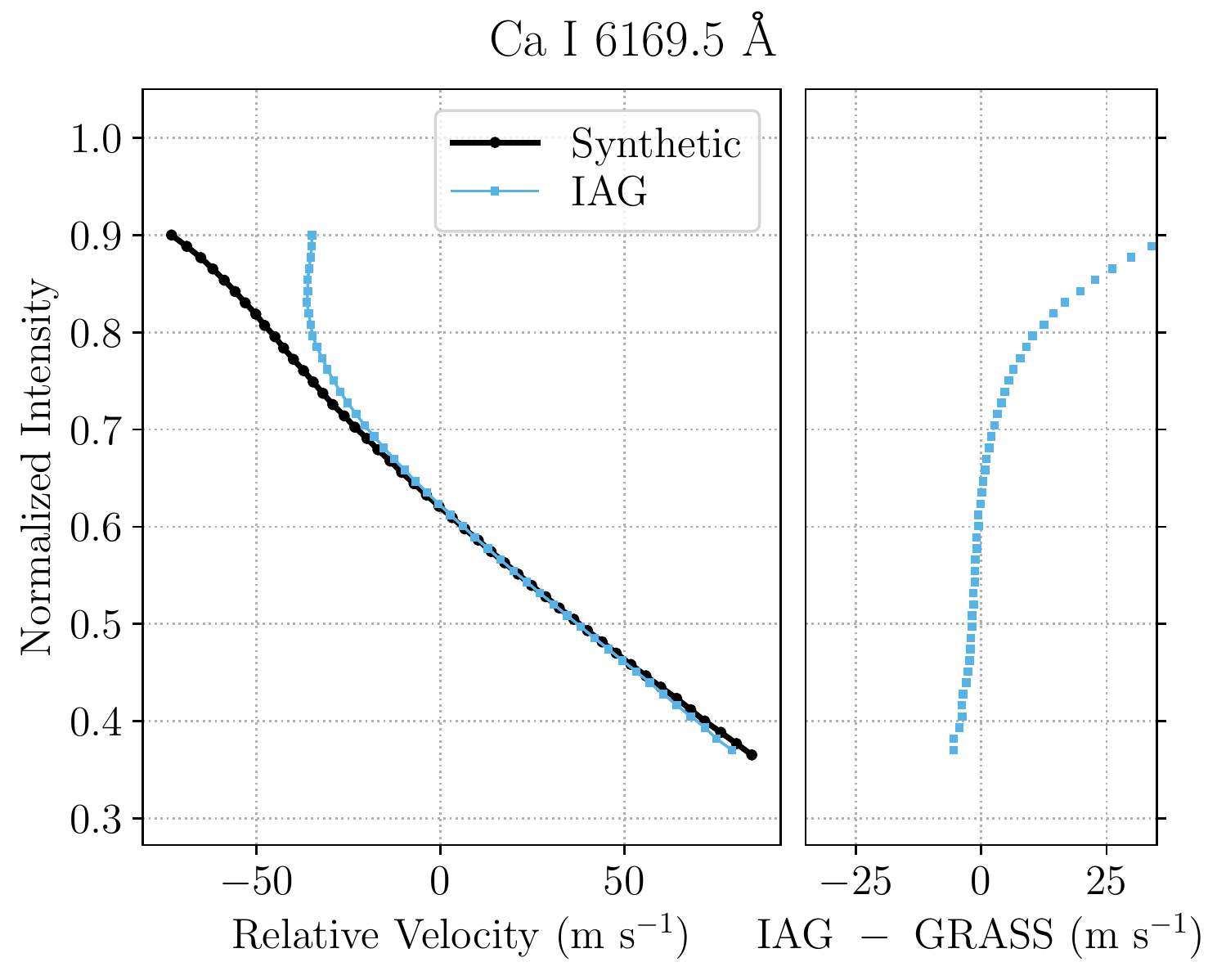}{0.45\textwidth}{}}
\caption{Same as previous figure.}
\label{fig:iag_bis_six}  
\end{figure*}

\begin{figure*}
\gridline{\fig{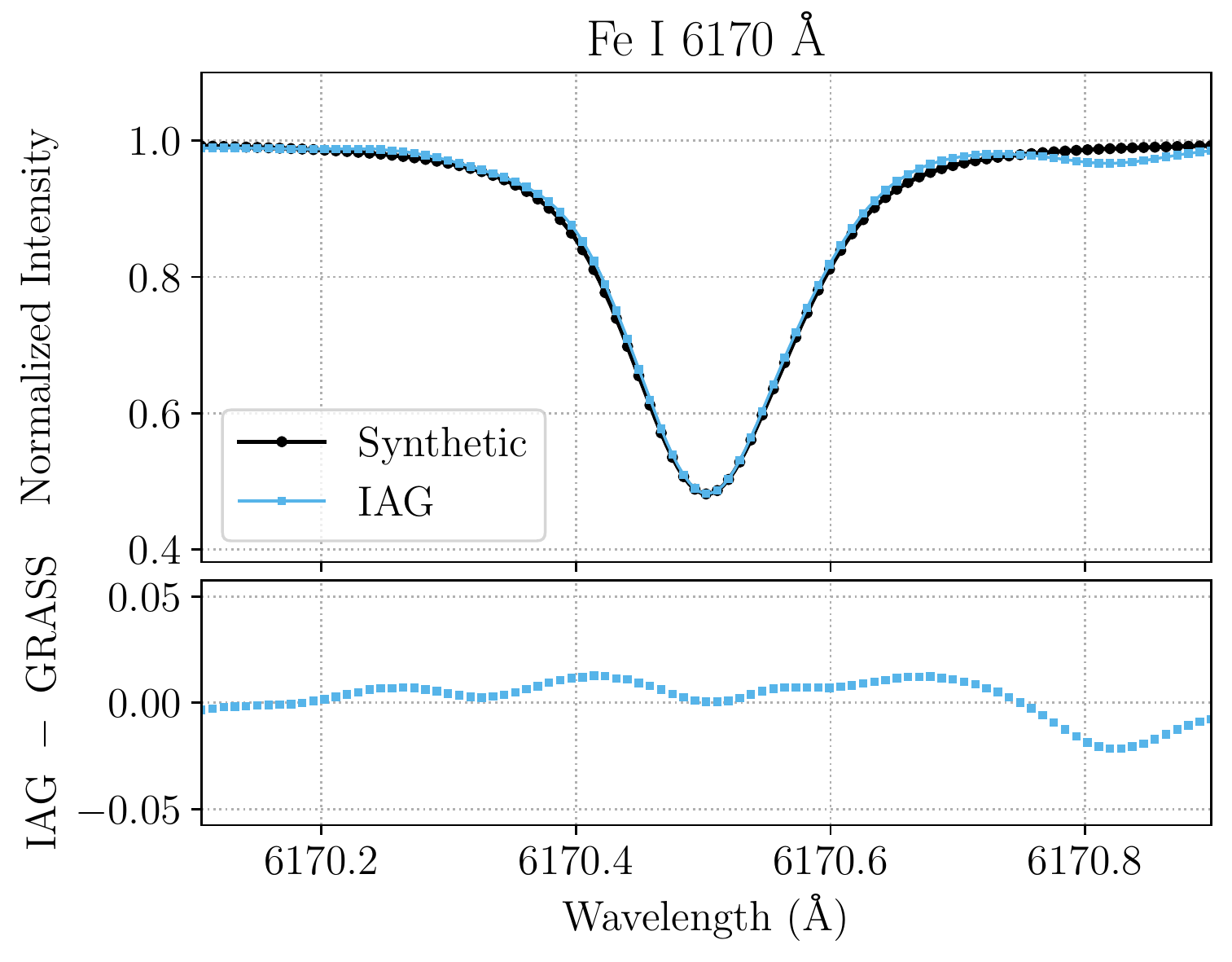}{0.455\textwidth}{}
          \fig{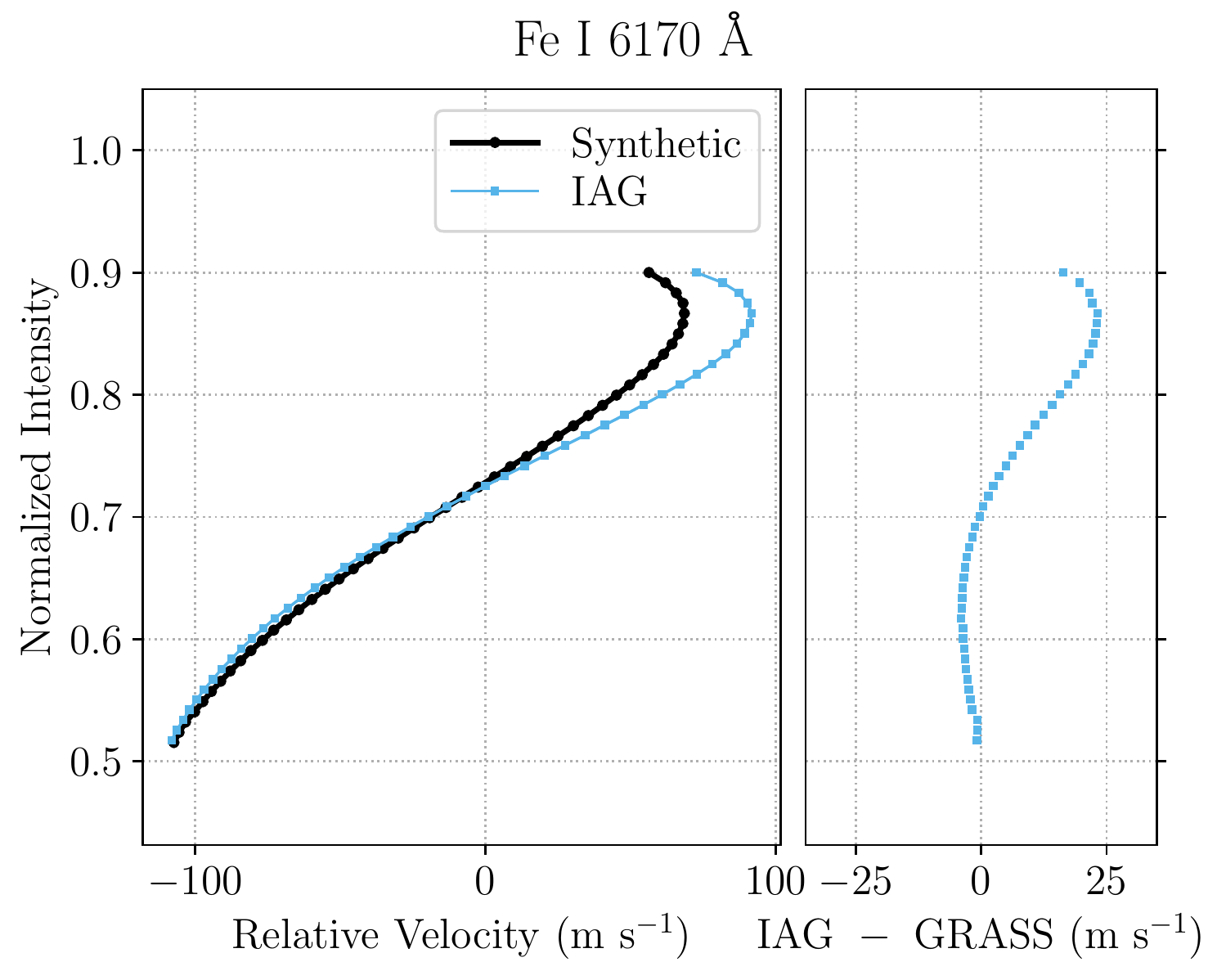}{0.45\textwidth}{}}
\gridline{\fig{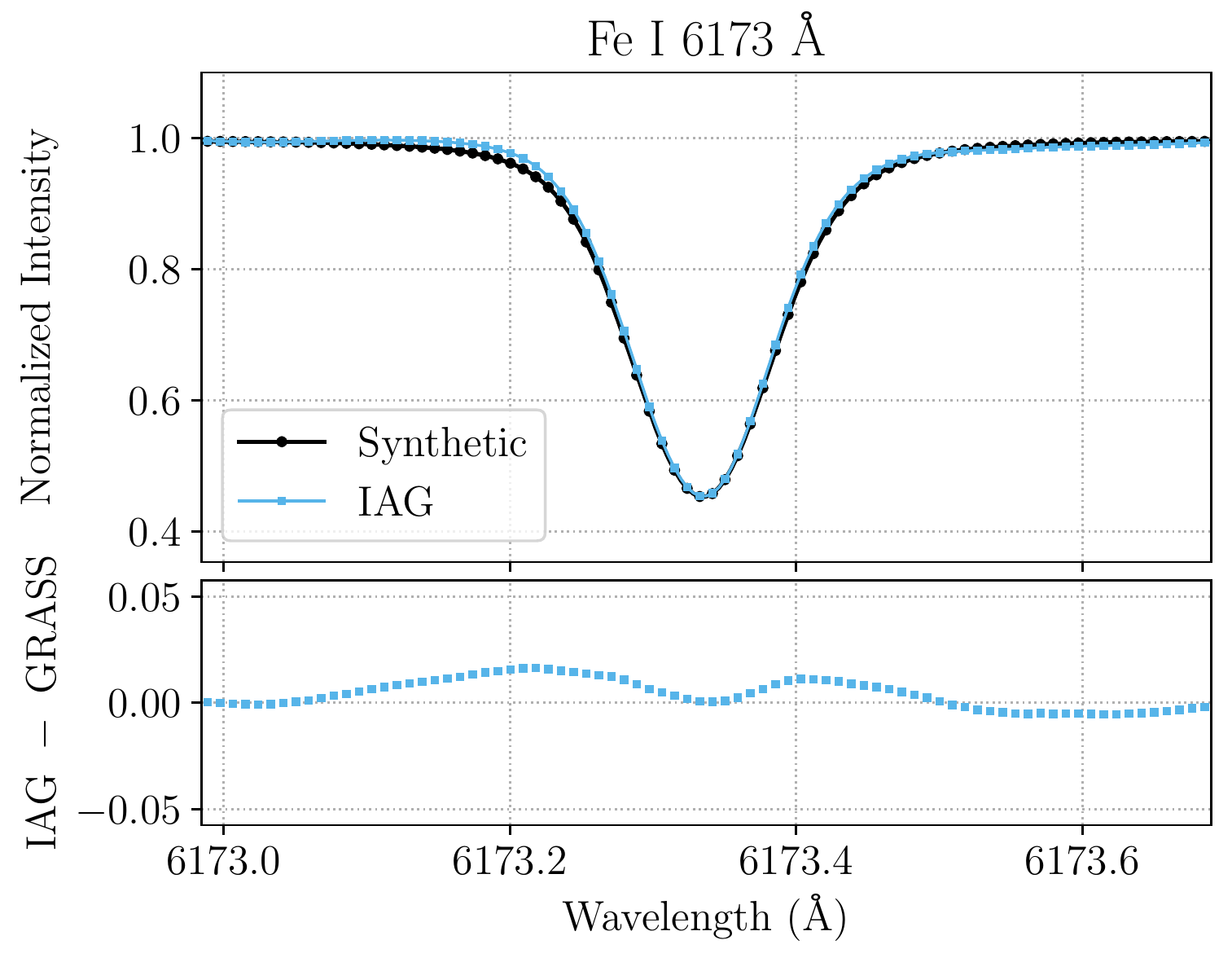}{0.455\textwidth}{}
          \fig{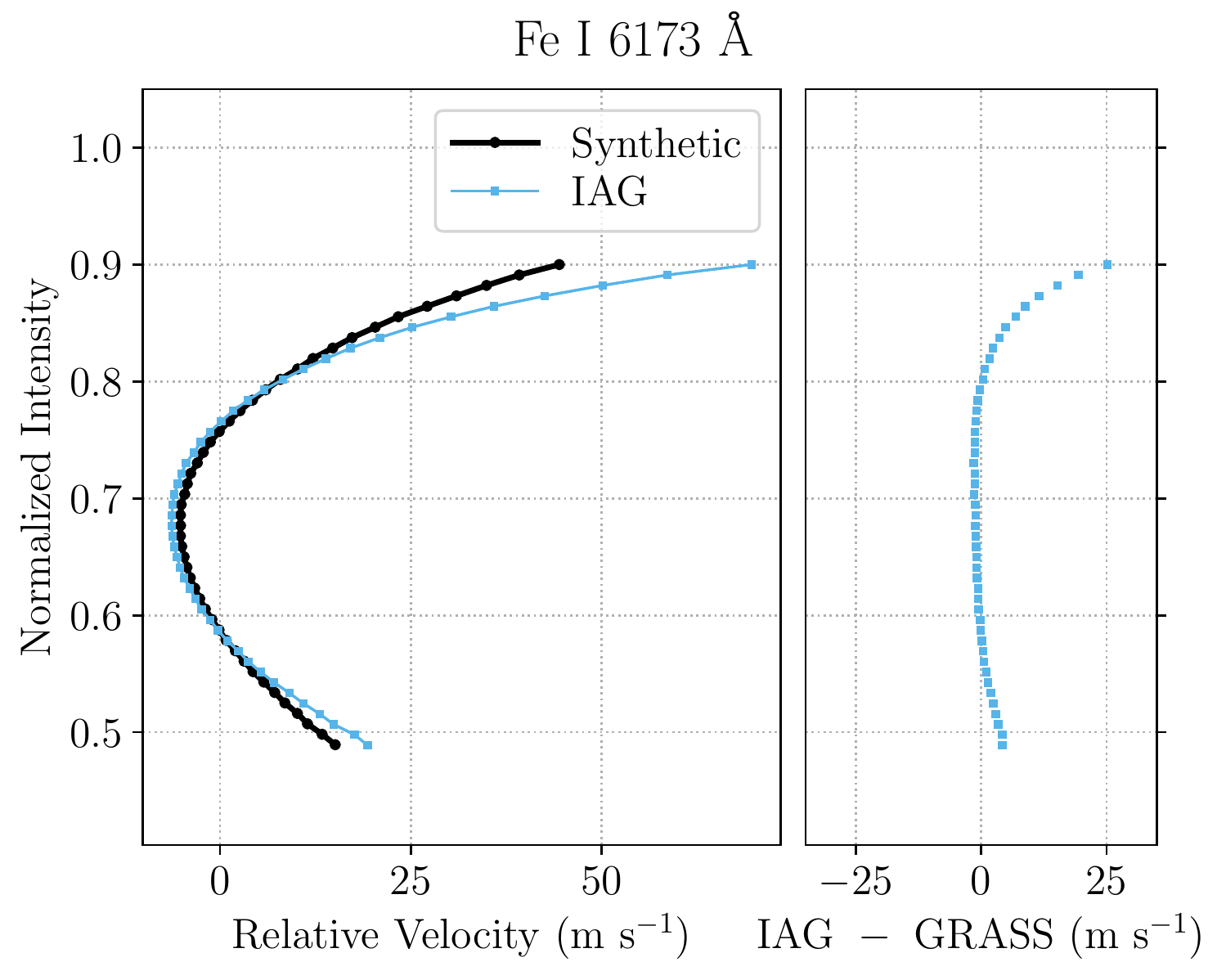}{0.45\textwidth}{}}
\gridline{\fig{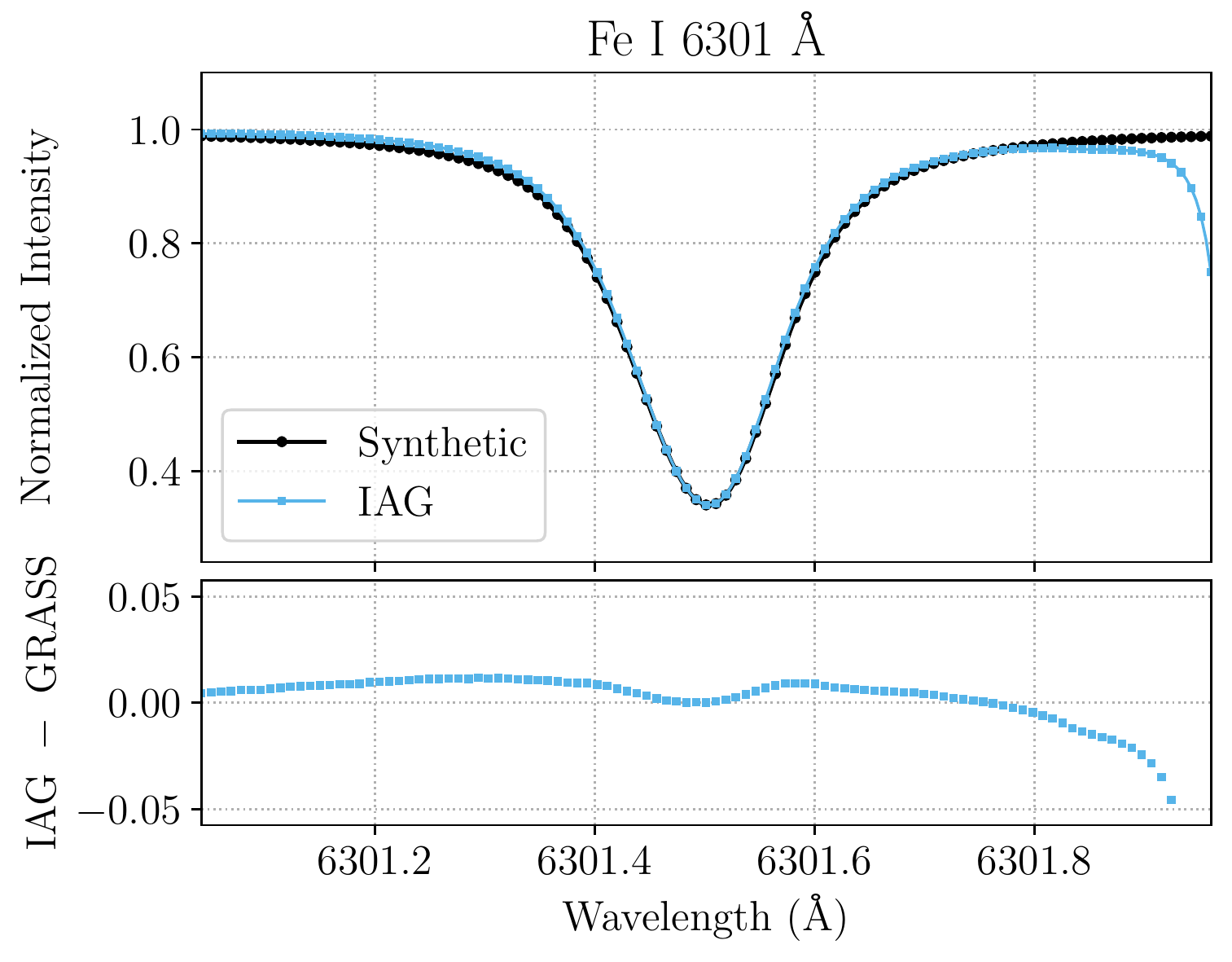}{0.455\textwidth}{}
          \fig{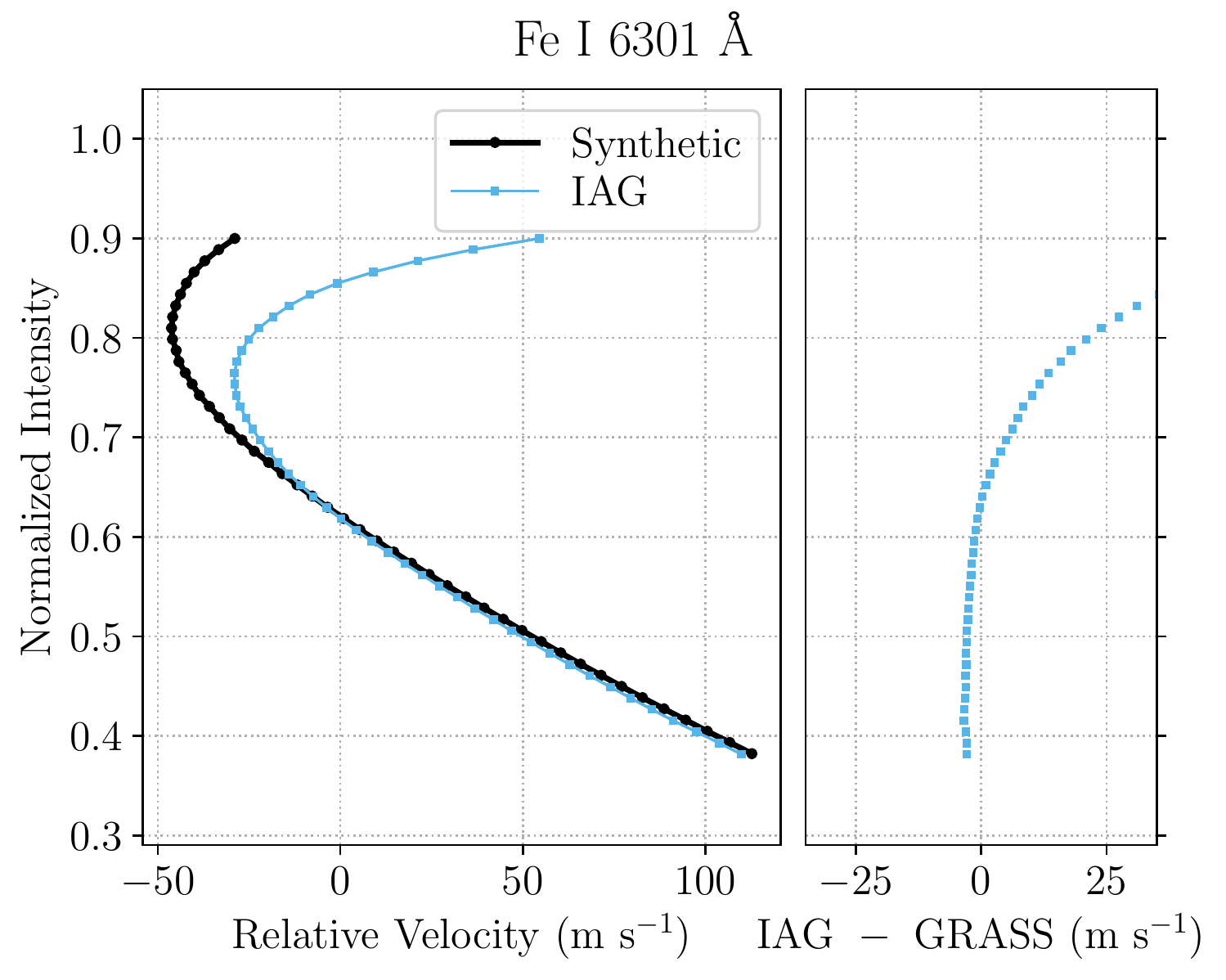}{0.45\textwidth}{}}
\caption{Same as previous figure.}
\label{fig:iag_bis_seven}  
\end{figure*}

\begin{figure*}
\gridline{\fig{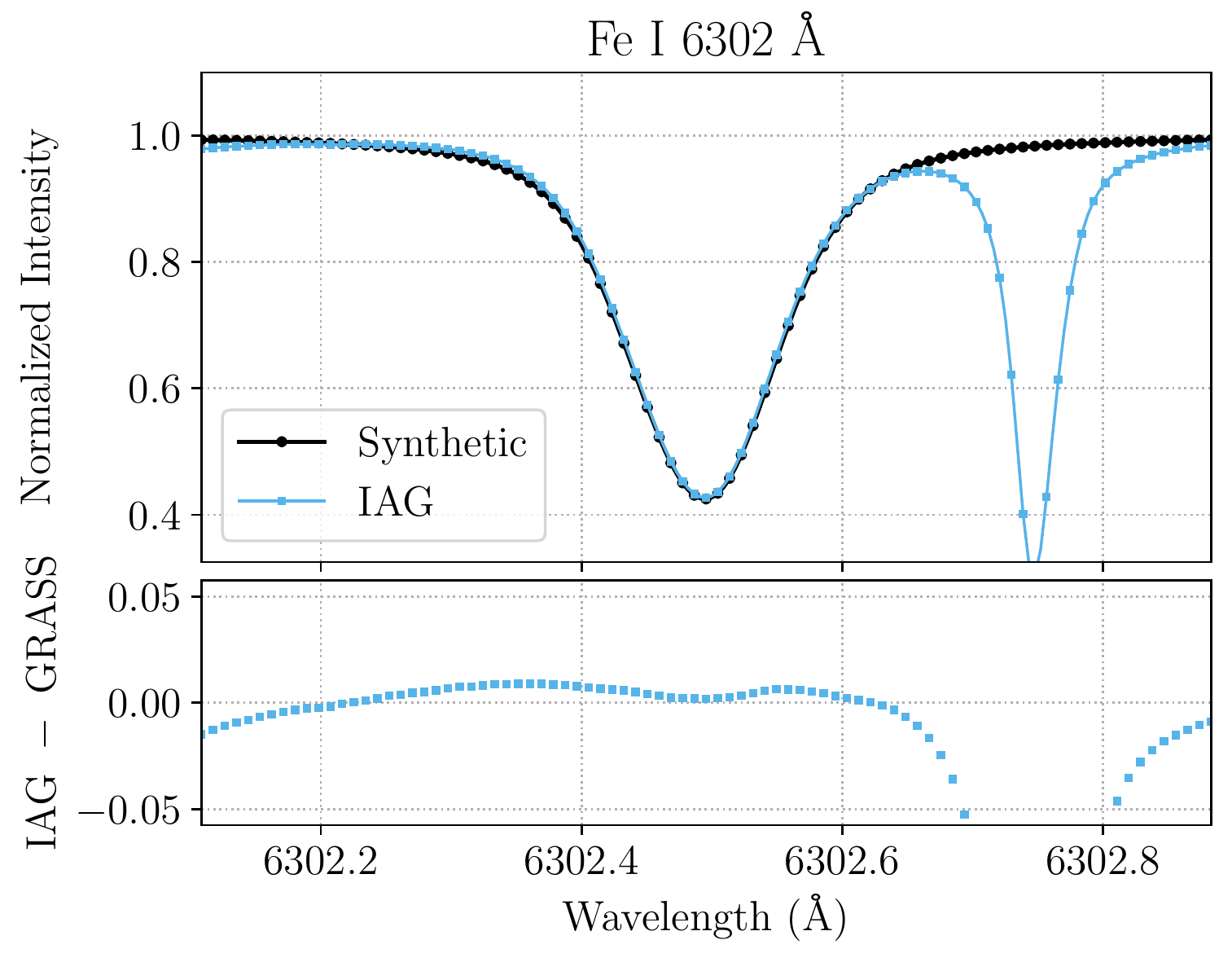}{0.455\textwidth}{}
          \fig{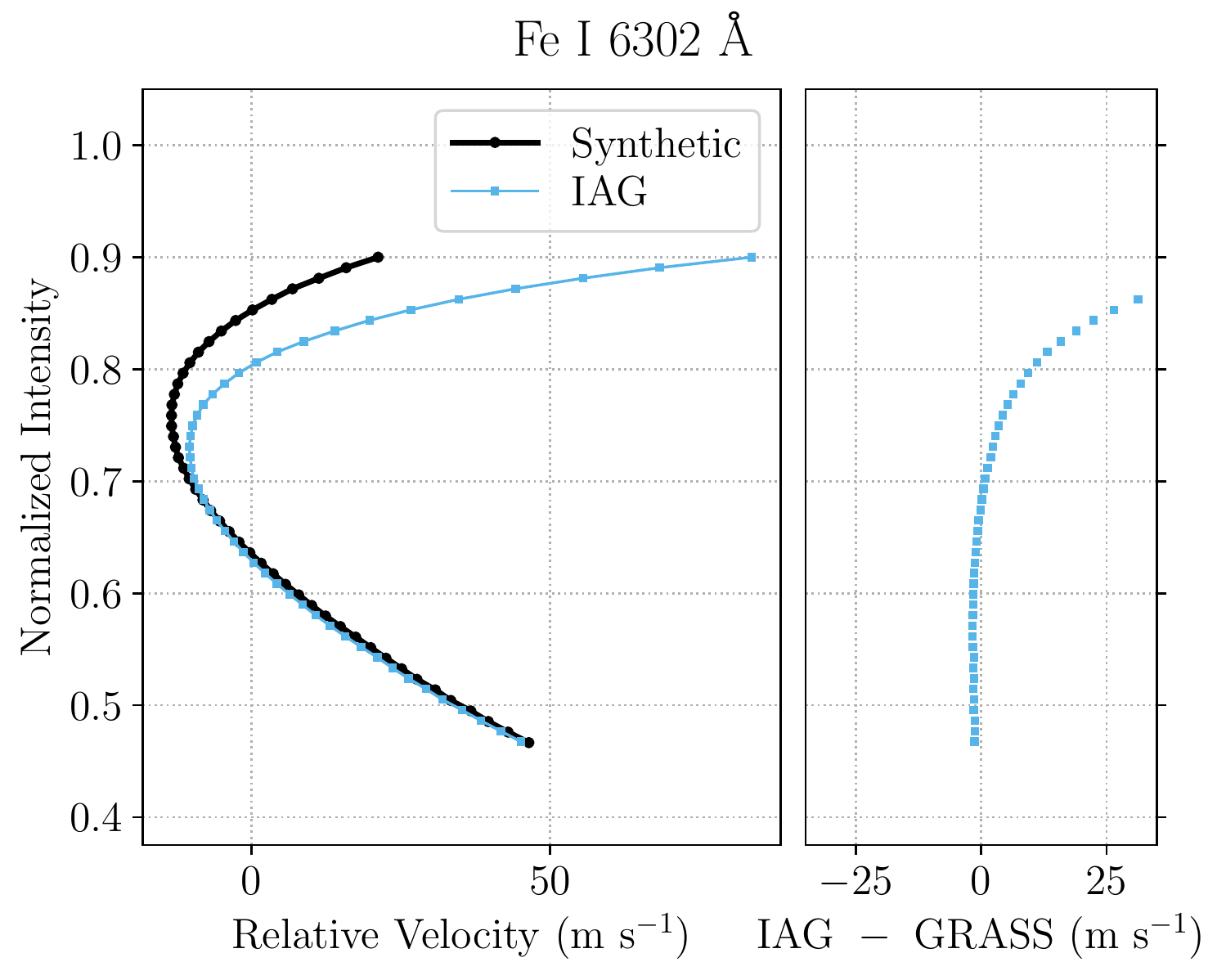}{0.45\textwidth}{}}
\caption{Same as previous figure.}
\label{fig:iag_bis_last}  
\end{figure*}

\end{document}